\documentclass[opre,nonblindrev]{informs3} 

\DoubleSpacedXI
\usepackage{amsfonts}
{\end{list}}

\def\Var{{\rm Var}}

\def\LR{{\rm LR}}
\def\TR{{\rm TruLR}}

\def\bA{{\bf A}}
\def\ba{{\bf a}}

\def\bi{{\bf i}}

\def\bX{{\bf X}}
\def\bx{{\bf x}}

\def\betheta{\boldsymbol\theta}
\def\bemu{\boldsymbol\mu}

\def\bmu{\mbox{\boldmath $\mu$}}

\def\bSigma{{\bf \Sigma}}

\def\btheta{\mbox{\boldmath $\theta$}}

\def\b1{{\bf 1}}

\def\blot{\quad {$\vcenter{\vbox{\hrule height.4pt
             \hbox{\vrule width.4pt height.9ex \kern.9ex \vrule
width.4pt}
             \hrule height.4pt}}$}}

\usepackage{amssymb,latexsym}
\usepackage{graphics}
\usepackage{amsmath,amsfonts,amssymb,amsbsy,dsfont}
\usepackage{graphicx}
\usepackage{epstopdf}
\usepackage{mathrsfs}
\usepackage{enumerate}
\usepackage{algorithm}
\usepackage{algorithmic}
\usepackage{booktabs}
\usepackage{float}
\usepackage{color}
\usepackage{xcolor}
\usepackage{multirow}
\usepackage{soul}

\usepackage{epstopdf}
\usepackage{subfigure}
\usepackage{cases}

\usepackage{natbib}
 \bibpunct[, ]{(}{)}{,}{a}{}{,}%
 \def\bibfont{\small}%
 \def\TR{{\rm TruLR}}
 \def\MC{{\rm MC}}

 \usepackage[normalem]{ulem}

\TheoremsNumberedThrough     

\EquationsNumberedThrough    

\begin{document}
\graphicspath{{fig/}}




\TITLE{New Bounds and Truncation Boundaries for Importance Sampling}

\ARTICLEAUTHORS{%
\AUTHOR{Yijuan Liang}
\AFF{School of Economics and Management, Southwest University, Chongqing 400715, China,\\ \EMAIL{yijuanliang@swu.edu.cn}}

\AUTHOR{Guangxin Jiang\footnotemark}  \footnotetext{Corresponding author}
\AFF{School of Management, Harbin Institute of Technology, Harbin, Heilongjiang 150001, China,\\ \EMAIL{gxjiang@hit.edu.cn}}

\AUTHOR{Michael C. Fu}
\AFF{The Robert H. Smith School of Business, Institute for Systems Research, University of Maryland, College Park, MD 20742, USA, \EMAIL{mfu@umd.edu}}
} 

\ABSTRACT{%
Importance sampling (IS) is a technique that enables statistical estimation of output performance at multiple input distributions from a single nominal input distribution.
IS is commonly used in Monte Carlo simulation for variance reduction and in machine learning applications for reusing historical data, but its effectiveness can be challenging to quantify.
In this work, we establish a new result showing the tightness of {\it polynomial} concentration bounds for classical IS likelihood ratio (LR) estimators in certain settings.
Then, to address a practical statistical challenge that IS faces regarding potentially high variance, we propose new truncation boundaries when using a truncated LR estimator, for which we establish upper concentration bounds that imply an {\it exponential} convergence rate.
Simulation experiments illustrate the contrasting convergence rates of the various LR estimators and the effectiveness of the newly proposed truncation-boundary LR estimators for examples from finance and machine learning.
}%


\HISTORY{}

\maketitle

%
\section{Introduction}\label{sec:Intro}

Importance sampling (IS) is a well-known variance reduction technique for Monte Carlo simulation (see \citealt{Asmussen2007,blanchet2012state,juneja2006rare}).
In the traditional setting, output from simulation at one set of input distributions can be used to estimate performance at another set of input distributions under appropriate conditions on the distributions.
For example, in option pricing, the input distributions could drive the dynamics of the underlying assets, and in queueing models, the input distributions might comprise the interarrival times, the service times, and routing parameters.
IS can also be used in settings where historical data has been gathered, and one wishes to exploit this information to infer something at some other values of input parameters. In particular, such applications are found in artificial intelligence (AI) and machine learning (ML). For example, in reinforcement learning (RL), if the policy takes the form of a parameterized distribution over actions, then the historical data corresponds to observed performance for one policy, for which IS can then be used to infer performance for another policy.

Recall that classical IS can be viewed as based on the following simple transformation:
$$
\int h(x) f(x) dx = \int h(x) \frac{f(x)}{f_0(x)} f_0(x) dx,
$$
where $h$ and $f$, and $f_0$ are appropriately well-defined functions.
For the right-hand integral of the equation to be well defined requires further conditions where $f_0$ vanishes.
Assuming such, if $f$ and $f_0$ are probability density functions (PDFs) such that $X \sim f$, then the above yields
$$
\mathds{E}[h(X)] = \mathds{E}[h(\tilde{X}) l(\tilde{X})],
$$
where $\tilde{X} \sim f_0$ and $l(x) \equiv {f(x)}/{f_0(x)}$ is the likelihood ratio (LR) function.
If one views $h$ as the output performance function and $X$ as the input random variable, then the above equation expresses the expected performance under input PDF $f$ as an expectation under input PDF $f_0$ weighted by the LR function.
In terms of estimation, the standard Monte Carlo estimator for the left-hand side from $X_i \sim f, i=1,...,n$, given by
$ \sum_{i=1}^n h(X_i)/n$ can be replaced by the standard Monte Carlo estimator for the right-hand side from $\tilde{X}_i \sim f_0, i=1,...,n$, given by
$$
\bar{h}_n^{\tiny \mbox{LR}} \equiv \frac{1}{n} \sum_{i=1}^{n} h(\tilde{X}_i) l(\tilde{X}_i),
$$
where the sample size $n$ is taken here to be the same for notational simplification.

In stochastic simulation settings, $X$ typically represents the vector of input random variables (r.v.s) with known (joint) PDF, whereas $h$ represents an output from the simulation whose distribution is not known explicitly but produced implicitly from the simulation as function of the inputs.
As a variance reduction technique, the effectiveness of the resulting IS estimator depends on the relationship between the nominal PDF $f_0$ and target PDF $f$, so that choosing a good $f_0$ is the primary focus.

The setting of this paper assumes that the nominal distribution (here represented by the PDF $f_0$) is {\it fixed} and samples from this distribution have already been obtained, with the objective being to estimate performance at some other target distribution (represented by $f$) without the need for further simulation.
This has also been called {\it green simulation}, since it reuses one set of input data to predict performance at other values of the input data without having to resimulate by simply applying the LR-based IS estimator.
The usefulness of the estimator for this purpose again depends on the relationship between the target and nominal distributions, and our first goal is to characterize the properties of the estimator
$\bar{h}_n^{\tiny \mbox{LR}}$ by providing concentration bounds.

Our second goal is motivated by the scenario where the target distribution may be ``far" from the nominal distribution,
so that the classical LR-based IS estimator suffers from excessively high variance.
To mitigate this, we propose a truncation-boundary likelihood ratio (TruLR) method, where a threshold called the {\em truncation boundary} is used to limit the magnitude of the LR based on the nominal input data. This approach avoids extreme values of the likelihood ratio and limits the variance of the estimator. However, the TruLR method introduces bias into the estimator due to the truncation boundary, so that selecting the truncation boundary carefully is crucial to balance bias and variance, with the goal being to reduce the overall mean squared error (MSE) of the estimator.

The choice of an appropriate truncation boundary should be informed by the input sample size. When the sample size is large, the variance of the estimator tends to decrease. In such cases, selecting a higher truncation boundary can accommodate a broader range of LR values, reducing the bias introduced by truncation and leading to a smaller MSE. Conversely, for a small sample size, the variance might dominate the MSE. Therefore, a lower truncation boundary may be preferred to limit the likelihood ratio values and mitigate the impact of variance on the estimator.

We explore the relationship between the truncation boundary and the sample size under different settings, which depend on the properties of the output function, illustrated by the following simplified versions of  examples that will later be tested in simulation experiments.

\begin{example}[Option pricing]
Let $X_T \sim f$ denote the price of the asset at option expiration $T$.
For a European call option on this asset at strike price $K$, the payoff function is given by $h(X_T)$, where $h(x) = (x-K)^+$.
Given a dataset of size $n$, $\{ \tilde{X}_{T,1},\tilde{X}_{T,2}, \ldots, \tilde{X}_{T,n} \}$ under $f_0$, e.g., generated by $n$ simulation replications or real data collected over $n$ separate periods (of length $T$),
the goal is to estimate the price of the option under some other probability distribution $X_T \sim f$.
\end{example}

\begin{example}[Policy evaluation for Markov decision processes]\label{exm:offline}
Let $\pi_0$ denote a given randomized stationary parametric policy for a Markov decision process (MDP) with state space $\mathcal{S}$, action space $\mathcal{A}$, and one-period reward function $r(\cdot,\cdot)$, where $r: \mathcal{S} \times \mathcal{A} \rightarrow \Re$ and $\pi: \mathcal{S} \rightarrow [0,1]$, and $\pi(s)$ gives the probability of choosing an action $a \in \mathcal{A}$.
Given a dataset of size $n$, ${\mathscr{X}_0}=\left\{\left(\tilde{s}_{t}, \tilde{a}_{t}, \tilde{r}_{t}\right)\right\}_{t=1}^n$ generated under policy $\pi_0$,
again obtained either from simulation or real data (reinforcement learning setting),
the off-policy evaluation problem aims to estimate the $Q$-function and/or value function for some other policy $\pi$ using the data from ${\mathscr{X}_0}$. 
\end{example}

In the option pricing example, the output function is unbounded, whereas in the MDP example, rewards are often assumed bounded in AI/ML problems such as multi-armed bandit models.
As a result, our algorithms and analysis will consider both the infinity($\infty$)-norm commonly used in the RL/MDP setting and the $p$-norm, appropriate for unbounded output functions with finite moments typically assumed in stochastic simulation models, e.g., bounded variance for asset returns.
Under these two settings, we derive concentration bounds for both the classical LR and TruLR estimators, specifically, probability bounds on the estimator's absolute error. For the classical LR estimator, the bound exhibits polynomial convergence rate. We then establish that the polynomial concentration is tight under appropriate conditions by deriving a so-called {\em anti-concentration} bound, i.e., demonstrating the existence of a case with polynomial convergence rate.

Next, we derive concentration bounds for the TruLR estimator under both norms.
The bounds are used to relate the truncation boundary and the sample size, from which an approximately optimal truncation boundary can be derived by minimizing the upper bound.
We then show that the TruLR estimator with the approximately optimal truncation boundary exhibits {\it exponential} convergence rate.

The closest work to ours is \cite{Metelli2021}, which derives a truncation boundary estimator for bounded functions in an RL setting and establishes corresponding concentration and anti-concentration bounds.
Our work advances their results by deriving specific ``optimized'' truncation boundaries in a general $\infty$-norm setting; extending the estimators and analysis to the $p$-norm setting; and providing a new analysis comparing the newly proposed TruLR estimator to the classical LR estimator.
In addition, the tightness results for the classical LR estimator are new for the $p$-norm.

In sum, our work makes the following contributions:
\begin{itemize}
\item
We derive both concentration and anti-concentration bounds for the classical LR estimators under both the $\infty$-norm and $p$-norm existence assumptions, demonstrating that the polynomial concentration of the LR method is tight under specific conditions.
\item
We introduce the new TruLR estimator and establish a concentration bound for it under the $\infty$-norm. We show that the TruLR estimators achieve exponential convergence, with the optimal truncation boundary determined for a fixed sample size.
\item
We further develop concentration bounds for the TruLR estimator under the $p$-norm setting, demonstrating exponential convergence with appropriate truncation boundaries.
\item
We empirically demonstrate the superiority of the new TruLR estimators over the classical LR estimators, based on simulation experiments on synthetic examples, offline policy evaluation for reinforcement learning, and option pricing.
\end{itemize}

The rest of this paper is organized as follows.
Section \ref{sec:liter} provides a review of the related literature.
In Section \ref{sec:LR}, we present the classical LR estimator and investigate its convergence rate.
Then in Section \ref{sec:TruLR}, we consider the truncated LR estimator and optimize truncation boundaries based on derived concentration inequalities, and analyze the convergence rates.
Section \ref{sec:num} contains the simulation experiments. Finally, Section \ref{sec:concl} concludes. 

\section{Literature Review}{\label{sec:liter}}

As alluded to earlier, our work is related to two branches of literature on reusing datasets via IS.
In the areas of statistics and stochastic simulation, IS is commonly used in Bayesian inference to estimate information (e.g., normalizing constants and moments) under the posterior distribution using data sampled from a known distribution, see, e.g., \cite{kloek1978bayesian} and \cite{geweke1989bayesian}. This idea is also applied to generate simulated samples from the posterior or predictive distribution (\citealt{rubin1988sir}).
\cite{feng2017green} introduced the term ``green simulation'' to denote the reuse of outputs from previous simulation experiments to estimate current system performance without resimulation. They  proposed LR and stochastic kriging estimators to reduce MSE.
To improve the computational efficiency for future experiments, \cite{feng2021green} proposed to implement green simulation by using a database Monte Carlo approach.
In addition to the estimation problem, green simulation has also been applied in optimization (\citealt{maggiar2018derivative}), input uncertainty quantification (\citealt{feng2019efficient}), and regression (\citealt{feng2020reusing}).
Another approach to reusing simulation data is simulation analytics (\citealt{Nelson2015}), which treats the simulation model as a data generator and applies data analytics tools to mine the data and estimate conditional statements.
\cite{HongJiang2019} introduced an offline simulation online application framework that reuses the simulation data to build a predictive model for online decision making, and the framework has been successfully employed in several applications, such as online risk monitoring (\citealt{JiangHongNelson2020}), online gradient estimation (\citealt{YunHong2019}), policy gradient estimation (\citealt{liu2020simulation, lin2024reusing}), and online ranking and selection (\citealt{ShenHong2021,LiuLiang2022}).

In the area of reinforcement learning, IS is used to evaluate policies.
\cite{cortes2010learning} provided theoretical and algorithmic results for importance weighting in learning from finite samples.
\cite{mahmood2014weighted} explored the benefits of weighted IS for off-policy learning with linear function approximation.
\cite{NEURIPS2018_6aed000a} and \cite{MetJMLR2020} applied the LR method to efficiently reuse trajectories and optimize policies for continuous control tasks. However, the variance of the LR estimator may sometimes be unacceptably high, so several approaches have been proposed to alleviate this issue, starting with truncated importance sampling in \cite{Ionides2008}.
\cite{vehtari2024pareto} introduced a method called Pareto smoothed importance, and \cite{papini2019optimistic} developed a multiple IS with truncation approach for off-policy estimation in reinforcement learning. \cite{Metelli2021} introduced a class of LR corrections based on smoothly shrinking the IS weights.
Other approaches to stabilize the LR estimator include clipping transformation (\citealt{martino2018comparison}), a doubly robust approach (\citealt{dudik2011doubly}), and the switch estimator (\citealt{wang2017optimal}).

We also briefly mention the use of IS for variance reduction. In contrast to the setting of data reuse, where the target distribution is assumed given and fixed, using IS for variance reduction requires constructing an effective IS distribution, which constitutes the key challenge. Good importance sampling distributions are typically obtained from large deviations theory (\citealt{budhiraja2019analysis}), and many techniques have been proposed based on the large derivation rate function (e.g., \citealt{bai2022rare}, \citealt{blanchet2012lyapunov}, and \citealt{blanchet2012efficient}). Another approach is the cross-entropy method (\citealt{rubinstein2004CE}), which chooses proper sampling distributions by minimizing the Kullback-Leibler divergence between the considered distribution and a zero-variance distribution. These methods have been successfully applied in many applications, such as financial engineering (\citealt{GlHeSh2000}), queueing (\citealt{BlLam2014}), and stochastic root-finding problem (\citealt{HeJiang2023}).

\section{The Likelihood Ratio (LR) Method for Importance Sampling}\label{sec:LR}
We consider different assumptions on the performance function corresponding to the two norms as defined here.

\begin{definition}[$\infty$-norm w.r.t.~a~measure] \label{def:inf-norm}
The $\infty$-norm of function $h:\mathcal{X} \rightarrow \Re$ with respect to measure $\mu$ is given by
$$
\| h \|_{\infty, \mu} = \sup \left\{ y: \int_{x \in \mathcal{X}} {\bf 1} \{ |h(x)|>y \} d\mu(x) >0 \right\}.
$$
\end{definition}

\begin{definition}[$p$-norm  w.r.t.~a~measure] \label{def:p-norm}
The $p$-norm ($p > 0$) of function $h:\mathcal{X} \rightarrow \Re$ with respect to measure $\mu$ is given by
$$
\| h \|_{p, \mu} = \left(\int_{x \in \mathcal{X}} |h(x)|^{p} d\mu(x)\right)^{1/p}.
$$
\end{definition}

Note that taking $\mu$ as Lebesgue measure gives the usual definitions of the $\infty$-norm and $p$-norm,
which in the former case corresponds to supremum if the function is bounded.

Let $\bX=(X_1,X_2,\ldots,X_d)^{\top}\in \mathcal{X}\subseteq \Re^d $ be a random vector with probability measure $\mathds{P}$.
Our interest will be the setting where the measures are parameterized probability measures $\mathds{P}_{\betheta}$, where $\btheta=(\theta_1, \theta_2, \ldots, \theta_k)^\top \in \Re^k$ determines the probability measures.
To quantify the difference between two probability measures, we define the following divergence between two parameterized probability measures \citep{papini2019optimistic}:

\begin{definition}[$\alpha$-divergence] \label{def:alpha-diverg}
For $\mathds{P}_{\betheta} \ll \mathds{P}_{\betheta_0}$, i.e., $\mathds{P}_{\betheta}$ is absolutely continuous w.r.t. $\mathds{P}_{\betheta_0}$,
\begin{equation*}
  I_{\alpha}(\mathds{P}_{\betheta} \|  \mathds{P}_{\betheta_0}) = \int_{\mathcal{X}} \left( \frac{d\mathds{P}_{\betheta}(\bx)}{d\mathds{P}_{\betheta_0}(\bx)} \right)^{\alpha} d\mathds{P}_{\betheta_0}(\bx), ~~\alpha \geq 1.
\end{equation*}
\end{definition}

With slight abuse of notation, we let  $l(\bx)\equiv {d\mathds{P}_{\betheta}(\bx)}/{d\mathds{P}_{\betheta_0}(\bx)}$ be the LR function. When $\bX\sim \mathds{P}_{\betheta}$ and $\tilde \bX \sim \mathds{P}_{\theta_0}$ are both continuous random vectors with respective PDFs $f(\bx)$ and $f_0(\bx)$, then $d\mathds{P}_{\betheta}(\bx) = f(\bx)d\bx$, $d\mathds{P}_{\betheta_0}(\bx) = f_0(\bx)d\bx$, and the definition of $l(\bx)$ coincides with the LR function defined in Section \ref{sec:Intro}, whereas the definition of $l(\bx)$ also applies to the LR function for discrete random vectors.
According to Definition \ref{def:alpha-diverg}, $\alpha$-divergence gives the $\alpha$th moment of $l(\tilde \bX)$ under the measure $P_{\betheta_0}$, which includes several common divergences, e.g., R\'{e}nyi divergence is $(\alpha-1)^{-1} \log I_{\alpha}(\mathds{P}_{\betheta} \| \mathds{P}_{\betheta_0})$ (\citealt{renyi1961measures}).

Suppose that we have the dataset $\{\tilde \bX_1, \tilde \bX_2, \ldots, \tilde \bX_n\}$, where $\tilde \bX_i \sim \mathds{P}_{\betheta_0}$, $i=1,2,\cdots,n$, and our goal is to estimate
\begin{equation}
\bar{h}(\btheta) = \mathds{E}[h(\bX)] =  \int_{\mathcal{X}} h(\bx)d\mathds{P}_{\betheta}(\bx).
\end{equation}
We estimate $\mathds{E}[h(\bX)]$, $\bX \sim \mathds{P}_{\betheta}$, using LR estimators of the form
\begin{equation}
\bar{h}_n^{\tiny \mbox{LR}} \equiv \frac{1}{n} \sum_{i=1}^{n} h(\tilde{\bX}_i) l(\tilde{\bX}_i),
\end{equation}
and derive concentration inequalities of the form
\begin{equation*}
 \Pr \left(\left|\bar h_n^{\LR} - \bar h\right|\leq \eta(n,\delta)\right)\geq 1- \delta, \quad \delta \in(0,1),
\end{equation*}
where $\eta$ is a function of $n$ and $\delta$.
For $\beta>0$ and $\gamma>0$, we say $\bar h_n^{\LR}$ exhibits {\em polynomial concentration} if $\eta(n,\delta) = \mathcal{O}(1/(n^\gamma\delta)^\beta)$, and {\em exponential concentration} if $\eta(n,\delta)=\mathcal{O}\left((\ln (1 / \delta) / n^\gamma)^\beta\right)$.

As the confidence level $1-\delta$ increases, indicating a more reliable bound, the required sample size $n$ must also increase to maintain accuracy. In the case of polynomial concentration, if $\delta$ is reduced by a factor of $k$, the sample size $n$ must be increased by a factor of $k^{1/\gamma}$ to preserve accuracy. In contrast, with exponential concentration, reducing $\delta$ by a factor of $k$ necessitates increasing the sample size $n$ by only a factor of $(\ln k)^{1/\gamma}$, which is significantly less than $k^{1/\gamma}$.

Next, we provide the concentration inequalities for the LR estimator under the existence of the $\infty$-norm and $p$-norm of $h$ w.r.t. the measure $\mathds{P}_{\betheta_0}$.
\begin{proposition}\label{theorem-IS-concentration}
If $\| h \|_{\infty, \mathds{P}_{\betheta_0}} < \infty$ and $I_{\alpha}(\mathds{P}_{\betheta} \|  \mathds{P}_{\betheta_0})< \infty$ for some $\alpha \in (1,2]$, then the LR estimator admits polynomial concentration, i.e.,
\begin{equation}\label{eq-IS-concentration}
\Pr \left(  | \bar {h}_n^{\LR} - \bar{h}(\btheta) | \leq  \| h \|_{\infty, \mathds{P}_{\betheta_0}} \left( \frac{4 I_{\alpha}(\mathds{P}_{\betheta} \|  \mathds{P}_{\betheta_0})}{\delta n^{\alpha-1}}\right)^{\frac{1}{\alpha}} \right) \geq 1-\delta, ~~\delta \in (0,1).
\end{equation}
\end{proposition}

\begin{proposition}\label{theorem-IS-bound-concentration}
If $\| h \|_{p, \mathds{P}_{\betheta_0}} < \infty$ for some $p > 2 $ and $I_{\alpha}(\mathds{P}_{\betheta} \|  \mathds{P}_{\betheta_0}) < \infty$ for some $\alpha \geq {2p}/{(p-2)}>2$, then the LR estimator admits polynomial concentration, i.e.,
\begin{equation}\label{eq-IS-bound-concentration}
\Pr \left(  | \bar {h}_n^{\LR} - \bar{h}(\btheta) | \leq
   \frac{2}{\sqrt{n \delta}}  \| h \|_{p, \mathds{P}_{\betheta_0}}  I_{\alpha}(\mathds{P}_{\betheta} \|  \mathds{P}_{\betheta_0})^{\frac{1}{\alpha}} \right) \geq 1-\delta, ~~\delta \in (0,1).
\end{equation}
\end{proposition}

Propositions \ref{theorem-IS-concentration} and \ref{theorem-IS-bound-concentration} demonstrate that the LR estimators exhibit polynomial concentration, with (upper bound) concentration rates $\mathcal{O}(n^{1/\alpha -1})$ and $\mathcal{O}(n^{-1/2})$, respectively.

Since Propositions \ref{theorem-IS-concentration} and \ref{theorem-IS-bound-concentration} provide upper bounds, the polynomial convergence rates could be conservative. However, the following lower bound results, establishing what are known as {\em anti-concentration bounds}, lead to the conclusion that the polynomial convergence is indeed tight.

\begin{proposition}\label{theorem-IS-concentration-lower}
There exist $h(\cdot)$ with $\| h \|_{\infty, \mathds{P}_{\betheta_0}} < \infty$
and probability measures $\mathds{P}_{\betheta}$ and $\mathds{P}_{\betheta_0}$ with $I_{\alpha}(\mathds{P}_{\betheta} \|  \mathds{P}_{\betheta_0}) < \infty$ for $\alpha \in (1,2]$ such that $\forall~\delta \in(0,e^{-1})$, $n \geq \max(1, e\delta ( I_{\alpha}(\mathds{P}_{\betheta} \|  \mathds{P}_{\betheta_0})-1)^{{1}/{(\alpha-1)}})$,
\begin{equation}\label{eq-IS-concentration-lower}
\Pr \left(  | \bar {h}_n^{\LR} - \bar{h}(\btheta) | \geq  \| h \|_{\infty, \mathds{P}_{\betheta_0}}   \left( \frac{ I_{\alpha}(\mathds{P}_{\betheta} \|  \mathds{P}_{\betheta_0})-1}{\delta n^{\alpha-1}}\right)^{\frac{1}{\alpha}} \left(1- \frac{ e \delta}{n}\right)^{\frac{n-1}{\alpha}}
  \right) \geq 1-\delta.
\end{equation}
\end{proposition}

\begin{proposition}\label{theorem-IS-bound-concentration-lower-1}
There exist $h(\cdot)$ with $\| h \|_{p, \mathds{P}_{\betheta_0}} < \infty$ for some $p>2$
and probability measures $\mathds{P}_{\betheta}$ and $\mathds{P}_{\betheta_0}$ with $I_{\alpha}(\mathds{P}_{\betheta} \|  \mathds{P}_{\betheta_0}) < \infty$
for some $ \alpha \geq {2p}/{(p-2)}>2$ such that $\forall~\delta \in(0,e^{-1})$, $n \geq \max(1, e\delta  I_{\alpha}(\mathds{P}_{\betheta} \|  \mathds{P}_{\betheta_0})^{{2}/{\alpha}})$,
\begin{equation}\label{eq-IS-bound-concentration-lower-1}
\Pr \left(  | \bar {h}_n^{\LR} - \bar{h}(\btheta) |  \geq    \frac{1}{\sqrt{n \delta}}  \|h \|_{p, \mathds{P}_{\betheta_0}}  I_{\alpha}(\mathds{P}_{\betheta} \|  \mathds{P}_{\betheta_0})^{\frac{1}{\alpha}} \left(1- \frac{ e \delta}{n}\right)^{\frac{n-1}{2}}
  \right) \geq 1-\delta.
\end{equation}
\end{proposition}

Propositions \ref{theorem-IS-concentration-lower} and \ref{theorem-IS-bound-concentration-lower-1} establish the existence of specific measures and performance functions, exhibiting the same convergence rate as the concentration (upper) bound multiplied by terms converging to the constants $e^{-e\delta/\alpha}$ and $e^{-e\delta/2}$, respectively. Thus, tightness follows by combining Proposition \ref{theorem-IS-concentration} with Proposition \ref{theorem-IS-concentration-lower}, and Proposition \ref{theorem-IS-bound-concentration} with Proposition \ref{theorem-IS-bound-concentration-lower-1}. \cite{Metelli2021} provided the results of Propositions \ref{theorem-IS-concentration} and \ref{theorem-IS-concentration-lower} under the condition that $h$ is a bounded function, a special case of our $\infty$-norm general setting.

\section{Truncation-Boundary Likelihood Ratio (TruLR) Method}\label{sec:TruLR}
In this section, we propose using the truncation-boundary likelihood ratio (TruLR) method to mitigate the high variance issue associated with the LR method. We define the truncated LR by
\begin{equation*}
    l'_n(\bx) \equiv \min\{l(\bx), \tau_n \},
\end{equation*}
where $\tau_n$ is the {\em truncation boundary}, which is a function of the size of the dataset $n$, corresponding to the number of replications in the stochastic simulation setting. Following \cite{Ionides2008}, the TruLR estimator of $\bar{h}(\btheta)$ is then given by
\begin{equation}\label{def-estimator-TR}
     \bar {h}_n^{\TR}  \equiv \frac{1}{n}\sum_{i=1}^n h(\tilde{\bX}_i)l'_n(\tilde{\bX}_i).
\end{equation}

Under some mild conditions, we can demonstrate that the TruLR estimator is asymptotically unbiased. The detailed proof is provided in \ref{appx:P5-proof}.

\begin{proposition}\label{theorem-TR-consistency}
Suppose that $\| h \|_{2, \mathds{P}_{\betheta}} < \infty$. Let $\tau_n \to \infty$ and $\tau_n/n\to 0$ as $n \to \infty$. Then $\bar {h}_n^{\TR}$ is asymptotically unbiased and its variance goes to $0$, thus it is weakly consistent.
\end{proposition}

Proposition \ref{theorem-TR-consistency} requires that the truncation boundary $\tau_n$ should increase without bound as the sample size $n$ grows, but at a sublinear rate.

We now derive ``optimal'' truncation boundaries $\tau_n$ by minimizing the concentration inequality upper bound of the TruLR estimator for both the $\infty$-norm and $p$-norm settings.

\subsection{Concentration Inequality and Truncation Boundary for $\infty$-Norm}\label{sc:concentration-bound-inf}
We begin by deriving a concentration inequality for the TruLR estimator under the $\infty$-norm. Let $B_n \equiv  \mathds{E}[\bar {h}_n^{\TR}] - \bar{h}(\betheta)$ and $V_n \equiv  \Var(\bar {h}_n^{\TR})$ represent the bias and the variance of the TruLR estimator, respectively. The following lemma provides bounds on the bias and variance (proof in \ref{appx:L1-proof}).
\begin{lemma}\label{theorem-TR-bound}
If $\| h \|_{\infty, \mathds{P}_{\betheta_0}} < \infty$ and $I_{\alpha}(\mathds{P}_{\betheta} \|  \mathds{P}_{\betheta_0})< \infty$ for some $\alpha \in (1,2]$, then
\begin{equation}\label{eq-TR-bound-bias}
  |B_n| \leq  \| h \|_{\infty, \mathds{P}_{\betheta_0}} {\tau_n}^{1-\alpha} I_{\alpha}(\mathds{P}_{\betheta} \|  \mathds{P}_{\betheta_0}),
\end{equation}
and
\begin{equation}\label{eq-TR-bound-variance}
 V_n \leq \frac{1}{n} \| h \|_{\infty, \mathds{P}_{\betheta_0}}^2  {\tau_n}^{2-\alpha} I_{\alpha}(\mathds{P}_{\betheta} \|  \mathds{P}_{\betheta_0}).
\end{equation}
\end{lemma}

Lemma \ref{theorem-TR-bound} suggests that increasing $\tau_n$ as a function of $n$ must balance bias and variance.

Building on Lemma \ref{theorem-TR-bound}, we can now establish a concentration bound for the TruLR estimator.
\begin{theorem}\label{theorem-TR-bound-concentration}
If $\| h \|_{\infty, \mathds{P}_{\betheta_0}} < \infty$ and $I_{\alpha}(\mathds{P}_{\betheta} \|  \mathds{P}_{\betheta_0})< \infty$ for some $\alpha \in (1,2]$, then
\begin{eqnarray}\label{eq-TR-bound-concentration}
   \Pr\Bigg(
     |\bar {h}_n^{\TR} - \bar{h}(\btheta)|\leq  &&  \| h \|_{\infty, \mathds{P}_{\betheta_0}}\sqrt{\frac{2  {\tau_n}^{2-\alpha}  \ln\frac{2}{\delta} }{n}I_{\alpha}(\mathds{P}_{\betheta} \|  \mathds{P}_{\betheta_0}) }
    + \| h \|_{\infty, \mathds{P}_{\betheta_0}}  \frac{\tau_n \ln\frac{2}{\delta}}{3n} \nonumber  \\
     &&+  \| h \|_{\infty, \mathds{P}_{\betheta_0}} {\tau_n}^{1-\alpha} I_{\alpha}(\mathds{P}_{\betheta} \|  \mathds{P}_{\betheta_0})
     \Bigg) \geq 1-\delta,~~ \delta \in (0,1).
\end{eqnarray}
\end{theorem}

\proof{Proof:} The proof is a straightforward application of Bernstein's inequality together with Lemma \ref{theorem-TR-bound}.
Since $|\bar {h}_n^{\TR}| \leq \tau_n \| h \|_{\infty, \mathds{P}_{\betheta_0}}$ a.s., applying Bernstein's inequality (\citealt{boucheron2013concentration}), we can state that with probability (w.p.) at least $1-\delta$ it holds that
\begin{eqnarray*}
  |\bar {h}_n^{\TR} - \mathds{E}[\bar {h}_n^{\TR} ]| & \leq & \sqrt{2 { \mathds{E}[\left(\bar {h}_n^{\TR} \right)^2]}  \ln\frac{2}{\delta}} + \frac{  \tau_n \| h \|_{\infty, \mathds{P}_{\betheta_0}} \ln\frac{2}{\delta}}{3n} \nonumber \\
   & \leq &  \| h \|_{\infty, \mathds{P}_{\betheta_0}} \sqrt{ \frac{2  {\tau_n}^{2-\alpha}  \ln\frac{2}{\delta} }{n} I_{\alpha}(\mathds{P}_{\betheta} \|  \mathds{P}_{\betheta_0}) } + \| h \|_{\infty, \mathds{P}_{\betheta_0}} \frac{\tau_n \ln\frac{2}{\delta}}{3n}, \label{eq-TR-bound-concentration-2}
\end{eqnarray*}
where the last inequality is obtained as a by-product during the proof of inequality \eqref{eq-TR-bound-variance} in Lemma \ref{theorem-TR-bound}. Combined with the bias term bound, i.e., inequality \eqref{eq-TR-bound-bias} in Lemma \ref{theorem-TR-bound}, we have
\begin{eqnarray*}
|\bar {h}_n^{\TR}  -\bar{h}(\btheta)| &=& | \bar {h}_n^{\TR} - \mathds{E}[\bar {h}_n^{\TR} ]  + \mathds{E}[\bar {h}_n^{\TR} ] - \bar{h}(\btheta)| \nonumber \\
   & \leq & | \bar {h}_{\TR} - \mathds{E}[\bar {h}_n^{\TR} ]|  + |\mathds{E}[\bar {h}_n^{\TR} ] - \bar{h}(\btheta)| \nonumber \\
   & \leq & \| h \|_{\infty, \mathds{P}_{\betheta_0}} \sqrt{\frac{2  {\tau_n}^{2-\alpha}  \ln\frac{2}{\delta} }{n}I_{\alpha}(\mathds{P}_{\betheta} \|  \mathds{P}_{\betheta_0})} +\| h \|_{\infty, \mathds{P}_{\betheta_0}} \frac{\tau_n \ln\frac{2}{\delta}}{3n} +\| h \|_{\infty, \mathds{P}_{\betheta_0}} {\tau_n}^{1-\alpha} I_{\alpha}(\mathds{P}_{\betheta} \|  \mathds{P}_{\betheta_0}).
\end{eqnarray*}
$\hfill\Box$
\endproof

Treating $\tau_n$ as a decision variable, we minimize the upper bound in \eqref{eq-TR-bound-concentration}.

\begin{theorem}\label{theorem-TR-bound-concentration-min1}
Under the same assumptions in Theorem \ref{theorem-TR-bound-concentration}, the upper bound in \eqref{eq-TR-bound-concentration} is uniquely minimized by selecting
$$ \tau_n^{*} = (x^{*} )^{\frac{2}{\alpha}}  \left( \frac{n I_{\alpha}(\mathds{P}_{\betheta} \|  \mathds{P}_{\betheta_0}) }{\ln\frac{2}{\delta}}\right)^{\frac{1}{\alpha}}~~{with}~~x^{*} =  \frac{3\sqrt{2}(\frac{\alpha}{2}-1) + 3\sqrt{2(\frac{\alpha}{2}-1)^2 + \frac{ 4}{3}(\alpha-1)}}{{2}},$$
and the TruLR estimator admits exponential concentration, i.e.,
\begin{eqnarray}
  \Pr\left( |\bar {h}_n^{\TR} - \bar{h}(\btheta)| \leq C_1^* \| h \|_{\infty, \mathds{P}_{\betheta_0}}\left( \frac{\ln\frac{2}{\delta}}{n} \right)^{1-\frac{1}{\alpha}} I_{\alpha}(\mathds{P}_{\betheta} \|  \mathds{P}_{\betheta_0})^{\frac{1}{\alpha}} \right) \geq 1-\delta,~~\delta\in(0,1), \label{eq-TR-bound-concentration-0}
\end{eqnarray}
where $C_1^* =  \sqrt{2}(x^{*})^{{2}/{\alpha}-1} +(1/3)(x^{*})^{{2}/{\alpha}} +(x^{*})^{{2}/{\alpha}-2}$.
\end{theorem}
\proof{Proof:}
To minimize the upper bound in \eqref{eq-TR-bound-concentration}, we set its first derivative to zero, leading to the following equation
\begin{equation}\label{eq-TR-bound-concentration-5}
    \sqrt{\frac{2    \ln\frac{2}{\delta} }{n}I_{\alpha}(\mathds{P}_{\betheta} \|  \mathds{P}_{\betheta_0}) } (1-\frac{\alpha}{2}) {\tau_n}^{\frac{\alpha}{2}} +  \frac{\ln\frac{2}{\delta}}{3n}\tau_n^{\alpha} + (1-\alpha) I_{\alpha}(\mathds{P}_{\betheta} \|  \mathds{P}_{\betheta_0})=0,
\end{equation}
which has a positive solution
\begin{equation}\label{eq-TR-bound-concentration-6}
    \tau_n^{\dag} = \left( \frac{\sqrt{2}(\frac{\alpha}{2}-1) + \sqrt{2(\frac{\alpha}{2}-1)^2 + \frac{ 4}{3}(\alpha-1)}}{\frac { 2}{3}}\right)^{\frac{2}{\alpha}}  \left( \frac{n I_{\alpha}(\mathds{P}_{\betheta} \|  \mathds{P}_{\betheta_0}) }{\ln\frac{2}{\delta}}\right)^{\frac{1}{\alpha}}.
\end{equation}


Next, we show that the solution $\tau_n^{\dag}$ given in \eqref{eq-TR-bound-concentration-6} provides the unique minimizer of the upper bound in \eqref{eq-TR-bound-concentration}. Suppose that a general truncation boundary has the form
$$\tau_n^{\frac{\alpha}{2}} = x \left( \frac{n I_{\alpha}(\mathds{P}_{\betheta} \|  \mathds{P}_{\betheta_0}) }{\ln\frac{2}{\delta}}\right)^{\frac{1}{2}},$$
with $x>0$. Substituting it into \eqref{eq-TR-bound-concentration}, the upper bound becomes
$$ g(x) \| h \|_{\infty, \mathds{P}_{\betheta_0}}\left( \frac{\ln\frac{2}{\delta}}{n} \right)^{1-\frac{1}{\alpha}} I_{\alpha}(\mathds{P}_{\betheta} \|  \mathds{P}_{\betheta_0})^{\frac{1}{\alpha}},$$
where
\begin{equation}\label{eq-TR-bound-concentration-8}
  g(x) = g(x;\alpha) = \sqrt{2} x^{\frac{2}{\alpha}-1} +\frac{1}{3}x^{\frac{2}{\alpha}} +x^{\frac{2}{\alpha}-2}.
\end{equation}

For simplicity, denote $k={2}/{\alpha}$, then $k \in [1,2]$ and
\begin{equation*}\label{eq-TR-bound-concentration-9}
  g(x) = g(x;k) = \sqrt{2} x^{k-1} +\frac{1}{3}x^{k} +x^{k-2},
\end{equation*}
whose second derivative is given by
\begin{equation*}\label{eq-TR-bound-concentration-10}
  g''(x) =x^{k-4}\left( \sqrt{2}(k-1)(k-2) x +\frac{1}{3}k(k-1)x^2 +(k-2)(k-3)\right) \equiv x^{k-4} g_1(x),
\end{equation*}
where $g_1(x)$ is a quadratic function of $x$. Given that $k \in [1,2]$ and $x>0$, the minimum value of $g_1(x)$ over $(0,\infty)$ is obtained at its axis of symmetry, i.e.,
\begin{eqnarray*}
  \min_{x>0} g_1(x) &=& (k-2)(k-3) -\frac{2(k-1)^2(k-2)^2 }{4\times\frac{1}{3}k(k-1)} \nonumber \\
   &=& \frac{2-k}{2k}(k^2-3k+6) \nonumber \\
   & = & \frac{2-k}{2k}( (k-\frac{3}{2})^2+\frac{15}{4} ) >0.
\end{eqnarray*}
Thus, $g''(x)>0$ and $g(x)$ is convex with a unique minimum value at point $x^{*} = 3{(\sqrt{2}({\alpha}/{2}-1) + \sqrt{2({\alpha}/{2}-1)^2 + { 4}(\alpha-1)/3})}/2 $. $x^{*}$ is increasing with $\alpha$ and its maximum value is $\sqrt{3}$.
$\hfill\Box$
\endproof

Theorem \ref{theorem-TR-bound-concentration-min1} implies that the absolute error of the TruLR estimator converges to zero at rate of at least $n^{-1+1/\alpha}$. Since the truncation boundary $\tau_n^*$ is of order $\mathcal{O}(n^{1/\alpha})$, it satisfies the conditions outlined in Proposition \ref{theorem-TR-consistency}, thus ensuring the consistency of the TruLR estimator.
More importantly, Theorem \ref{theorem-TR-bound-concentration-min1} demonstrates that with an appropriately chosen truncation boundary, the TruLR estimator admits exponential concentration, as opposed to the polynomial concentration of the classical LR estimator. 


Taking $x^{*}=1$ in Theorem \ref{theorem-TR-bound-concentration-min1} leads to the following corollary corresponding to Theorem 1 in \cite{papini2019optimistic}.
\begin{corollary}\label{theorem-TR-bound-concentration-min2}
Under the same assumptions in Theorem \ref{theorem-TR-bound-concentration}, for
$$\tau_n^{\dag} =  \left( \frac{n I_{\alpha}(\mathds{P}_{\betheta} \|  \mathds{P}_{\betheta_0}) }{\ln\frac{2}{\delta}}\right)^{\frac{1}{\alpha}},$$
the TruLR estimator admits exponential concentration, i.e.,
\begin{equation}\label{eq-TR-bound-concentration-1}
  \Pr \left( |\bar {h}_n^{\TR} - \bar{h}(\btheta)| \leq  \| h \|_{\infty, \mathds{P}_{\betheta_0}}  \left( \sqrt{2}+\frac{4}{3} \right)   \left( \frac{\ln\frac{2}{\delta}}{n} \right)^{1-\frac{1}{\alpha}} I_{\alpha}(\mathds{P}_{\betheta} \|  \mathds{P}_{\betheta_0})^{\frac{1}{\alpha}} \right) \geq 1-\delta,~~\delta\in(0,1).
\end{equation}
\end{corollary}

Although the TruLR estimator with the truncation boundary in Corollary \ref{theorem-TR-bound-concentration-min2} also admits exponential concentration, the upper bound on the absolute error of the TruLR estimator is larger than the corresponding bound in Theorem \ref{theorem-TR-bound-concentration-min1}.

\subsection{Concentration inequality and truncation boundary for $p$-Norm}\label{sc:concentration-bound-p}

We now consider the $p$-norm setting.
Following a similar approach as in Section \ref{sc:concentration-bound-inf}, we begin by bounding the bias $B_n$ and variance $V_n$ of the TruLR estimator in the following lemma (proof in \ref{appx:L2-proof}).
\begin{lemma}\label{theorem-TR-bound-moment}
If $\| h \|_{p, \mathds{P}_{\betheta_0}} < \infty$ for some $p > 2 $ and $I_{\alpha}(\mathds{P}_{\betheta} \|  \mathds{P}_{\betheta_0}) < \infty$ for some $ \alpha\in ({p}/{(p-1)}, {2p}/{(p-2)})$, then
\begin{equation}\label{eq-TR-bound-moment-bias}
  |B_n| \leq  \| h \|_{p, \mathds{P}_{\betheta_0}}  {\tau_n}^{1-\alpha+\frac{\alpha}{p}} I_{\alpha}(\mathds{P}_{\betheta} \|  \mathds{P}_{\betheta_0})^{1-\frac{1}{p}},
\end{equation}
and
\begin{equation}\label{eq-TR-bound-moment-variance}
  V_n   \leq \frac{1}{n} \| h \|_{p, \mathds{P}_{\betheta_0}}^2 {\tau_n}^{2-\alpha+\frac{2\alpha}{p}} I_{\alpha}(\mathds{P}_{\betheta} \|  \mathds{P}_{\betheta_0})^{1-\frac{2}{p}}.
\end{equation}
\end{lemma}

Since the upper bounds for $B_n$ and $V_n$ depend on $p$, the concentration inequality upper bound for  the TruLR estimator also depends on $p$.

To derive the concentration bound, we need an additional assumption.
\begin{assumption}[Moment Generating Function Condition]\label{assump-MGF}
The moment generating function of $h(\tilde{\bX}), \tilde{\bX}\sim \mathds{P}_{\betheta_0}$, exists in a neighborhood of zero, i.e., $\exists \lambda_B >0$, s.t. $\mathds{E}[  e^{\lambda h(\tilde{\bX})  } ]   < \infty$, $\forall$ $|\lambda|< \lambda_B$.
\end{assumption}

Note that Assumption \ref{assump-MGF} implies the existence of $p$-norm, i.e., $\| h \|_{p, \mathds{P}_{\betheta_0}} < \infty$ for any $p\geq 1$. Under this assumption, we establish the concentration bound in the following theorem.
\begin{theorem}\label{theorem-TR-bound-moment-concentration-MGF}
If Assumption \ref{assump-MGF} holds, and $I_{\alpha}(\mathds{P}_{\betheta} \|  \mathds{P}_{\betheta_0})<\infty$ for some $ \alpha\in ({p}/{(p-1)}, {2p}/{(p-2)})$, where $p>2$, then $\forall$ $\sigma^2 \geq$ $\Var(h(\tilde{\bX})l'_n(\tilde{\bX}))$, $\exists$ $\Lambda\geq 0$ such that
\begin{equation}
  \mathds{E}\left[  e^{ \lambda (h(\tilde{\bX})l'_n(\tilde{\bX}) -\mathds{E}[h(\tilde{\bX})l'_n(\tilde{\bX})] )}\right]
    \leq  e^{\frac{1}{2} \lambda^2  \sigma^2}, \quad \forall|\lambda| \leq \frac{1}{\Lambda }, \label{eq-TR-bound-moment-concentration-MGF-1-0}
\end{equation}
for which the following holds:
\begin{eqnarray}\label{eq-TR-bound-moment-concentration-MGF-1}
 \Pr\Bigg(
     |\bar {h}_n^{\TR} - \bar{h}(\btheta)| \leq  \max\left( \sqrt{ \frac{2 \sigma^2 }{n} \ln \frac{2}{\delta} } , \frac{2\Lambda}{n} \ln \frac{2}{\delta} \right)
  && + \| h \|_{p, \mathds{P}_{\betheta_0}} {\tau_n}^{1-\alpha+\frac{\alpha}{p}} I_{\alpha}(\mathds{P}_{\betheta} \|  \mathds{P}_{\betheta_0})^{1-\frac{1}{p}}
     \Bigg) \nonumber \\
&&      \geq 1-\delta,~~\delta \in (0,1).
\end{eqnarray}
\end{theorem}

\proof{Proof:} First, we show that the MGF of $h(\tilde{\bX})l'_n(\tilde{\bX})$ exists under measure $\mathds{P}_{\betheta_0}$ in a neighborhood of zero.
Recall that $l'_n(\tilde{\bX})  \leq \tau_n$ by definition,
\begin{equation*}
    e^{ \lambda h(\tilde{\bX})l'_n(\tilde{\bX})  } \leq   e^{ | \lambda h(\tilde{\bX})l'_n(\tilde{\bX})|  }
      \leq e^{ \tau_n |{\lambda h(\tilde{\bX})}| } \leq  e^{ \lambda \tau_n h(\tilde{\bX}) } + e^{ -\lambda \tau_n h(\tilde{\bX}) }.
\end{equation*}
Hence the MGF of $h(\tilde{\bX})l'_n(\tilde{\bX})$ also exists for $|\lambda \tau_n|< \lambda_B$, where $\lambda_B$ is given by Assumption \ref{assump-MGF}.

Next, applying a Taylor series expansion to the MGF of $h(\tilde{\bX})l'_n(\tilde{\bX})$, we have
\begin{equation}\label{eq-TR-bound-moment-concentration-MGF-40}
   \ln \mathds{E}\left[  e^{ \lambda ( h(\tilde{\bX})l'_n(\tilde{\bX}) -\mathds{E}[h(\tilde{\bX})l'_n(\tilde{\bX}) ] )}\right]
    = \frac{1}{2} \lambda^2 \Var(h(\tilde{\bX})l'_n(\tilde{\bX}) )   +o(\lambda^2).
\end{equation}
Based on \eqref{eq-TR-bound-moment-concentration-MGF-40}, for $\sigma^2 \geq \Var(h(\tilde{\bX})l'_n(\tilde{\bX}))$, the following inequality holds for sufficiently small $\lambda$, i.e., $\exists$ $\Lambda \geq 0$ depending on $h(\tilde{\bX})l'_n(\tilde{\bX})$ such that
\begin{equation}\label{eq-TR-bound-moment-concentration-MGF-4}
  \mathds{E}\left[  e^{ \lambda ( h(\tilde{\bX})l'_n(\tilde{\bX})  -\mathds{E}[h(\tilde{\bX})l'_n(\tilde{\bX})] )}\right] \leq e^{\frac{1}{2} \lambda^2  \sigma^2}, \quad {\forall} |\lambda |\leq  \frac{1}{\Lambda }.
\end{equation}

Now, using Markov's inequality, we get for any $t>0$ and $\lambda \in[0,{1}/{\Lambda}] $
\begin{eqnarray}
    \Pr\left( \bar {h}_n^{\TR} - \mathds{E}[\bar {h}_n^{\TR} ] \geq t \right)
   &=& \Pr \left( \frac{1}{n}\sum_{i=1}^n( h(\tilde{\bX}_i)l'_n(\tilde{\bX}_i)- \mathds{E}[h(\tilde{\bX}_i)l'_n(\tilde{\bX}_i)])  \geq t \right) \nonumber \\
   &\leq&   e^{-\lambda n t} \prod_{i=1}^n \mathds{E}\left[  e^{ \lambda ( h(\tilde{\bX}_i)l'_n(\tilde{\bX}_i) -\mathds{E}[h(\tilde{\bX}_i)l'_n(\tilde{\bX}_i) ] )} \right] \nonumber \\
   &\leq& e^{-\lambda nt  + \frac{1}{2} n \lambda^2 \sigma^2 }, \label{eq-TR-bound-moment-concentration-MGF-5}
\end{eqnarray}
where the last line follows from \eqref{eq-TR-bound-moment-concentration-MGF-4}.
If ${t}/{\sigma^2} \leq {1}/{\Lambda}$, the bound in \eqref{eq-TR-bound-moment-concentration-MGF-5} obtains its minimum value at $\lambda={t}/{\sigma^2}$ as $e^{-{nt^2}/(2\sigma^2)}$.
If ${t}/{\sigma^2} \geq {1}/{\Lambda}$, which implies that $\sigma^2 \leq  t \Lambda$, then the bound in \eqref{eq-TR-bound-moment-concentration-MGF-5} obtains its minimum value at $t={1}/{\Lambda}$ as $e^{- nt/{\Lambda}  +  n  \sigma^2 /{(2\Lambda^2)}} \leq e^{- nt/{\Lambda}  +  n  t\Lambda/{(2\Lambda^2)}} = e^{-nt/{(2\Lambda)}}$. Thus choosing $\lambda=\min\{{t}/{\sigma^2},{1}/{\Lambda}\}$ yields
\begin{equation}\label{eq-TR-bound-moment-concentration-MGF-7}
 \Pr\left( \bar {h}_n^{\TR} - \mathds{E}[\bar {h}_n^{\TR} ] \geq t \right) \leq e^{- \min \left\{\frac{nt^2}{2\sigma^2}, \frac{nt}{2\Lambda }  \right\}}.
\end{equation}
%
Following similar steps, we obtain the bound, for any $t>0$,
\begin{equation}\label{eq-TR-bound-moment-concentration-MGF-9}
  \Pr\left( \bar {h}_n^{\TR} - \mathds{E}[\bar {h}_n^{\TR}] \leq -t \right) \leq e^{- \min \left\{\frac{nt^2}{2\sigma^2}, \frac{nt}{2\Lambda }  \right\}}.
\end{equation}

Combining inequalities \eqref{eq-TR-bound-moment-concentration-MGF-7} and \eqref{eq-TR-bound-moment-concentration-MGF-9}, we have
\begin{equation*}\label{eq-TR-bound-moment-concentration-MGF-10}
  \Pr\left( | \bar {h}_n^{\TR} - \mathds{E}[\bar {h}_n^{\TR} ] | \leq t \right) \geq  1- 2e^{- \min \left\{\frac{nt^2}{2\sigma^2}, \frac{nt}{2\Lambda }  \right\}}.
\end{equation*}
Setting $2e^{- \min\{nt^2/(2\sigma^2), nt/(2\Lambda)\}} =\delta$, we obtain
\begin{equation}\label{eq-TR-bound-moment-concentration-MGF-11}
  \Pr\left( |\bar {h}_n^{\TR} - \mathds{E}[\bar {h}_n^{\TR}]| \leq  \max\left( \sqrt{\frac{2 \sigma^2}{n} \ln \frac{1}{\delta} } , \frac{2\Lambda}{n} \ln \frac{1}{\delta} \right)  \right) \geq 1-\delta.
\end{equation}
Writing
\begin{equation*}\label{eq-TR-bound-moment-concentration-MGF-12}
  |\bar {h}_n^{\TR} - \bar{h}(\betheta)| = |\bar {h}_n^{\TR} - \mathds{E}[\bar {h}_n^{\TR}]  + \mathds{E}[\bar {h}_n^{\TR}] - \bar{h}(\btheta)|,
\end{equation*}
then \eqref{eq-TR-bound-moment-concentration-MGF-11} combined with the bias term bound, i.e., \eqref{eq-TR-bound-moment-bias} in Lemma \ref{theorem-TR-bound-moment}, gives \eqref{eq-TR-bound-moment-concentration-MGF-1}.
$\hfill\Box$
\endproof

The first term in the upper bound of (\ref{eq-TR-bound-moment-concentration-MGF-1}) serves as a concentration bound for $ \bar {h}_n^{\TR} $, implying the following convergence rates (see \eqref{eq-TR-bound-moment-concentration-MGF-11}):
If $n \geq ({2\Lambda^2}/{\sigma^2}) \ln({2}/{\delta})$, then the term $ \sqrt{({2 \sigma^2 }/{n}) \ln ({2}/{\delta}) }$ is dominant, implying a generally slower convergence rate $\mathcal{O}(n^{-1/2})$.

Similar to Theorem \ref{theorem-TR-bound-concentration} in Section \ref{sc:concentration-bound-inf},
the deviation of the TruLR estimator from the true value in Theorem \ref{theorem-TR-bound-moment-concentration-MGF} is bounded by terms that include both the truncation boundary $\tau_n$ and the sample size $n$. However, the upper bound in Theorem \ref{theorem-TR-bound-moment-concentration-MGF} also incorporates the norm $p$ along with $\alpha$ and $I_{\alpha}(\mathds{P}_{\betheta} \|  \mathds{P}_{\betheta_0})$.
To minimize this bound, we optimize the choice of $\tau_n$ in \eqref{eq-TR-bound-moment-concentration-MGF-1}, formalized in the following theorem.

\begin{theorem}\label{theorem-TR-bound-moment-concentration-MGF-min1}
Under the same assumptions in Theorem \ref{theorem-TR-bound-moment-concentration-MGF}, for $n \geq ({2\Lambda^2}/{\sigma^2}) \ln ({2}/{\delta}) $, the upper bound in \eqref{eq-TR-bound-moment-concentration-MGF-1} is uniquely minimized by selecting
\begin{equation*}
\tau_n^{*}=(x^{*})^{\frac{2}{\alpha} }  \left( \frac{n I_{\alpha}(\mathds{P}_{\betheta} \|  \mathds{P}_{\betheta_0}) }{\ln\frac{2}{\delta}}\right)^{\frac{1}{\alpha}}~~{with}~~x^{*}=\frac{\alpha-1-\frac{\alpha}{p}}{\sqrt{2}(1-\frac{\alpha}{2}+\frac{\alpha}{p}) },
\end{equation*}
and the TruLR estimator admits exponential concentration, i.e.,
\begin{eqnarray}
  \Pr\left(|\bar {h}_n^{\TR} - \bar{h}(\betheta)| \leq  C_2^*\| h \|_{p,\mathds{P}_{\betheta_0}}  I_{\alpha}(\mathds{P}_{\betheta} \|  \mathds{P}_{\betheta_0})^{\frac{1}{\alpha}} \left( \frac{   \ln \frac{2}{\delta}}{n}     \right)^{1-\frac{1}{\alpha}-\frac{1}{p}}\right) \geq 1-\delta,~~\delta \in (0,1), \label{eq-TR-bound-moment-concentration-MGF-2}
\end{eqnarray}
where $C_2^* = \sqrt{2}(x^{*})^{{2}/{\alpha} + {2}/{p} -1} + (x^{*})^{{2}/{\alpha}+{2}/{p}-2}$.
\end{theorem}
\proof{Proof:}

If $n$ is large enough such that $\sqrt{ {(2 \sigma^2 }/{n}) \ln ({2}/{\delta}) } \geq (2\Lambda/{n}) \ln({2}/{\delta}) $, which is equivalent to $n \geq ({2\Lambda^2}/{\sigma^2}) \ln({2}/{\delta}) $, the upper bound of absolute error in (\ref{eq-TR-bound-moment-concentration-MGF-1}) becomes
\begin{equation}\label{eq-TR-bound-moment-concentration-MGF-13}
   \sqrt{ \frac{2 \sigma^2 }{n} \ln \frac{2}{\delta} }  + \| h \|_{p,\mathds{P}_{\betheta_0}} {\tau_n}^{1-\alpha+\frac{\alpha}{p}}  I_{\alpha}(\mathds{P}_{\betheta} \|  \mathds{P}_{\betheta_0})^{1-\frac{1}{p}}.
\end{equation}
As proved before, $\sigma^2$ is actually an upper bound for $\Var(h(\tilde{\bX})l'_n(\tilde{\bX}))$, so can be taken as the bound shown in \eqref{eq-TR-bound-moment-variance}, then the bound in \eqref{eq-TR-bound-moment-concentration-MGF-13} becomes as
\begin{equation}\label{eq-TR-bound-moment-concentration-MGF-14}
  \sqrt{\frac{2 \ln \frac{2}{\delta} }{n} \|h \|_{p,\mathds{P}_{\betheta_0}}^2 {\tau_n}^{2-\alpha+\frac{2\alpha}{p}}  I_{\alpha}(\mathds{P}_{\betheta} \|  \mathds{P}_{\betheta_0})^{1-\frac{2}{p}}} + \| h \|_{p,\mathds{P}_{\betheta_0}}  {\tau_n}^{1-\alpha+\frac{\alpha}{p}}  I_{\alpha}(\mathds{P}_{\betheta} \|  \mathds{P}_{\betheta_0})^{1-\frac{1}{p}}.
\end{equation}

Note $ {p}/{(p-1)} < \alpha < {2p}/{(p-2)}$ and $p > 2 $, then $1-{\alpha}/{2}+{\alpha}/{p}>0$ and $1-\alpha+{\alpha}/{p}<0$, which means the first term in \eqref{eq-TR-bound-moment-concentration-MGF-14} is increasing with respect to $\tau_n$ while the second term is decreasing. Thus, \eqref{eq-TR-bound-moment-concentration-MGF-14} is nonnegative, continuous with respect to  $\tau_n$, and its limit is $+\infty$ as $\tau_n$ approaches 0 or $+\infty$. Minimizing \eqref{eq-TR-bound-moment-concentration-MGF-14} by solving for the zero of its first derivative gives
\begin{equation*}
\tau_n^{\dag}= (x^{*})^{\frac{2}{\alpha}} \left( \frac{  I_{\alpha}(\mathds{P}_{\betheta} \|  \mathds{P}_{\betheta_0}) n}{\ln \frac{2}{\delta}}\right)^{\frac{1}{\alpha}},~~\text{where}~~x^{*}=\frac{-(1-\alpha+\frac{\alpha}{p})}{\sqrt{2}(1-\frac{\alpha}{2}+\frac{\alpha}{p} )},
\end{equation*}
with corresponding minimum value
\begin{equation*}
  \left(\sqrt{2} (x^{*})^{\frac{2}{\alpha} + \frac{2}{p} -1} + (x^{*})^{\frac{2}{\alpha}+\frac{2}{p}-2} \right)\| h \|_{p,\mathds{P}_{\betheta_0}}  I_{\alpha}(\mathds{P}_{\betheta} \|  \mathds{P}_{\betheta_0})^{\frac{1}{\alpha}} \left( \frac{   \ln \frac{2}{\delta}}{n}     \right)^{1-\frac{1}{\alpha}-\frac{1}{p}}.
\end{equation*}

To show that the solution $\tau_n^{\dag}$ gives the unique minimum value of \eqref{eq-TR-bound-moment-concentration-MGF-14}, suppose a general truncation boundary has the form
\begin{equation*}
   \tau_n^{\frac{\alpha}{2}} = x \left( \frac{n  I_{\alpha}(\mathds{P}_{\betheta} \|  \mathds{P}_{\betheta_0}) }{\ln\frac{2}{\delta}}\right)^{\frac{1}{2}},
\end{equation*}
with $x>0$, then \eqref{eq-TR-bound-moment-concentration-MGF-14} becomes
\begin{equation*}
    g(x)\| h\|_{p,\mathds{P}_{\betheta_0}} \left( \frac{ \ln\frac{2}{\delta}}{n} \right)^{1-\frac{1}{\alpha}-\frac{1}{p}}  I_{\alpha}(\mathds{P}_{\betheta} \|  \mathds{P}_{\betheta_0})^{\frac{1}{\alpha}},
\end{equation*}
where
\begin{equation}
  g(x) = g(x;\alpha,p) = \sqrt{2} x^{\frac{2}{\alpha}+\frac{2}{p}-1}+x^{\frac{2}{\alpha}+\frac{2}{p}-2}.
\end{equation}
For simplicity, denote $k={2}/{\alpha}+{2}/{p}$. Because $p > 2$ and ${p}/({p-1}) < \alpha < {2p}/({p-2})$, then $k \in (1,2) $ and
\begin{equation}
  g(x) = g(x;k) =   \sqrt{2} x^{k-1} +x^{k-2},
\end{equation}
which has first derivative $g'(x) =x^{k-3}\left(  \sqrt{2}(k-1) x +(k-2)\right)$, and second derivative $g''(x) = (k-2)x^{k-4}\left(  \sqrt{2}(k-1) x +(k-3)\right)$. Since the optimal value $x^{*}$ satisfies $g'(x^{*}) = 0$, $ \sqrt{2}(k-1) x^{*} +(k-2) = 0$, so $g''(x^{*}) = (2-k)(x^{*})^{k-4}>0$, which proves the uniqueness of $x^{*}$.
$\hfill\Box$

\begin{remark} 
For the condition $n \geq ({2\Lambda^2}/{\sigma^2}) \ln({2}/{\delta})$ in Theorem \ref{theorem-TR-bound-moment-concentration-MGF-min1}, selecting $\Lambda$ as small {as} possible gives $n \geq 0$ if $h(\tilde{\bX})l'_n(\tilde{\bX})$ is bounded or Gaussian ($\Lambda=0$) and $n \geq 2 \ln({2}/{\delta})$ if it is exponential ($\Lambda=\sigma$). For details, see \ref{appx:remark}.
\end{remark}


Similar to Corollary \ref{theorem-TR-bound-concentration-min2} in Section \ref{sc:concentration-bound-inf}, we can choose $x^{*}=1$ and obtain the following corollary.

\begin{corollary}\label{theorem-TR-bound-moment-concentration-MGF-min2}
Under the same assumptions in Theorem \ref{theorem-TR-bound-moment-concentration-MGF}, for $n \geq ({2\Lambda^2}/{\sigma^2}) {\ln ({2}/{\delta})} $ and
$$\tau_n^{\dag} =   \left( \frac{n I_{\alpha}(\mathds{P}_{\betheta} \|  \mathds{P}_{\betheta_0})}{\ln{\frac{2}{\delta}}}\right)^{\frac{1}{\alpha}},$$
the TruLR estimator admits exponential concentration, i.e.,
\begin{equation}\label{eq-TR-bound-moment-concentration-MGF-3}
  \Pr\left(|\bar {h}_n^{\TR} - \bar{h}(\betheta)| \leq \| h \|_{p,\mathds{P}_{\betheta_0}} (\sqrt{2}+1)I_{\alpha}(\mathds{P}_{\betheta} \|  \mathds{P}_{\betheta_0})^{\frac{1}{\alpha}} \left( \frac{   \ln \frac{2}{\delta}}{n}     \right)^{1-\frac{1}{\alpha}-\frac{1}{p}} \right) \geq 1-\delta,~~\delta \in(0,1).
\end{equation}
\end{corollary}

From Theorem \ref{theorem-TR-bound-moment-concentration-MGF-min1} and Corollary \ref{theorem-TR-bound-moment-concentration-MGF-min2} (under the assumptions therein), we observe the following:
\begin{itemize}
\item
The upper bound on the absolute error of the TruLR estimator converges to zero  at rate $\mathcal{O}(n^{-1+1/\alpha+1/p})$, compared with the results in Theorem \ref{theorem-TR-bound-concentration-min1} and Corollary \ref{theorem-TR-bound-concentration-min2}, which is slower by a factor of $1/p$, whereas the truncation boundary increases at the same rate $\mathcal{O}(n^{1/\alpha})$.
\item
The upper bound in Corollary \ref{theorem-TR-bound-moment-concentration-MGF-min2} is slightly looser than that in Theorem \ref{theorem-TR-bound-moment-concentration-MGF-min1}, because the truncation boundary in Corollary \ref{theorem-TR-bound-moment-concentration-MGF-min2} is simpler, as it only depends on the values of $\delta$, $\alpha$, and $I_{\alpha}(\mathds{P}_{\betheta} \| \mathds{P}_{\betheta_0})$, whereas the truncation boundary in Theorem \ref{theorem-TR-bound-moment-concentration-MGF-min1} also requires the estimation of an additional parameter $x^{*}$, which includes the norm $p$.
\item
The TruLR estimators exhibit exponential concentration when appropriate truncation boundaries are chosen, whereas the classical (untruncated) LR estimator only achieves polynomial concentration in Propositions \ref{theorem-IS-bound-concentration} and \ref{theorem-IS-bound-concentration-lower-1}.
\end{itemize}

The previous results require the existence of the MGF of $h(\tilde{\bX}), \tilde{\bX}\sim \mathds{P}_{\betheta_0}$ (i.e., Assumption \ref{assump-MGF}), but this assumption may not be easily verifiable, so we offer an alternative condition under which we derive an appropriate truncation boundary in Proposition \ref{theorem-TR-bound-moment-concentration-min1} (proof in \ref{appx:P6-proof}).

\begin{definition}[Bernstein's moment condition]\label{def:bernstein}
Random variable $Y$ satisfies Bernstein moment condition with $b >0$ if
$$ \mathds{E}[ |Y|^k] \leq \frac{1}{2} \mathds{E}[ Y^2] k!  b^{k-2},~~\forall k\geq 2.$$
\end{definition}

\begin{proposition}\label{theorem-TR-bound-moment-concentration-min1}
If $h(\tilde{\bX}), \tilde{\bX}\sim \mathds{P}_{\betheta_0}$, satisfies Bernstein's moment condition for parameter $b_h>0$, $\| h \|_{p,\mathds{P}_{\betheta_0}} < \infty$ for some $p > 2 $, and $I_{\alpha}(\mathds{P}_{\betheta_1} \| \mathds{P}_{\betheta_0})$ is finite for some $\alpha >1$ ($ \frac{p}{p-1} \leq \alpha \leq \frac{2p}{p-2}$), then for
$$\tau_n^{*} = (x^{*} )^{\frac{2}{\alpha}}  \left( \frac{n I_{\alpha}(\mathds{P}_{\betheta} \|  \mathds{P}_{\betheta_0}) }{\ln\frac{2}{\delta}}\right)^{\frac{1}{\alpha}},$$
with
$$x^{*} = \frac{-\sqrt{2}(1-\frac{\alpha}{2}+\frac{\alpha}{p}) + \sqrt{2(1-\frac{\alpha}{2}+\frac{\alpha}{p})^2 - 4b(1+\frac{\alpha}{p})(1-\alpha+\frac{\alpha}{p}))}}{ 2b(1+\frac{\alpha}{p})},$$
where  $b=b_h I_{\alpha}(\mathds{P}_{\betheta} \|  \mathds{P}_{\betheta_0})^{\frac{1}{p}} / \|h \|_{p,\mathds{P}_{\betheta_0}}$, the TruLR estimator admits exponential concentration, i.e.,
\begin{equation}
  \Pr \Bigg(|\bar {h}_n^{\TR} - \bar{h}(\betheta)|   \leq   C_3^* \| h \|_{p,\mathds{P}_{\betheta_0}}  \left( \frac{\ln\frac{2}{\delta}}{n} \right)^{1-\frac{1}{\alpha}-\frac{1}{p}} I_{\alpha}(\mathds{P}_{\betheta} \|  \mathds{P}_{\betheta_0})^{\frac{1}{\alpha}}\Bigg) \geq 1-\delta,~~\delta \in(0,1), \label{eq-TR-bound-moment-concentration-0}
\end{equation}
where $C_3^* =  \sqrt{2} (x^{*})^{{2}/{\alpha}+{2}/{p}-1} + b (x^{*})^{{2}/{\alpha}+{2}/{p}} +(x^{*})^{{2}/{\alpha}+{2}/{p}-2} $.
\end{proposition}

Similar to Theorem \ref{theorem-TR-bound-moment-concentration-MGF-min1} and Corollary \ref{theorem-TR-bound-moment-concentration-MGF-min2}, Proposition \ref{theorem-TR-bound-moment-concentration-min1} also implies that the TruLR estimator error converges to zero at rate (at least) $\mathcal{O}(n^{-1+1/\alpha+1/p})$ with exponential concentration, again as the truncation boundary grows at rate $\mathcal{O}(n^{1/\alpha})$.

\section{Simulation Experiments}\label{sec:num}
In this section, we conduct simulation experiments to test the performance of the TruLR estimators on a set of synthetic examples, followed by examples of offline RL policy evaluation and estimating the value of a financial portfolio.

\subsection{Synthetic Examples}
We consider two synthetic examples, where the (output) performance function $h(X)$ is simply the input, i.e., $h(X)=X$, where the single r.v. $X$ follows a beta distribution and a normal distribution. For the beta distribution, which has bounded support, we use the TruLR estimators proposed in Section \ref{sc:concentration-bound-inf}.
For the unbounded normal distributions, we use the TruLR estimators proposed in Section \ref{sc:concentration-bound-p}. \ref{appx:chi2} gives another example for the chi-squared distribution.

\subsubsection{Beta Distribution.}
\begin{figure}[htpb]
\begin{minipage}[t]{0.48\linewidth}
\centering
\includegraphics[width=8.2cm]{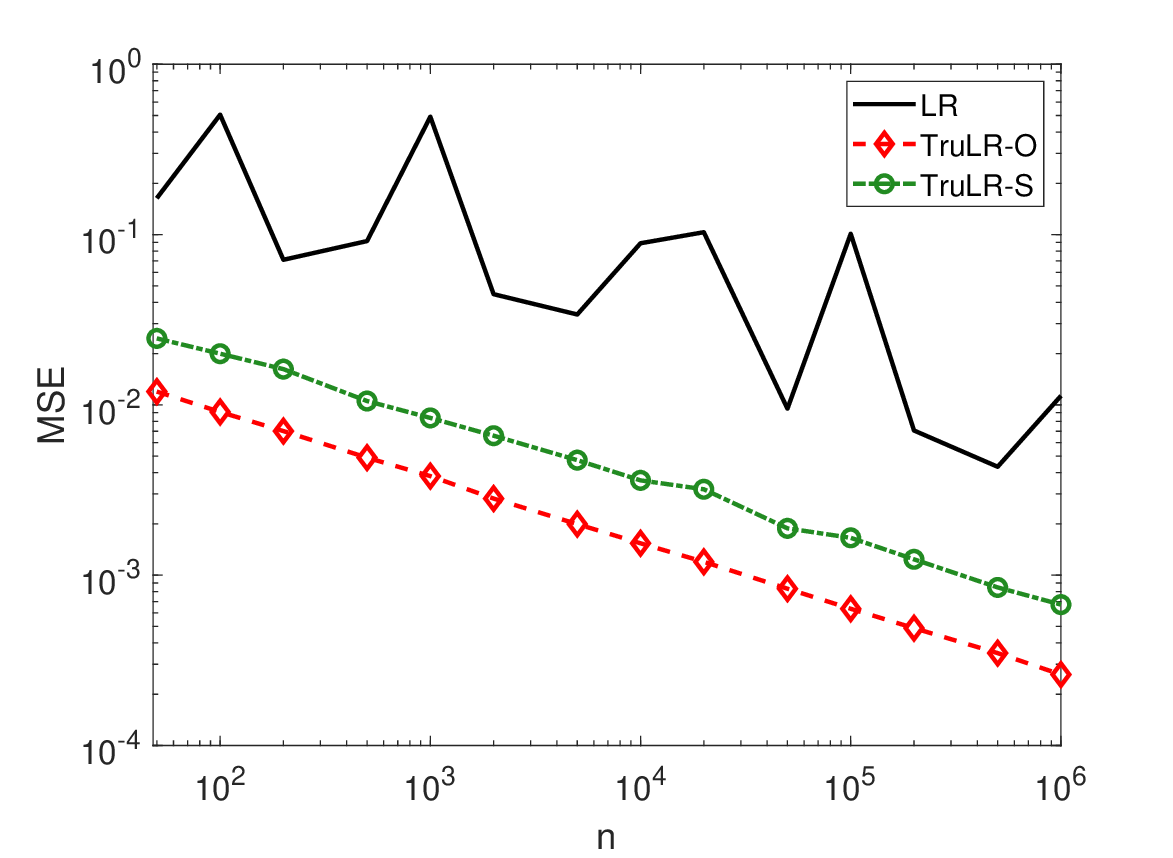}
\end{minipage}
\begin{minipage}[t]{0.48\linewidth}
\centering
\includegraphics[width=8.2cm]{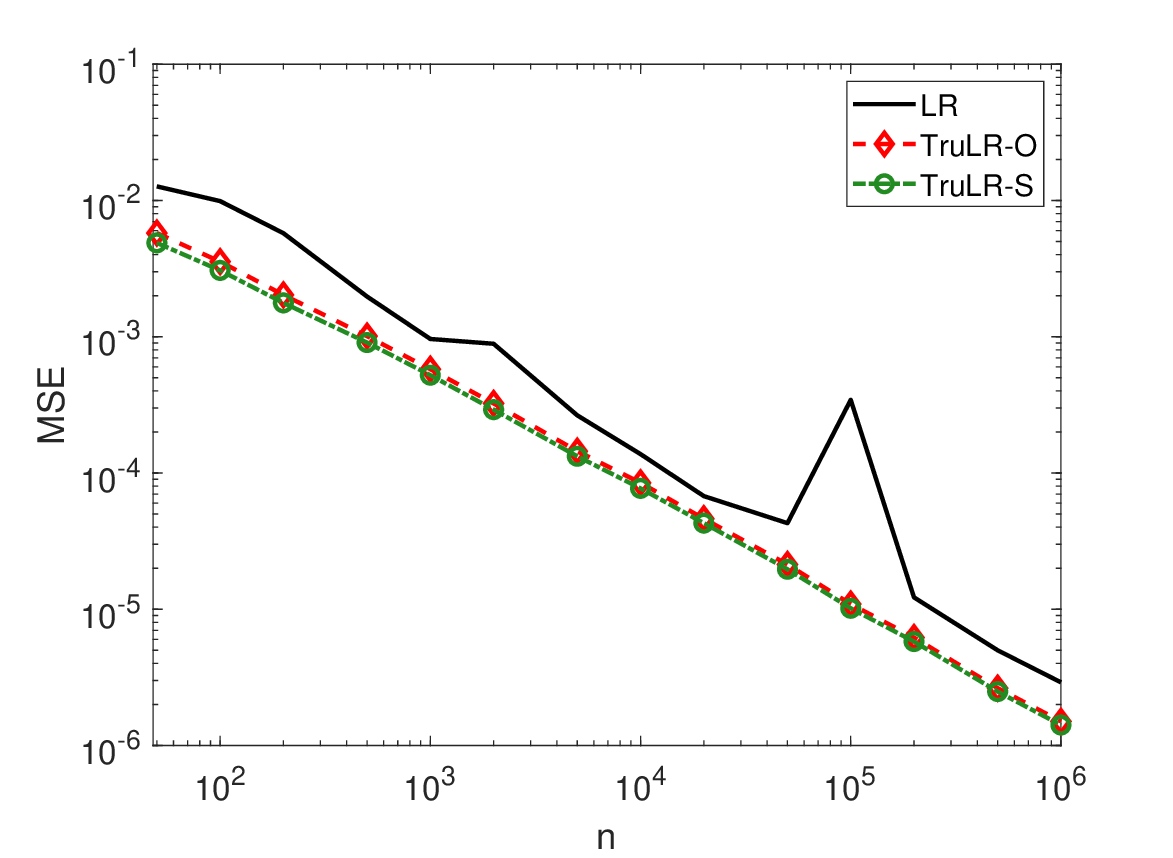}
\end{minipage}
\caption{\baselineskip10pt MSEs of LR and TruLR estimators for beta distribution. Left graph is for case (i) $(a_0,b_0)=(90,120)$, $(a,b)=(16,21)$, $\alpha=1.2$; right graph is for case (ii) $(a_0,b_0)=(44,22)$, $(a,b)=(20,10)$, $\alpha=1.8$. Each MSE is computed based on 5000 independent replications ($\delta=0.01$).}
\label{fig:beta}
\end{figure}

We consider a beta distributed random variable which is defined in the interval $(0,1)$ with PDF $f(x)=1/B(a,b)x^{a-1}(1-x)^{b-1}$, where $B(a,b)=\Gamma(a)\Gamma(b)/\Gamma(a+b)$ is the beta function, $\Gamma(x)=\int_0^{\infty} t^{x-1}e^{-t}dt$ is the gamma function, parameters $a>0$ and $b>0$.
Given the data generated under $\btheta_0=(a_0,b_0)$, we want to estimate $\mathds{E}[X]$ for $\btheta=(a,b)$. Notice that $I_{\alpha}(\mathds{P}_{\betheta} \| \mathds{P}_{\betheta_0}) = \left( B(a_0,b_0)/B(a,b) \right)^{\alpha-1}\left( B(a_\alpha,b_\alpha)/B(a,b) \right)$, $a_\alpha = \alpha a+(1-\alpha) a_0$ and $b_\alpha = \alpha b+(1-\alpha) b_0$. Ensuring that $a_\alpha>0$ and $b_{\alpha}>0$ imposes certain requirements on the relationship between $a$, $a_0$, $b$, $b_0$, and $\alpha$.
We consider two settings: (i) $(a_0,b_0)=(90,120)$, $(a,b)=(16,21)$, $\alpha=1.2$; (ii) $(a_0,b_0)=(44,22)$, $(a,b)=(20,10)$, $\alpha=1.8$. We denote the results obtained using the TruLR estimator with the optimal truncation boundary from Theorem \ref{theorem-TR-bound-concentration-min1} as ``TruLR-O'', and the results using the TruLR estimator with the simple truncation boundary from Corollary \ref{theorem-TR-bound-concentration-min2} as ``TruLR-S'', which in the example is equivalent to the TruLR estimator proposed in \cite{papini2019optimistic}. The results obtained using the classical LR estimator are denoted as ``LR''. We vary the number of samples $n$ from $50$ to $10^6$, and the MSEs of these estimators are graphed in Figure \ref{fig:beta}. Since the LR estimator is unbiased, its MSE is just its variance.

From Figure \ref{fig:beta}, we observe that both the TruLR-O and TruLR-S estimators can significantly improve the accuracy of the estimation. In particular, for setting (i), where the distance between the nominal $\btheta_0$ and the target $\btheta$ is relatively large, the MSEs of the LR estimators are quite large and highly variable compared with the TruLR-O and TruLR-S estimators. For example, for sample size $n=50000$, the MSE of the LR estimator is around 0.01, whereas the MSEs of the TruLR-O and TruLR-S estimators are approximately 0.0008 and 0.0019, respectively, corresponding to a reduction in MSE by approximately 12.5 and 5.3 times, respectively. Additionally, the TruLR-O estimator outperforms the TruLR-S estimator in this setting.
In setting (ii), where the distance between the nominal $\btheta_0$ and the target $\btheta$ is smaller compared to setting (i), the TruLR-O and TruLR-S estimators perform similarly, and the improvement over the LR estimator is not as significant. When $n=50000$, the MSE of the LR estimator is approximately $4\times10^{-5}$, compared to around $2\times10^{-5}$ for the MSEs of the TruLR-O and TruLR-S estimators.

\begin{figure}[htpb]
\begin{minipage}[t]{0.48\linewidth}
\centering
\includegraphics[width=8.2cm]{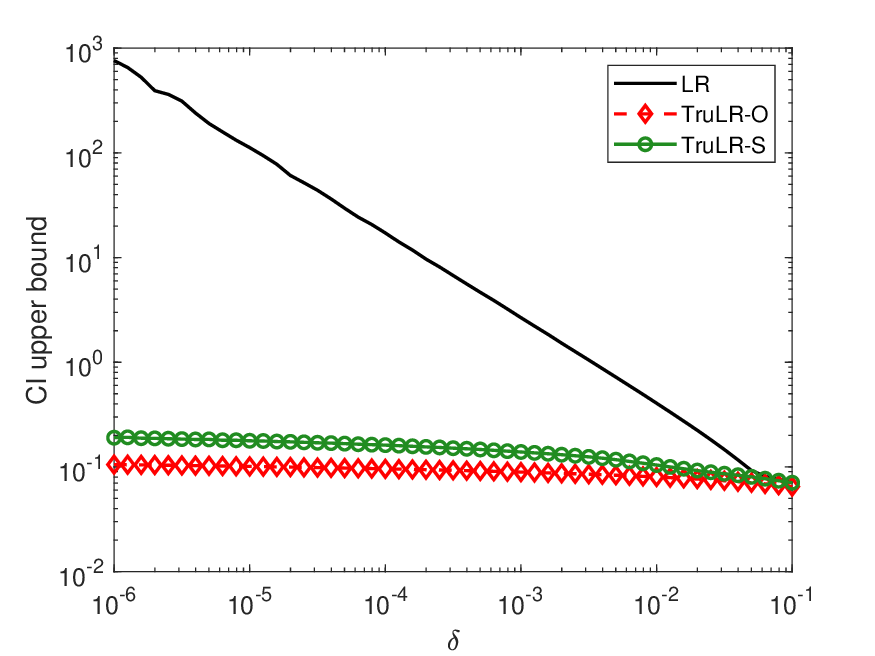}
\end{minipage}
\begin{minipage}[t]{0.48\linewidth}
\centering
\includegraphics[width=8.2cm]{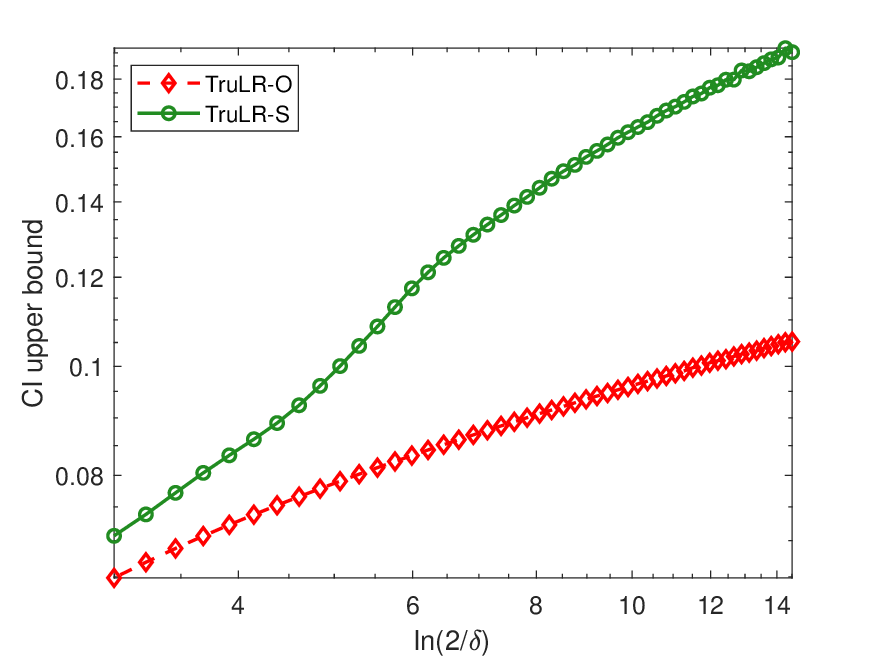}
\end{minipage}
\caption{\baselineskip10ptError upper bounds  for LR and TruLR estimators in Case (i) of the beta distribution: $(a_0,b_0)=(90,120)$, $(a,b)=(16,21)$, $\alpha=1.2$, $n=5000$. Each upper bound is computed based on $10^8$ independent replications. The left graph is the log-log plot of the error upper bound versus $\delta$, whereas the right graph is the log-log plot of the error upper bound versus $\ln(2/\delta)$.}
\label{fig:beta_CI}
\end{figure}

Next, we empirically test how well the LR estimator and TruLR estimators follow the polynomial concentration and exponential concentration, respectively. Specifically, we vary $\delta$ from $10^{-1}$ to $10^{-6}$ in Case (i) and estimate $(1-\delta)$-quantile confidence interval (CI) upper bounds of estimator errors based on $10^{8}$ independent replications. The resulting plots of the estimated upper bounds with respect to $\delta$ are presented in Figure \ref{fig:beta_CI}.
Theoretical analysis predicts that the error upper bound for LR estimators follows a polynomial concentration, $C_1(2/\delta)^{\beta_1}$.
In contrast, the error upper bound for TruLR estimators adheres to an exponential concentration, $C_2(\ln(2/\delta))^{\beta_2}$, thus a linear relationship is expected in the log-log plot of the error upper bound versus $\ln(2/\delta)$ {in} the right graph.

\subsubsection{Normal Distribution.}\label{test:normal}
We consider a normally distributed random variable $X \sim \mathcal{N}(\mu,\sigma^2)$. Since the output is unbounded, the results of Section \ref{sc:concentration-bound-inf} do not apply, we employ the truncation boundaries in Section \ref{sc:concentration-bound-p}, as both the moment generating function condition (Assumption \ref{assump-MGF}) and Bernstein's moment condition (Definition \ref{def:bernstein}) are satisfied. Given the data generated under $\btheta_0=(\mu_0,\sigma_0)$, we want to estimate $\mathds{E}[X]$ under $\btheta=(\mu,\sigma)$, $\sigma \geq 0$. For the normal distribution,
$I_{\alpha}(\mathds{P}_{\betheta} \| \mathds{P}_{\betheta_0}) =  \exp\big(\alpha(\alpha)(\mu_0-\mu)^2/(2\sigma^2_\alpha)\big)\sigma^{1-\alpha}\sigma_0^{\alpha}/\sigma_\alpha $, where $\sigma_{\alpha}^2 = (1-\alpha)\sigma^2 + \alpha\sigma_0^2>0$, and again the requirement that $\sigma_{\alpha}\geq 0$ imposes constraints on the values of $\alpha$ and $\sigma$.

\begin{figure}[htpb]
\begin{minipage}[t]{0.48\linewidth}
\centering
\includegraphics[width=8.2cm]{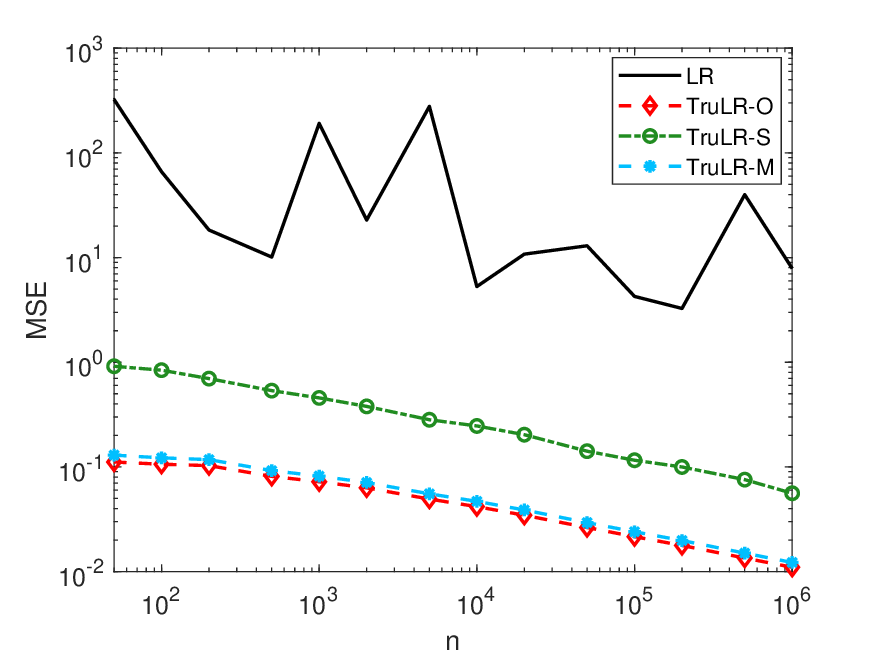}
\end{minipage}
\begin{minipage}[t]{0.48\linewidth}
\centering
\includegraphics[width=8.2cm]{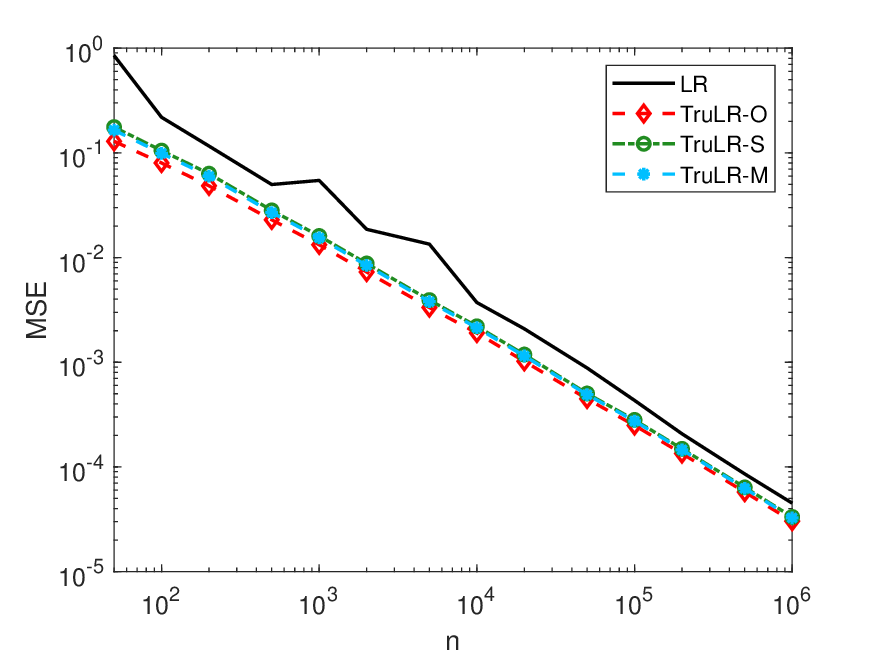}
\end{minipage}
\caption{\baselineskip10pt MSEs of LR and TruLR estimators for normal distribution. Left graph is for case (i) $(\mu_0,\sigma_0)=(0,1.7)$, $(\mu,\sigma)=(0.2,4)$, $\alpha=1.2$, $p=40$; right graph is for case (ii) $(\mu_0,\sigma_0)=(1,1.5)$, $(\mu,\sigma)=(0.6,2)$, $\alpha=2.1$, $p=4$. Each MSE is computed based on $5000$ independent replications ($\delta=0.01$).}
\label{fig:normal}
\end{figure}

In this example, we consider two settings: (i) $(\mu_0,\sigma_0)=(0,1.7)$, $(\mu,\sigma)=(0.2,4)$, $\alpha=1.2$; (ii) $(\mu_0,\sigma_0)=(1,1.5)$, $(\mu,\sigma)=(0.6,2)$, $\alpha=2.1$. In addition, we set $p=40$ for setting (i) and $p=4$ for setting (ii) to satisfy the constraint relationship between $\alpha$ and $p$ in Theorem \ref{theorem-TR-bound-moment-concentration-MGF-min1}, Corollary \ref{theorem-TR-bound-moment-concentration-MGF-min2}, and Proposition \ref{theorem-TR-bound-moment-concentration-min1}.
Since the normal distribution satisfies the moment generating function condition (Assumption \ref{assump-MGF}), we denote the TruLR estimator with the optimal truncation boundary in Theorem \ref{theorem-TR-bound-moment-concentration-MGF-min1} as ``TruLR-M''. The TruLR estimator with the simple truncation boundary in Corollary \ref{theorem-TR-bound-moment-concentration-MGF-min2} is denoted as ``TruLR-S''. For a normal random variable $X$, the Bernstein's constant $b_h$ in Proposition \ref{theorem-TR-bound-moment-concentration-min1} can be chosen to equal its standard deviation (\citealt{zhang2021concentration}). Thus, we denote the TruLR estimator with the optimal truncation boundary in Proposition \ref{theorem-TR-bound-moment-concentration-min1} using this Bernstein's constant as ``TruLR-O''. Then, we vary the number of sample $n$ from $50$ to $10^6$, and the MSEs of these estimators are shown in Figure \ref{fig:normal}.
For general distributions, an approximation method is provided in \ref{appx:para} to find a Bernstein's constant $b_h$, and we find that our  truncation boundary method is not sensitive to the choice of Bernstein's constant $b_h$; see \ref{appx:normal} for more simulation results.

We obtain results similar to those in the beta distribution example. When the distance between the nominal $\btheta_0$ and the target $\btheta$ is large, as in setting (i), the TruLR estimators significantly improve the accuracy of the estimation, yielding a smaller MSE compared to the LR estimators.
However, when the distance between the nominal $\btheta_0$ and the target $\btheta$ is small, as in setting (ii), this improvement diminish. In addition, the TruLR-O estimators are comparable to the TruLR-M estimator in terms of estimation accuracy. For example, when $n=50000$, the MSEs of the TruLR-O and TruLR-M are 0.026 {and} 0.029, respectively, and their MSEs are smaller than that of the TruLR-S, which is 0.141.
In setting (i), while TruLR-S shows some improvement in estimation accuracy compared to LR, it does not perform as well as TruLR-O or TruLR-M. However, in setting (ii), TruLR-S performs similarly to the other two TruLR estimators in terms of estimation accuracy. For example, when $n=50000$, the MSEs of the TruLR-O, TruLR-M and TruLR-S are 0.00045, 0.00049 and 0.00050, respectively.

\begin{figure}[htpb]
\begin{minipage}[t]{0.48\linewidth}
\centering
\includegraphics[width=8.2cm]{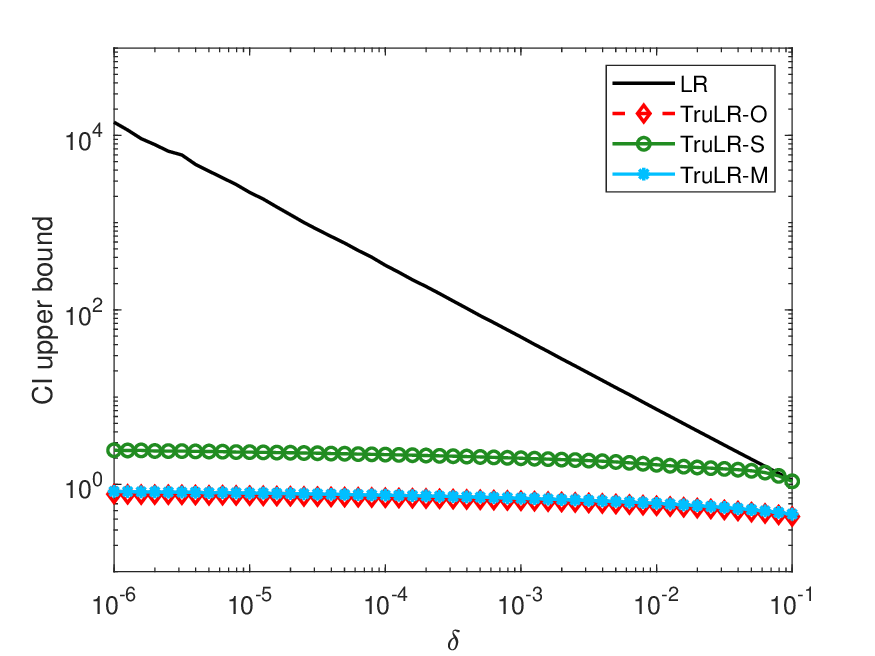}
\end{minipage}
\begin{minipage}[t]{0.48\linewidth}
\centering
\includegraphics[width=8.2cm]{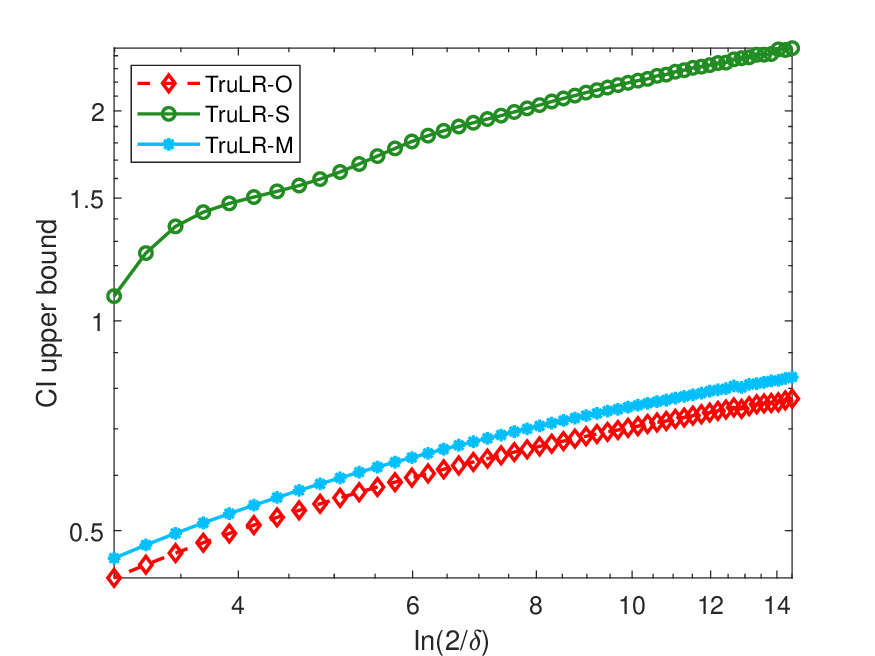}
\end{minipage}
\caption{\baselineskip10pt Error upper bounds for LR and TruLR estimators in Case (i) of the normal distribution: $(\mu_0,\sigma_0)=(0,1.7)$, $(\mu,\sigma)=(0.2,4)$, $\alpha=1.2$, $p=40$, $n=5000$. Each upper bound is computed based on $10^8$ independent replications. The left graph is the log-log plot of the error upper bound versus $\delta$, whereas the right graph is the log-log plot of the error upper bound versus $\ln(2/\delta)$.}
\label{fig:norm_CI}
\end{figure}

Next, we examine the concentration properties of the LR and TruLR estimators, and show the results in Figure \ref{fig:norm_CI}. Similar to the beta distribution example, the error upper bounds of the LR estimator appear as a straight line in the left log-log plot, whereas those of the TruLR estimators exhibit linear trends in the right log-log plot against $\ln(2/\delta)$.

\subsection{Offline Policy Evaluation}
Offline policy evaluation is a commonly used method for solving contextual multi-armed bandit problems. As in the setting of Example \ref{exm:offline} in Section \ref{sec:Intro}, we consider a randomized policy where, upon observing a context ${\bx}_t$, we generate an action from policy $\tilde{a}_{t} \sim \pi_{\theta_0}\left(\cdot | \bx_{t}\right)$, where the parameter $\theta_0$ determines the current policy. When we switch to another policy $\pi_{\theta}$, we aim to use the data collected under policy $\pi_{\theta_0}$ to estimate the value $v(\pi_{\theta})$ {defined as}
\begin{equation}\label{eq:CMABvalue_theta}
v\left(\pi_{\theta}\right)=\frac{1}{|\mathscr{X}_{\text{eval}}|} \sum_{\bx_i \in \mathscr{X}_{\text{eval}}} \sum_{{a} \in \bA} \pi_{\theta}({a} | \bx_i) r\left(\bx_i, a_i,{a}\right),
\end{equation}
where $\mathscr{X}_{\text{eval}}$ is the context space for evaluating the policy $\pi_{\theta}$, $\bA$ is the action space, $r(\bx_i, a_i, a)$ is the one-period reward for action $a$ corresponding to context ${\bx_i}$ for (unknown) target $a_i \in \bA$.

We use the Letter Recognition classification dataset\footnote{https://archive.ics.uci.edu/dataset/59/letter+recognition} from the UCI Machine Learning Repository as the contextual multi-armed bandit example, which is the same as in \cite{Metelli2021}. Specifically, the letter dataset contains a total of 20,000 instances, each with 16 features. The label for each instance is a single letter from the 26 letters of the English alphabet.
We represent the dataset as $\mathscr{X} = \{({\bx}_1,{a}_1),\ldots,({\bx}_n,{a}_n)\}$, where ${\bx}_i$ denotes the 16-dimensional feature vector of instance $i$, and ${a}_i$ denotes the true label of instance $i$, with ${a}_i\in\bA=\{1,2,\ldots,26\}$.
Each sample $({\bx}_i,{a}_i)$ corresponds to a binary reward given by $r({\bx}_i,{a}_i,\tilde{a})$, where $\tilde{a}\in\bA$ is the action and
\begin{equation}\label{eq:rewardCMAB1}
r\left({\bx}_i, {a}_i,\tilde{a}\right)=\left\{\begin{array}{l}
1,~~\tilde{a}={a}_i, \\
0,~~\tilde{a} \neq {a}_i,
\end{array}\right.
\end{equation}
i.e., correct classification receives a reward of $1$, and incorrect classifications receive no reward.

We divide the dataset $\mathscr{X}$ into a training set $\mathscr{X}_{\text{train}}$ (30\%) and an evaluation set $\mathscr{X}_{\text{eval}}$ (70\%). The training set $\mathscr{X}_{\text{train}}$ is used to train a classifier $\mathcal{C}(\cdot)$, which provides classification predictions $\mathcal{C}({\bx}_i)$ for any ${\bx}_i$. We define (randomized) policy $\pi_{\theta_0}(\tilde{a}| {\bx})$ as follows:
\begin{equation}\label{eq:policyb}
\pi_{\theta_0}(\tilde{a} | {\bx}) = \left\{\begin{aligned}
{\theta_0} + \frac{1 - {\theta_0}}{26}, &~~\text{if } \tilde{a} = \mathcal{C}({\bx}), \\
\frac{1 - {\theta_0}}{26},~~~ &~~\text{if } \tilde{a} \in \bA \setminus \mathcal{C}({\bx}),
\end{aligned}\right.
\end{equation}
where ${\theta_0}\in[0,1]$. The policy chooses the letter selected by the classifier $\mathcal{C}$ with probability ${\theta_0} + {(1 - {\theta_0})}/{26}$ and randomly selects one of the remaining 25 letters with equal probability ${(1 - {\theta_0})}/{26}$.
For another policy (usually called a target policy) $\pi_{\theta}(a| \bx)$, our goal is to use the data collected by the behavior policy $\pi_{\theta_0}(\tilde{a}| \bx)$ to evaluate the value of target policy $v\left(\pi_{\theta}\right)$.

\begin{figure}[htpb!]
\begin{minipage}[t]{0.48\linewidth}
\centering
\includegraphics[width=8.2cm]{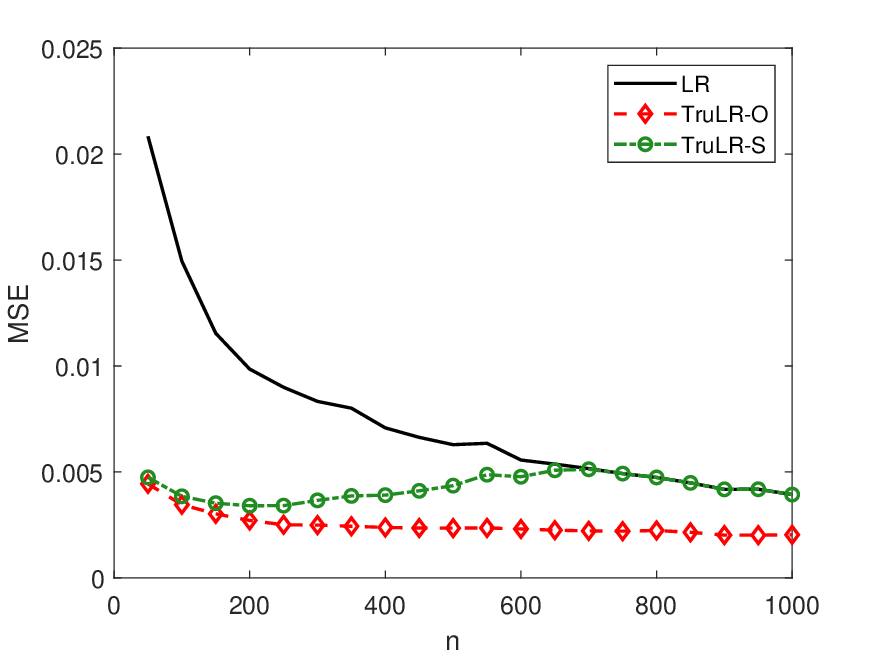}
\end{minipage}
\begin{minipage}[t]{0.48\linewidth}
\centering
\includegraphics[width=8.2cm]{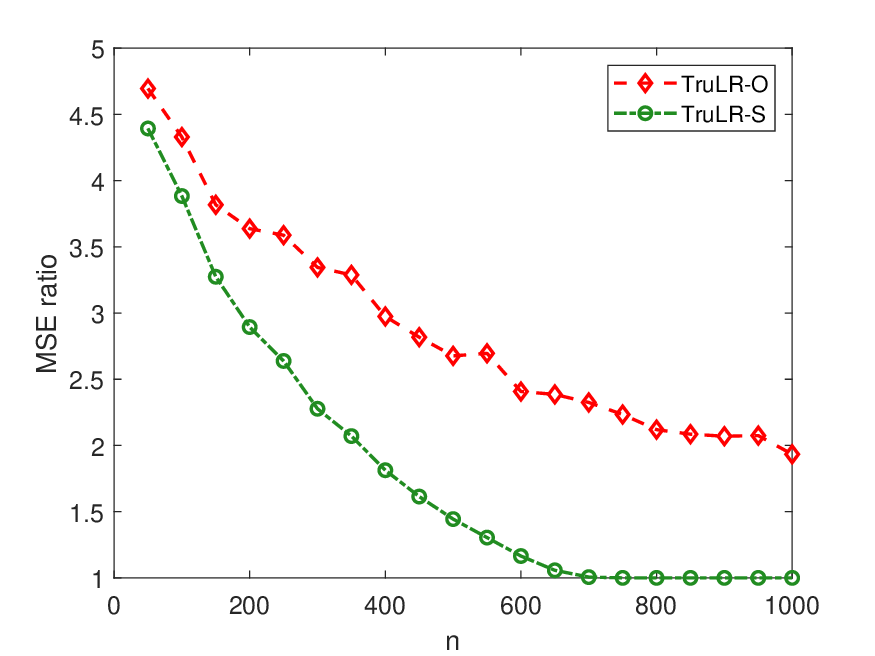}
\end{minipage}
\caption{\baselineskip10pt Results for Letter Recognition classification problem with binary reward ($\theta_0=0.5$, $\theta=0.99$ and $\alpha = 1.3$). The left graph plots the MSEs of the LR, TruLR-O, and TruLR-S estimators, whereas the right graph plots the MSE ratios of the LR estimator to the TruLR-O and TruLR-S estimators. Each MSE is estimated based on $5000$ independent replications ($\delta=0.01$).}
\label{fig:CMAB01}
\end{figure}

As the reward given by \eqref{eq:rewardCMAB1} is bounded, we can apply the truncation boundaries derived by Theorem \ref{theorem-TR-bound-concentration-min1} and Corollary \ref{theorem-TR-bound-concentration-min2}. For $\theta_0=0.5$, $\theta=0.99$, $\delta=0.01$, and $\alpha = 1.3$, we obtain Figure \ref{fig:CMAB01} and Table \ref{tab:CMAB01}, from which we observe: (i) When the sample size is small ($ n\leq400$), both TruLR-O and TruLR-S have higher estimation accuracy in policy evaluation compared to the classical LR method; (ii) Among the two truncation methods, TruLR-O is better than TruLR-S. When the sample size is large, TruLR-O consistently outperforms the classical LR, whereas TruLR-S performs similarly to LR.

\begin{table}[htpb!]
\scriptsize
\centering
		\caption{\baselineskip10pt Mean and standard error of MSE of LR and TruLR estimators for Letter Recognition classification problem with binary reward (based on $5000$ independent replications).}\label{tab:CMAB01}
\begin{tabular}{cccccccccccc}
\toprule
\multicolumn{2}{c}{$n$}         & 100      & 200      & 300      & 400      & 500      & 600      & 700      & 800      & 900      & 1000     \\
\midrule
\multirow{2}{*}{LR}    & mean & 1.49E-02 & 9.85E-03 & 8.33E-03 & 7.08E-03 & 6.29E-03 & 5.56E-03 & 5.16E-03 & 4.74E-03 & 4.18E-03 & 3.93E-03 \\
                       & std err  & 1.12E-03 & 3.91E-04 & 2.26E-04 & 2.24E-04 & 2.07E-04 & 2.15E-04 & 1.22E-04 & 1.62E-04 & 9.45E-05 & 7.72E-05 \\
                       \hline
\multirow{2}{*}{TruLR-O} & mean & 3.45E-03 & 2.71E-03 & 2.49E-03 & 2.38E-03 & 2.35E-03 & 2.31E-03 & 2.22E-03 & 2.24E-03 & 2.02E-03 & 2.03E-03 \\
                       & std err   & 9.00E-05 & 5.74E-05 & 4.52E-05 &3.52E-05 & 5.97E-05 & 4.76E-05 & 3.71E-05 & 3.35E-05 & 3.12E-05 & 3.54E-05 \\
                       \hline
\multirow{2}{*}{TruLR-S} & mean & 3.85E-03 & 3.40E-03 & 3.66E-03 & 3.91E-03 & 4.35E-03 & 4.77E-03 & 5.13E-03 & 4.74E-03 & 4.18E-03 & 3.93E-03 \\
                       & std err   & 9.27E-05 &7.35E-05 & 7.37E-05 & 8.87E-04 & 1.30E-04 & 1.77E-04 & 1.19E-04 & 1.62E-04 & 9.45E-05 & 7.72E-05 \\
                       \bottomrule
\end{tabular}
\end{table}

\begin{figure}[htpb!]
\begin{minipage}[t]{0.48\linewidth}
\centering
\includegraphics[width=8.2cm]{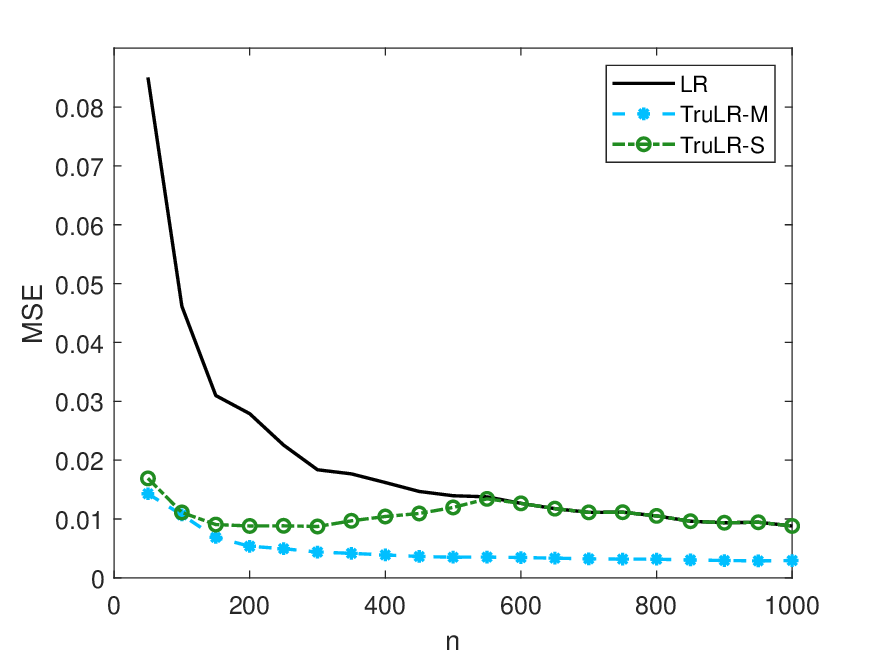}
\end{minipage}
\begin{minipage}[t]{0.48\linewidth}
\centering
\includegraphics[width=8.2cm]{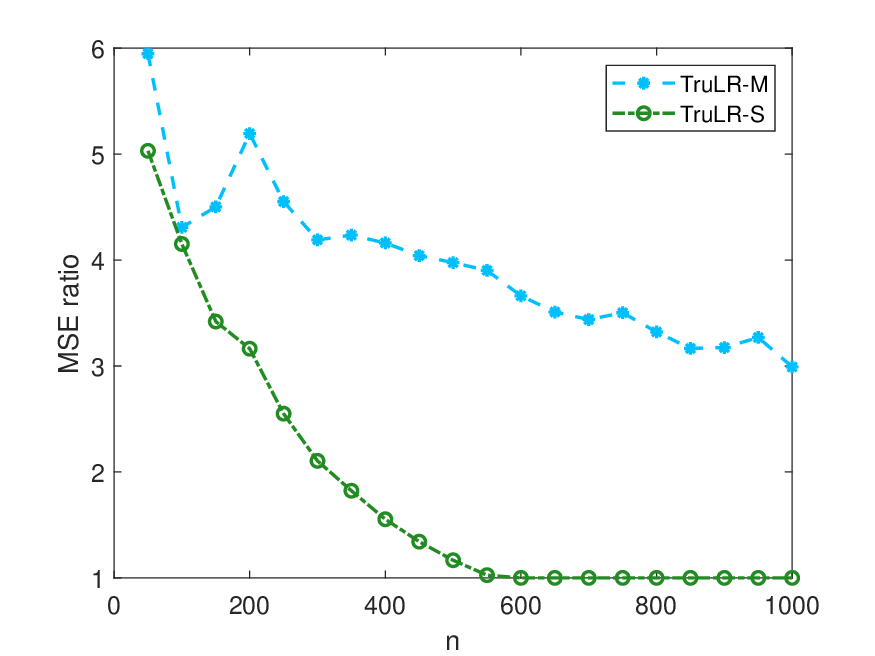}
\end{minipage}
\caption{\baselineskip10pt Results  for Letter Recognition classification problem with normal reward ($\theta_0=0.5$, $\theta=0.99$ and $\alpha = 1.2$). The left graph plots the MSEs of the LR, TruLR-M, and TruLR-S estimators, whereas the right graph plots the MSE ratios of the LR estimator to the TruLR-M and TruLR-S estimators. Each MSE is estimated based on $5000$ independent replications ($\delta=0.01$).}
\label{fig:CMAB02}
\end{figure}

Next, we modify the deterministic binary reward \eqref{eq:rewardCMAB1} by adding a zero-mean Gaussian noise:
\begin{equation}\label{eq:rewardCMAB2}
r\left(\bx_i, a_i,\tilde{a}\right) \sim \left\{\begin{array}{l}
\mathcal{N}(1,0.5^2),~~\tilde{a}=a_i, \\
\mathcal{N}(0,0.5^2),~~\tilde{a} \neq a_i.
\end{array}\right.
\end{equation}
Since rewards are no longer bounded, we use the truncation boundaries proposed in Theorem \ref{theorem-TR-bound-moment-concentration-MGF-min1} and Corollary \ref{theorem-TR-bound-moment-concentration-MGF-min2}. For $\theta_0=0.5$, $\theta=0.99$, $\delta=0.01$, and $\alpha = 1.2$, we obtain Figure \ref{fig:CMAB02} and Table \ref{tab:CMAB02}, which show similar results to those with the binary reward: TruLR-M performs better than TruLR-S and the classical LR, and for small sample size, TruLR-S performs better than the classical LR.

\begin{table}[htpb!]
\scriptsize
\centering
		\caption{\baselineskip10pt Mean and standard error of MSE of LR and TruLR estimators for Letter Recognition classification problem with normal reward (based on $5000$ independent replications). }\label{tab:CMAB02}
\begin{tabular}{cccccccccccc}
\toprule
\multicolumn{2}{c}{$n$}         & 100      & 200      & 300      & 400      & 500      & 600      & 700      & 800      & 900      & 1000     \\
\midrule
\multirow{2}{*}{LR}    & mean & 4.61E-02 & 2.79E-02 & 1.84E-02 & 1.62E-02 & 1.40E-02 & 1.27E-02 & 1.11E-02 & 1.05E-02 & 9.37E-03 & 8.82E-03 \\
                       & std err   & 2.21E-03 &1.28E-03 & 4.66E-04 & 5.30E-04 & 4.02E-04 & 3.19E-04 & 2.53E-04 & 2.62E-04 & 2.13E-04 & 2.48E-04 \\
                       \hline
\multirow{2}{*}{TruLR-M} & mean & 1.07E-02 & 5.37E-03 & 4.38E-03 & 3.89E-03 & 3.51E-03 & 3.46E-03 & 3.24E-03 & 3.17E-03 & 2.95E-03 & 2.95E-03 \\
                       & std err  & 1.63E-04 & 1.02E-04 & 9.02E-05 & 7.10E-05 & 7.65E-05 & 5.73E-05 & 5.38E-05 & 4.58E-05 & 4.83E-05 & 6.56E-05 \\
                       \hline
\multirow{2}{*}{TruLR-S} & mean & 1.11E-02 & 8.82E-03 & 8.73E-03 & 1.04E-02 & 1.20E-02 & 1.27E-02 & 1.11E-02 & 1.05E-02 & 9.37E-03 & 8.82E-03 \\
                       & std err  & 2.09E-04 & 2.54E-04 & 2.09E-04 & 3.02E-04 & 3.44E-04 & 3.19E-04 & 2.53E-04 & 2.62E-04 & 2.13E-04 & 2.48E-04\\
                       \bottomrule
\end{tabular}
\end{table}

\subsection{Portfolio Pricing}
In this example, we price a portfolio of Asian options, specifically $N$ arithmetic-average call Asian options that are monitored at the same fixed times, i.e., $t_m=m \Delta t,m=1,2,\cdots, M, \Delta t = T/M$, where $T$ is the expiration date of (all) the options and $M$ is the total number of monitored points. Each Asian option in the portfolio is written on an underlying asset driven by a geometric Brownian motion $S_i(t), i=1,2,\cdots, N$, with value at observation point $t_m$ given by
\begin{equation}\label{eq-portfolio-S}
  S_i(t_m)=S_i(t_{m-1})\exp\left( \left(r-\frac{1}{2}\sigma_i^2\right) \Delta t + \sigma_i (W_i(t_m)- W_i(t_{m-1})) \right),~ i=1,\cdots,N,~ m=1,\cdots,M,
\end{equation}
where $r$ is the risk-free interest rate, $\sigma_i$ is the volatility of underlying asset $i$, and $W_i(t)$, $i=1,2,\cdots, N$ are mutually independent standard Brownian motions. Denote $X^{(i)}_{m}=\ln(S_i(t_m))$ as the log-price of underlying asset $i$ at time $t_m$, and let $\bX^{(i)} = (X^{(i)}_{1},X^{(i)}_{2},\ldots,X^{(i)}_{M})^\top$ be the vector of input random variables for the $i$th Asian option. Let $\btheta^{(i)}=(S_i(0), \sigma_i)^{\top}$ be the parameters that affect $\bX^{(i)}$, i.e., the input random vectors determined by the initial price and the volatility of the underlying asset.

The discounted payoff function (i.e., the performance function) in pricing the $i$th arithmetic-average Asian option is
\begin{eqnarray*}\label{eq-portfolio-performance}
  h^{(i)}(\bX^{(i)}) &=&  e^{-rT} \max\left(\frac{1}{M} \sum_{m=1}^M S_i(t_m) - K_i, 0\right)= e^{-rT} \left(\frac{1}{M} \sum_{m=1}^M e^{X^{(i)}_{m}} - K_i\right)^+,
\end{eqnarray*}
where $K_i$ is the strike price for the $i$th Asian option,
Then the price of the $i$th Asian option is $v^{(i)}(\btheta^{(i)}) = \mathds{E}[ h^{(i)}(\bX^{(i)}) ]$,
and the price of the portfolio is given by
$$\bar{h}(\betheta^{(1)},\betheta^{(2)},\cdots,\betheta^{(N)}) = \sum_{i=1}^N v^{(i)}(\betheta^{(i)}) =\sum_{i=1}^N \mathds{E}\left[ h^{(i)}(\bX^{(i)}) \right].$$

We consider a portfolio of three Asian call options, $N=3$, with different strike prices $K_1=100$, $K_2=45$, and $K_3=80$, but the same expiration $T=0.25$ year (=13 weeks). We monitor the price of underlying assets every week, i.e., $\Delta t=1/52$ and $M=13$. Let $r=0.05$. Additionally, we assume that all the parameters of the input random variables are recalibrated once a week, and denote $\btheta_j^{(i)}(i=1,\cdots, N)$ as the input parameters recalibrated at the $j$th week. For the $j$th week, we conduct simulation experiments to price the portfolio, and collect these simulation data denoted by $\mathcal{X}_j=\{\tilde{\bX}_{j,s}^{(i)}\}(i=1,\cdots,N, s=1,\cdots,n)$. Suppose that we have recalibrated the input parameters and collect the simulation data for the past four weeks, and now we want to reuse these data to price the portfolio for the current week. In this example, we compare our TruLR-M and TruLR-S estimators with the classical LR estimator. 
All the parameter settings of the TruLR-M and TruLR-S estimators are provided in Table \ref{Tab:para-AsianPort}, and other details are provided in \ref{appx:port_pricing}.
\begin{table}[htpb!]
\scriptsize
\centering
		\caption{\baselineskip10pt Parameters for the Asian option portfolio.}\label{Tab:para-AsianPort}
\begin{tabular}{cccccccccccr}
\toprule
& $\btheta_0^{(i)}$      & $\btheta_1^{(i)}$      & $\btheta_2^{(i)}$      & $\btheta_3^{(i)}$     & $\btheta^{(i)}$        \\
\midrule
$S_1(0)$ & 100 & 102.5 & 105 & 107.5 & 110  \\
$S_2(0)$  & 55 & 65 & 65 & 60 & 55  \\
$S_3(0)$  & 80 & 85 & 75 & 85 & 80  \\
\hline
 $\sigma_1$ & 0.18 & 0.18 & 0.24 & 0.24 & 0.36  \\
 $\sigma_2$  & 0.70 & 0.70 & 0.70 & 0.70 & 0.25  \\
$\sigma_3$  & 0.10 & 0.15 & 0.20 & 0.25 & 0.30  \\
\hline
$\alpha_1$ & 1.2 & 1.2 & 1.6 &1.6 &  \\
$\alpha_2$  & 1.7 & 1.7 & 1.7 & 1.7 & \\
$\alpha_3$  & 1.1 & 1.2 & 1.2 & 1.6 & \\
 \bottomrule
\end{tabular}
\end{table}

\begin{figure}[htpb!]
\centering
\includegraphics[width=9.2cm]{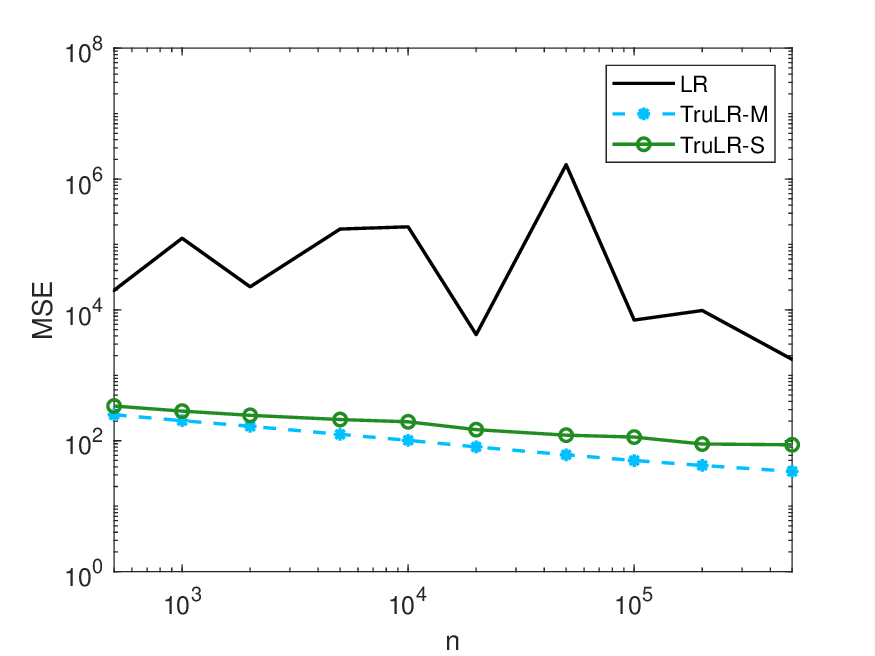}
\caption{\baselineskip10pt MSEs of the LR and TruLR estimators for pricing a portfolio of 3 Asian options with the same expiration $T=0.25$, but different strike prices $K_1=100$, $K_2=45$, and $K_3=80$. Let $r=0.05$, $\delta=0.01$, and other parameter are reported in Table \ref{Tab:para-AsianPort}. Each MSE is estimated based on $100000$ independent replications.}
\label{fig:Asian}
\end{figure}

Varying sample size $n$ from $500$ to $500000$, the MSEs and MSE ratios of these estimators are shown in Figure \ref{fig:Asian} and Table \ref{Tab:Asian}, from which we observe: (1) The classical LR estimator {exhibits highly variable behavior compared with TruLR estimators} (e.g., $n=1000, 5000, 10000, 50000$). (2) The TruLR-M and TruLR-S estimators can greatly improve the performance of the LR estimators, and the TruLR-M estimator consistently outperforms the TruLR-S estimator. For example, taking $n=100000$, the MSEs of the TruLR-M and TruLR-S estimators are $50$ and $114$, respectively, while the MSE of the LR estimator is {close to $7000$}, corresponding to a reduction in MSE over the LR estimators of approximately $140$ and $61$ times, respectively. 

\section{Concluding Remarks}\label{sec:concl}

We began with a new result establishing the tightness of {\it polynomial} concentration bounds for the classical LR-based IS estimators.
Specifically, we derived concentration and anti-concentration bounds under both the $\infty$-norm and $p$-norm.
Then we introduced a new truncation-boundary estimator, for which we established upper concentration bounds for both norms, which imply an {\it exponential} convergence rate.
For practical implementation, we derived appropriate truncation boundaries to construct effective TruLR estimators.
Simulation experiments on synthetic examples, offline RL policy evaluation, and option pricing demonstrate significant reductions in MSE for the TruLR estimators compared to the classical LR estimator.

Ongoing work includes applying many of the ideas here to the LR method {in other settings, including} stochastic gradient estimation, which also experiences variance challenges in certain settings, {and multiple importance sampling.}

\bibliographystyle{apalike}
\SingleSpacedXI
\bibliography{RSTL}

\begin{thebibliography}{}

\bibitem[Asmussen and Glynn, 2007]{Asmussen2007}
Asmussen, S. and Glynn, P.~W. (2007).
\newblock {\em Stochastic Simulation: Algorithms and Analysis}.
\newblock Springer Science \& Business Media, New York, NY.

\bibitem[Bai et~al., 2022]{bai2022rare}
Bai, Y., Huang, Z., Lam, H., and Zhao, D. (2022).
\newblock Rare-event simulation for neural network and random forest
  predictors.
\newblock {\em ACM Transactions on Modeling and Computer Simulation},
  32(3):1--33.

\bibitem[Blanchet and Lam, 2014]{BlLam2014}
Blanchet, J. and Lam, H. (2014).
\newblock Rare-event simulation for many-server queues.
\newblock {\em Mathematics of Operations Research}, 39(4):1142--1178.

\bibitem[Blanchet et~al., 2012a]{blanchet2012lyapunov}
Blanchet, J.~H., Glynn, P.~W., and Leder, K. (2012a).
\newblock On {L}yapunov inequalities and subsolutions for efficient importance
  sampling.
\newblock {\em ACM Transactions on Modeling and Computer Simulation},
  22(3):1--27.

\bibitem[Blanchet and Lam, 2012]{blanchet2012state}
Blanchet, J.~H. and Lam, H. (2012).
\newblock State-dependent importance sampling for rare-event simulation: An
  overview and recent advances.
\newblock {\em Surveys in Operations Research and Management Science},
  17(1):38--59.

\bibitem[Blanchet et~al., 2012b]{blanchet2012efficient}
Blanchet, J.~H., Lam, H., and Zwart, B. (2012b).
\newblock Efficient rare-event simulation for perpetuities.
\newblock {\em Stochastic Processes and Their Applications},
  122(10):3361--3392.

\bibitem[Boucheron et~al., 2013]{boucheron2013concentration}
Boucheron, S., Lugosi, G., and Massart, P. (2013).
\newblock {\em Concentration Inequalities: A Nonasymptotic Theory of
  Independence}.
\newblock Oxford University Press, Oxford, UK.

\bibitem[Budhiraja and Dupuis, 2019]{budhiraja2019analysis}
Budhiraja, A. and Dupuis, P. (2019).
\newblock {\em Analysis and Approximation of Rare Events: Representations and
  Weak Convergence Methods}.
\newblock Springer, New York, NY.

\bibitem[Cortes et~al., 2010]{cortes2010learning}
Cortes, C., Mansour, Y., and Mohri, M. (2010).
\newblock Learning bounds for importance weighting.
\newblock In {\em Advances in Neural Information Processing Systems 23}, pages
  442--450.

\bibitem[Dud{\'\i}k et~al., 2011]{dudik2011doubly}
Dud{\'\i}k, M., Langford, J., and Li, L. (2011).
\newblock Doubly robust policy evaluation and learning.
\newblock In {\em Proceedings of the 28th International Conference on Machine
  Learning}, pages 1097--1104.

\bibitem[Feng and Jiang, 2020]{feng2020reusing}
Feng, B. and Jiang, G. (2020).
\newblock Reusing simulation outputs of repeated experiments via likelihood
  ratio regression.
\newblock In Bae, K.-H., Feng, B., Kim, S., Lazarova-Molnar, S., Zheng, Z.,
  Roeder, T., and Thiesing, R., editors, {\em Proceedings of the 2020 Winter
  Simulation Conference}, pages 11--22, Piscataway, New Jersey. IEEE.

\bibitem[Feng and Song, 2019]{feng2019efficient}
Feng, M. and Song, E. (2019).
\newblock Efficient input uncertainty quantification via green simulation using
  sample path likelihood ratios.
\newblock In Mustafee, N., Bae, K.-H., Lazarova-Molnar, S., Rabe, M.,
  Szab{\'o}, C., Haas, P., and Son, Y.-J., editors, {\em Proceedings of the
  2019 Winter Simulation Conference}, pages 3693--3704, Piscataway, New Jersey.
  IEEE.

\bibitem[Feng and Staum, 2017]{feng2017green}
Feng, M. and Staum, J. (2017).
\newblock Green simulation: Reusing the output of repeated experiments.
\newblock {\em ACM Transactions on Modeling and Computer Simulation},
  27(4):1--28.

\bibitem[Feng and Staum, 2021]{feng2021green}
Feng, M. and Staum, J. (2021).
\newblock Green simulation with database {M}onte {C}arlo.
\newblock {\em ACM Transactions on Modeling and Computer Simulation},
  31(1):1--26.

\bibitem[Geweke, 1989]{geweke1989bayesian}
Geweke, J. (1989).
\newblock Bayesian inference in econometric models using {M}onte {C}arlo
  integration.
\newblock {\em Econometrica}, 57:1317--1339.

\bibitem[Glasserman et~al., 2000]{GlHeSh2000}
Glasserman, P., Heidelberger, P., and Shahabuddin, P. (2000).
\newblock Variance reduction techniques for estimating {V}alue-at-{R}isk.
\newblock {\em Management Science}, 46(10):1349--1363.

\bibitem[He et~al., 2024]{HeJiang2023}
He, S., Jiang, G., Lam, H., and Fu, M.~C. (2024).
\newblock Adaptive importance sampling for efficient stochastic root finding
  and quantile estimation.
\newblock {\em Operations Research}, 72(6):2612--2630.

\bibitem[Hong and Jiang, 2019]{HongJiang2019}
Hong, L.~J. and Jiang, G. (2019).
\newblock Offline simulation online application: {A} new framework of
  simulation-based decision making.
\newblock {\em Asia-Pacific Journal of Operational Research}, 36(6):No.1940015.

\bibitem[Ionides, 2008]{Ionides2008}
Ionides, E.~L. (2008).
\newblock Truncated importance sampling.
\newblock {\em Journal of Computational and Graphical Statistics},
  17(2):295--311.

\bibitem[Jiang et~al., 2020]{JiangHongNelson2020}
Jiang, G., Hong, L.~J., and Nelson, B.~L. (2020).
\newblock Online risk monitoring using offline simulation.
\newblock {\em INFORMS Journal on Computing}, 32(2):356--375.

\bibitem[Juneja and Shahabuddin, 2006]{juneja2006rare}
Juneja, S. and Shahabuddin, P. (2006).
\newblock Rare-event simulation techniques: {A}n introduction and recent
  advances.
\newblock In Henderson, S.~G. and Nelson, B.~L., editors, {\em Handbooks in
  {O}perations {R}esearch and {M}anagement {S}cience}, volume~13, chapter~11,
  pages 291 -- 350. Elsevier, Amsterdam, The Netherlands.

\bibitem[Kloek and van Dijk, 1978]{kloek1978bayesian}
Kloek, T. and van Dijk, H.~K. (1978).
\newblock Bayesian estimates of system equation parameters: {A}n application of
  integration by {M}onte {C}arlo.
\newblock {\em Econometrica}, 46:1--19.

\bibitem[Lin et~al., 2025]{lin2024reusing}
Lin, Y., Wang, Y., and Zhou, E. (2025).
\newblock Reusing historical trajectories in natural policy gradient via
  importance sampling: Convergence and convergence rate.
\newblock {\em Operations Research}, Forthcoming.

\bibitem[Lin and Bai, 2011]{lin2011probability}
Lin, Z. and Bai, Z. (2011).
\newblock {\em Probability Inequalities}.
\newblock Springer Science \& Business Media, New York, NY.

\bibitem[Liu et~al., 2022]{LiuLiang2022}
Liu, H., Liang, L., Lee, L.~H., and Chew, E.~P. (2022).
\newblock Unifying offline and online simulation for online decision-making.
\newblock {\em IISE Transactions}, 54(10):923--935.

\bibitem[Liu and Zhou, 2020]{liu2020simulation}
Liu, T. and Zhou, E. (2020).
\newblock Simulation optimization by reusing past replications: Don’t be
  afraid of dependence.
\newblock In Bae, K.-H., Feng, B., Kim, S., Lazarova-Molnar, S., Zheng, Z.,
  Roeder, T., and Thiesing, R., editors, {\em Proceedings of the 2020 Winter
  Simulation Conference}, pages 2923--2934, Piscataway, New Jersey. IEEE.

\bibitem[Maggiar et~al., 2018]{maggiar2018derivative}
Maggiar, A., Wachter, A., Dolinskaya, I.~S., and Staum, J. (2018).
\newblock A derivative-free trust-region algorithm for the optimization of
  functions smoothed via {G}aussian convolution using adaptive multiple
  importance sampling.
\newblock {\em SIAM Journal on Optimization}, 28(2):1478--1507.

\bibitem[Mahmood et~al., 2014]{mahmood2014weighted}
Mahmood, A.~R., van Hasselt, H., and Sutton, R.~S. (2014).
\newblock Weighted importance sampling for off-policy learning with linear
  function approximation.
\newblock In {\em Advances in Neural Information Processing Systems 27}, pages
  3014--3022.

\bibitem[Martino et~al., 2018]{martino2018comparison}
Martino, L., Elvira, V., Miguez, J., Artes-Rodriguez, A., and Djuric, P.~M.
  (2018).
\newblock A comparison of clipping strategies for importance sampling.
\newblock In {\em 2018 IEEE Workshop on Statistical Signal Processing}, pages
  560--564. IEEE.

\bibitem[Metelli et~al., 2018]{NEURIPS2018_6aed000a}
Metelli, A.~M., Papini, M., Faccio, F., and Restelli, M. (2018).
\newblock Policy optimization via importance sampling.
\newblock In {\em Advances in Neural Information Processing Systems 31}, pages
  5447--5459.

\bibitem[Metelli et~al., 2020]{MetJMLR2020}
Metelli, A.~M., Papini, M., Montali, N., and Restelli, M. (2020).
\newblock Importance sampling techniques for policy optimization.
\newblock {\em Journal of Machine Learning Research}, 21(141):1--75.

\bibitem[Metelli et~al., 2021]{Metelli2021}
Metelli, A.~M., Russo, A., and Restelli, M. (2021).
\newblock Subgaussian and differentiable importance sampling for off-policy
  evaluation and learning.
\newblock In {\em Advances in Neural Information Processing Systems 34}, pages
  8119--8132.

\bibitem[Nelson, 2016]{Nelson2015}
Nelson, B.~L. (2016).
\newblock `{S}ome tactical problems in digital simulation' for the next ten
  years.
\newblock {\em Journal of Simulation}, 10(1):2--11.

\bibitem[Papini et~al., 2019]{papini2019optimistic}
Papini, M., Metelli, A.~M., Lupo, L., and Restelli, M. (2019).
\newblock Optimistic policy optimization via multiple importance sampling.
\newblock In {\em International Conference on Machine Learning}, pages
  4989--4999.

\bibitem[Pinelis, 2015]{pinelis2015best}
Pinelis, I. (2015).
\newblock Best possible bounds of the von {B}ahr--{E}sseen type.
\newblock {\em Annals of Functional Analysis}, 6(4):1--29.

\bibitem[R{\'e}nyi, 1961]{renyi1961measures}
R{\'e}nyi, A. (1961).
\newblock On measures of entropy and information.
\newblock In {\em Proceedings of the Fourth Berkeley Symposium on Mathematical
  Statistics and Probability, Volume 1: Contributions to the Theory of
  Statistics}, volume~4, pages 547--562. University of California Press.

\bibitem[Rubin, 1988]{rubin1988sir}
Rubin, D.~B. (1988).
\newblock Using the {SIR} algorithm to simulate posterior distributions.
\newblock In Bernardo, J.~M., DeGroot, M.~H., Lindley, D.~V., and Smith, A.
  F.~M., editors, {\em Bayesian Statistics 3}, pages 395--402. Oxford
  University Press.

\bibitem[Rubinstein and Kroese, 2004]{rubinstein2004CE}
Rubinstein, R.~Y. and Kroese, D.~P. (2004).
\newblock {\em The Cross-Entropy Method: {A} Unified Approach to Combinatorial
  Optimization, {M}onte-{C}arlo Simulation, and Machine Learning}.
\newblock Springer-Verlag, New York, NY.

\bibitem[Shen et~al., 2021]{ShenHong2021}
Shen, H., Hong, L.~J., and Zhang, X. (2021).
\newblock Ranking and selection with covariates for personalized decision
  making.
\newblock {\em INFORMS Journal on Computing}, 32(2):356--375.

\bibitem[Vehtari et~al., 2024]{vehtari2024pareto}
Vehtari, A., Simpson, D., Gelman, A., Yao, Y., and Gabry, J. (2024).
\newblock Pareto smoothed importance sampling.
\newblock {\em Journal of Machine Learning Research}, 25(72):1--58.

\bibitem[Vershynin, 2010]{vershynin2010introduction}
Vershynin, R. (2010).
\newblock Introduction to the non-asymptotic analysis of random matrices.
\newblock {\em arXiv preprint arXiv:1011.3027}.

\bibitem[Wang et~al., 2017]{wang2017optimal}
Wang, Y.-X., Agarwal, A., and Dud{\'\i}k, M. (2017).
\newblock Optimal and adaptive off-policy evaluation in contextual bandits.
\newblock In {\em Proceedings of the 34th International Conference on Machine
  Learning}, pages 3589--3597.

\bibitem[Yun et~al., 2019]{YunHong2019}
Yun, X., Hong, L.~J., Jiang, G., and Wang, S. (2019).
\newblock On {G}amma estimation via matrix kriging.
\newblock {\em Naval Research Logistics}, 66(5):393--410.

\bibitem[Zhang and Chen, 2021]{zhang2021concentration}
Zhang, H. and Chen, S. (2021).
\newblock Concentration inequalities for statistical inference.
\newblock {\em Communications in Mathematical Research}, 37(1):1--85.

\end{thebibliography}

\newpage
\DoubleSpacedXI
\ECSwitch
\ECHead{Appendices}
\setcounter{equation}{0}
\setcounter{lemma}{0}
\setcounter{algorithm}{0}
\setcounter{subsection}{0}
\setcounter{assumption}{0}
\setcounter{theorem}{1}
\setcounter{proposition}{0}
\renewcommand{\theequation}{A.\arabic{equation}}
\renewcommand{\thelemma}{A.\arabic{lemma}}
\renewcommand{\thesubsection}{\thesection.\arabic{subsection}}
\renewcommand{\thealgorithm}{A.\arabic{algorithm}}
\renewcommand{\theassumption}{A.\arabic{assumption}}
\renewcommand{\thetheorem}{A.\arabic{theorem}}
\renewcommand{\theproposition}{A.\arabic{proposition}}

\section{Proof of Proposition \ref{theorem-IS-concentration}}\label{appx:P1-proof}

Recall that $\bX \sim \mathds{P}_{\betheta}$ and $\tilde{\bX} \sim \mathds{P}_{\betheta_0}$. For $\tilde{\bX}_i, i=1,...,n$ i.i.d samples from probability measure $\mathds{P}_{\betheta_0}$,
\begin{eqnarray*}
  \mathds{E}[\bar {h}_n^{\LR}] & = & \mathds{E}\left[\frac{1}{n} \sum_{i=1}^{n} h(\tilde{\bX}_i) l(\tilde{\bX}_i) \right] \\
   & = & \mathds{E}[h(\tilde{\bX}) l(\tilde{\bX}) ] \\
   & = & \int_{\mathcal{X}} h(\bx) \frac{d\mathds{P}_{\betheta}(\bx)}{d\mathds{P}_{\betheta_0}(\bx)} d\mathds{P}_{\betheta_0}(\bx) \\
   & = & \int_{\mathcal{X}} h(\bx) d\mathds{P}_{\betheta}(\bx) \\
   & = & \mathds{E}[h(\bX)] = \bar{h}(\btheta).
\end{eqnarray*}
Thus, $\bar{h}_n^{\LR}$ is an unbiased estimator of $\bar{h}(\btheta)$.

\proof{Proof of Proposition \ref{theorem-IS-concentration}:} The proof is motivated by Theorem A.1 of \cite{Metelli2021}.
First, by the definition of $\bar {h}_n^{\LR}$,
\begin{eqnarray}
  \mathds{E}[ |\bar{h}_n^{\LR}-\bar{h}(\btheta)| ^{\alpha}] &=&\mathds{E} \left[ \left|\frac{1}{n}\sum_{i=1}^n h(\tilde{\bX}_i)\l(\tilde{\bX}_i) -  \bar{h}(\btheta) \right| ^{\alpha}\right] \nonumber \\
  & = & \frac{1}{n^{\alpha}} \mathds{E}\left[ \left| \sum_{i=1}^n \Big( h(\tilde{\bX}_i)\l(\tilde{\bX}_i) - \bar{h}(\btheta)\Big) \right| ^{\alpha} \right] \label{eq-IS-concentration-2}
\end{eqnarray}

Let
$$S_n \equiv \sum_{i=1}^n \Big( h(\tilde{\bX}_i)\l(\tilde{\bX}_i) - \bar{h}(\btheta) \Big) $$
and
$$Y_i \equiv S_i - S_{i-1} = h(\tilde{\bX}_i)\l(\tilde{\bX}_i) - \bar{h}(\btheta),$$
with the convention $S_0 \equiv 0$.
Note that $\mathds{E}\left[ \left| Y_i  \right|\right]<\infty$ and $\mathds{E}[Y_i|S_{i-1}] = 0$. That is, $\{S_i\}_{i=1}^n$ is a v-martingale (see \citealt{pinelis2015best}). Then by Proposition 1.8 of \cite{pinelis2015best}, for $\alpha \in (1,2]$, we have
\begin{eqnarray}
\mathds{E}\left[ \left| \sum_{i=1}^n \Big( h(\tilde{\bX}_i)\l(\tilde{\bX}_i) - \bar{h}(\btheta)\Big) \right| ^{\alpha} \right]
&\leq &   \mathds{E}\left[ |  Y_1 | ^{\alpha} \right]  + 2^{2-\alpha}\sum_{i=2}^n \mathds{E}_{\betheta_0}\left[ |  Y_i | ^{\alpha} \right] \nonumber \\
&& \leq  n 2^{2-\alpha} \mathds{E}\left[ |  Y_i | ^{\alpha} \right] \nonumber \\
&& =  n 2^{2-\alpha}\mathds{E}\left[ \Big|  h(\tilde{\bX}_i)\l(\tilde{\bX}_i) - \bar{h}(\btheta) \Big| ^{\alpha} \right]. \label{eq-IS-concentration-3}
\end{eqnarray}

In addition, note $\alpha \in (1,2]$ and by the $c_r$-inequality (Section 9.1, \citealt{lin2011probability}),
\begin{eqnarray}
    \mathds{E}\left[ \Big|  h(\tilde{\bX}_i)\l(\tilde{\bX}_i) - \bar{h}(\btheta) \Big| ^{\alpha} \right] 
    &\leq&  2^{\alpha-1 } \mathds{E}\left[ \Big| h(\tilde{\bX}_i)\l(\tilde{\bX}_i) \Big| ^{\alpha} \right]  +  2^{\alpha-1 } \mathds{E}\left[ |  \bar{h}(\btheta) | ^{\alpha} \right]  \nonumber \\
    && = 2^{\alpha-1 } \mathds{E}\left[ \Big|  h(\tilde{\bX}_i)\l(\tilde{\bX}_i)  \Big| ^{\alpha} \right]  +  2^{\alpha-1 } |  \bar{h}(\btheta) | ^{\alpha}  \nonumber \\
    && = 2^{\alpha-1 } \mathds{E}\left[ \Big|   h(\tilde{\bX}_i)\l(\tilde{\bX}_i) \Big| ^{\alpha} \right]  +  2^{\alpha-1 } \Big|  \mathds{E}\left[ h(\tilde{\bX}_i)\l(\tilde{\bX}_i)\right]  \Big| ^{\alpha}  \nonumber \\
    && \leq 2^{\alpha-1 } \mathds{E}\left[ \Big|   h(\tilde{\bX}_i)\l(\tilde{\bX}_i) \Big| ^{\alpha} \right]  +  2^{\alpha-1 }\mathds{E}\left[ \Big|   h(\tilde{\bX}_i)\l(\tilde{\bX}_i) \Big| ^{\alpha} \right]  \label{eq-IS-concentration-4}  \\
    && = 2^{\alpha } \mathds{E}\left[ \Big|   h(\tilde{\bX}_i)\l(\tilde{\bX}_i) \Big| ^{\alpha} \right]\nonumber \\
    && \leq  2^{\alpha}\| h \|_{\infty, \mathds{P}_{\betheta_0}}^{\alpha}  \mathds{E}[  \l^{\alpha}(\tilde{\bX}_i)  ] \nonumber \\
    &&  = 2^{\alpha } \| h \|_{\infty, \mathds{P}_{\betheta_0}}^{\alpha}  I_{\alpha}(\mathds{P}_{\betheta} \|  \mathds{P}_{\betheta_0}), \label{eq-IS-concentration-5}
\end{eqnarray}
where \eqref{eq-IS-concentration-4} is due to the convexity of $|x|^{\alpha}( 1< \alpha \leq 2)$ and Jensen's inequality.

Then combining \eqref{eq-IS-concentration-2}, \eqref{eq-IS-concentration-3} and \eqref{eq-IS-concentration-5}, we have
\begin{equation*}
     \mathds{E}[|\bar{h}_n^{\LR}-\bar{h}(\btheta) | ^{\alpha}] \leq \frac{4}{n^{\alpha - 1}} \| h \|_{\infty, \mathds{P}_{\betheta_0}}^{\alpha}  I_{\alpha}(\mathds{P}_{\betheta} \|  \mathds{P}_{\betheta_0}).
\end{equation*}

Now we can derive the concentration inequality:
\begin{eqnarray}
  \Pr( |\bar{h}_n^{\LR}-\bar{h}(\btheta) |\geq \epsilon ) & = & \Pr( |\bar{h}_n^{\LR}-\bar{h}(\btheta)  |^{\alpha} \geq \epsilon^{\alpha} ) \nonumber \label{eq-IS-concentration-6} \\
   & \leq & \frac{\mathds{E}[ |\bar{h}_n^{\LR}-\bar{h}(\btheta)  | ^{\alpha}]}{\epsilon^{\alpha}}   \label{eq-IS-concentration-7} \\
   & \leq & \frac{4}{n^{\alpha-1} \epsilon^{\alpha} }  \| h \|_{\infty, \mathds{P}_{\betheta_0}}^{\alpha}  I_{\alpha}(\mathds{P}_{\betheta} \|  \mathds{P}_{\betheta_0}) \nonumber \label{eq-IS-concentration-8},
\end{eqnarray}
where \eqref{eq-IS-concentration-7} follows from Markov's inequality. Setting the right-hand side of the last inequality equal to $\delta$ gives the result.
$\hfill\Box$
\endproof

\section{Proof of Proposition \ref{theorem-IS-bound-concentration}}\label{appx:P2-proof}
To prove Proposition \ref{theorem-IS-bound-concentration}, we first derive the variance of the LR estimator $\bar{h}_n^{\text{LR}}$ in the following lemma.
\begin{lemma}\label{theorem-IS-bound-variance}
If $\| h \|_{p, \mathds{P}_{\betheta_0}} < \infty$ holds for some $ p > 2 $ and $I_{\alpha}(\mathds{P}_{\betheta} \|  \mathds{P}_{\betheta_0})$ is finite for some $ \alpha \geq \frac{2p}{p-2}>2$, then the variance of LR estimator $\bar{h}_n^{\text{LR}}$ can be upper bounded by
\begin{equation}\label{eq-IS-bound-variance}
  \Var(\bar {h}_n^{\LR}) \leq \frac{1}{n} \| h \|_{p,\mathds{P}_{\betheta_0}}^2  I_{\alpha}(\mathds{P}_{\betheta} \|  \mathds{P}_{\betheta_0})^{\frac{2}{\alpha}}.
\end{equation}
\end{lemma}
\proof{Proof:}
Note that
\begin{eqnarray}
  \Var(\bar {h}_n^{\LR}) &=& \frac{1}{n} \Var(h(\tilde{\bX}_i)\l(\tilde{\bX}_i) ) \nonumber \\
  & \leq &  \frac{1}{n} \mathds{E} [ h^2(\tilde{\bX})\l^2(\tilde{\bX}) ]  \nonumber \\
   & \leq &  \frac{1}{n} \left( \mathds{E}[ |h(\tilde{\bX}) |^{2\tilde{p}} ] \right)^{\frac{1}{\tilde{p}}}
   \left(\mathds{E}[  \l^{2\tilde{q}}(\tilde{\bX})  ] \right)^{\frac{1}{\tilde{q}}}  \label{eq-IS-bound-variance-1}\\
   & = & \frac{1}{n} \| h \|_{p, \mathds{P}_{\betheta_0}}^2  I_{2\tilde{q}}(\mathds{P}_{\betheta} \| \mathds{P}_{\betheta_0})^{\frac{1}{\tilde{q}}}  \label{eq-IS-bound-variance-2}\\
   & \leq & \frac{1}{n} \| h \|_{p, \mathds{P}_{\betheta_0}}^2  I_{ \alpha }(\mathds{P}_{\betheta} \| \mathds{P}_{\betheta_0})^{\frac{2}{\alpha}},  \label{eq-IS-bound-variance-3}
\end{eqnarray}
where \eqref{eq-IS-bound-variance-1} is an application of H\"{o}lder inequality with $\frac{1}{\tilde{p}} + \frac{1}{\tilde{q}} = 1$, \eqref{eq-IS-bound-variance-2} is obtained by picking $\tilde{p} = \frac{p}{2}$, \eqref{eq-IS-bound-variance-3} follows from the monotonicity of the $L_r$-norm (Section 8.4, \citealt{lin2011probability}) if $\alpha \geq 2\tilde{q}$. Recall that $p > 2$, thus $\alpha \geq 2\tilde{q} = \frac{2\tilde{p}}{\tilde{p}-1} = \frac{2p}{p-2} >2$.
$\hfill\Box$
\endproof

\proof{Proof of Proposition \ref{theorem-IS-bound-concentration}:} The proof is similar to that in {Proposition} \ref{theorem-IS-concentration}. Pick $r \in (1,2] $, by the definition of $\bar {h}_n^{\LR}$,
\begin{eqnarray}
  \mathds{E}[ |\bar{h}_n^{\LR}-\bar{h}(\btheta)| ^{r}] &=&\mathds{E}\left[ \left|\frac{1}{n}\sum_{i=1}^n \left( h(\tilde{\bX}_i)\l(\tilde{\bX}_i) -  \bar{h}(\btheta) \right) \right| ^{r}\right] = \frac{1}{n^{r}} \mathds{E}\left[ \left| \sum_{i=1}^n \Big(  h(\tilde{\bX}_i)\l(\tilde{\bX}_i) -  \bar{h}(\btheta)\Big) \right| ^{r} \right].  \nonumber \\ \label{eq-IS-bound-concentration-1}
\end{eqnarray}

Let
$S_n \equiv \sum_{i=1}^n \Big(  h(\tilde{\bX}_i)\l(\tilde{\bX}_i) -  \bar{h}(\btheta)\Big)$,
 {$S_0 \equiv 0$}, and
$$Y_i \equiv S_i - S_{i-1} =  h(\tilde{\bX}_i)\l(\tilde{\bX}_i) -  \bar{h}(\btheta).$$
Note that $\mathds{E}\left[ \left| Y_i  \right|\right]<\infty$ and $\mathds{E}[Y_i|S_{i-1}] = 0$. That is, $\{S_i\}_{i=1}^n$ is a v-martingale (see \citealt{pinelis2015best}). Then by Proposition 1.8 of \cite{pinelis2015best}, and $r \in (1,2]$, we have
\begin{equation}
\mathds{E}\left[ \left| \sum_{i=1}^n \Big(  h(\tilde{\bX}_i)\l(\tilde{\bX}_i) -  \bar{h}(\btheta)\Big) \right| ^{r} \right]
\leq  n 2^{2-r}\mathds{E}\left[ \Big|   h(\tilde{\bX}_i)\l(\tilde{\bX}_i) -  \bar{h}(\btheta) \Big| ^{r} \right]. \label{eq-IS-bound-concentration-2}
\end{equation}

In addition, for $r \in (1,2]$, by the $c_r$-inequality (Section 9.1, \cite{lin2011probability}) and Jensen's inequality,
\begin{equation}
    \mathds{E}\left[ \Big|   h(\tilde{\bX}_i)\l(\tilde{\bX}_i) -  \bar{h}(\btheta) \Big| ^{r} \right] \leq 2^{r } \mathds{E}\left[ \Big|  h(\tilde{\bX}_i)\l(\tilde{\bX}_i)  \Big| ^{r} \right]. \label{eq-IS-bound-concentration-3}
\end{equation}
Note that
\begin{eqnarray}
  \mathds{E}\left[ \Big|  h(\tilde{\bX}_i)\l(\tilde{\bX}_i)\Big| ^{r} \right] 
   & \leq &  \left( \mathds{E}[ {|h(\tilde{\bX}_i)|^{r \tilde{p}}}  ] \right)^{\frac{1}{\tilde{p}}}
   \left(\mathds{E}[  \l^{r \tilde{q}}(\tilde{\bX}_i)  ] \right)^{\frac{1}{\tilde{q}}}  \label{eq-IS-bound-concentration-4}\\
   & = &  \| h \|_{p,\mathds{P}_{\betheta_0}}^r  I_{r\tilde{q}}(\mathds{P}_{\betheta} \| \mathds{P}_{\betheta_0})^{\frac{1}{\tilde{q}}}  \label{eq-IS-bound-concentration-5}\\
   & \leq &  \| h \|_{p,\mathds{P}_{\betheta_0}}^r  I_{ \alpha }(\mathds{P}_{\betheta} \| \mathds{P}_{\betheta_0})^{\frac{r}{\alpha}},  \label{eq-IS-bound-concentration-6}
\end{eqnarray}
where \eqref{eq-IS-bound-concentration-4} is an application of H\"{o}lder inequality with $\frac{1}{\tilde{p}} + \frac{1}{\tilde{q}} = 1$, \eqref{eq-IS-bound-concentration-5} is obtained by picking $\tilde{p} = \frac{p}{r}$, \eqref{eq-IS-bound-concentration-6} follows from the monotonicity of the $L_r$-norm (\citealt{lin2011probability}) if $\alpha \geq r\tilde{q}$. Recall that $p > 2\geq r$, thus $\alpha \geq r\tilde{q} = \frac{r\tilde{p}}{\tilde{p}-1} = \frac{rp}{p-r} > r$.

Then combining \eqref{eq-IS-bound-concentration-1} - \eqref{eq-IS-bound-concentration-6}, we have
\begin{equation*}
     \mathds{E}[ |\bar{h}_n^{\LR}-\bar{h}(\btheta) | ^{r}] \leq \frac{4}{n^{r - 1}} \| h \|_{p,\mathds{P}_{\betheta_0}}^r  I_{ \alpha }(\mathds{P}_{\betheta} \| \mathds{P}_{\betheta_0})^{\frac{r}{\alpha}}.
\end{equation*}

Now we can derive the concentration inequality:
\begin{eqnarray}
  \Pr( |\bar{h}_n^{\LR}-\bar{h}(\btheta) |\geq \epsilon ) & = & \Pr( |\bar{h}_n^{\LR}-\bar{h}(\btheta) |^{r} \geq \epsilon^{r} ) \nonumber \label {eq-IS-bound-concentration-7} \\
   & \leq & \frac{\mathds{E}[ |\bar{h}_n^{\LR}-\bar{h}(\btheta) | ^{r}]}{\epsilon^{r}}   \label{eq-IS-bound-concentration-8}  \\
   & \leq & \frac{4}{n^{r - 1}\epsilon^{r}} \|h \|_{p,\mathds{P}_{\betheta_0}}^r  I_{ \alpha }(\mathds{P}_{\betheta} \| \mathds{P}_{\betheta_0})^{\frac{r}{\alpha}}, \nonumber \label{eq-IS-bound-concentration-9}
\end{eqnarray}
where \eqref{eq-IS-bound-concentration-8} follows from Markov's inequality. By setting the right-hand side of the last {inequality} equal to $\delta$, we get the following result:
\begin{equation*}\label{eq-IS-bound-concentration-10}
  \epsilon = \left( \frac{4}{n^{r - 1}\delta} \right)^{\frac{1}{r}}  \| h \|_{p,\mathds{P}_{\betheta_0}}  I_{ \alpha }(\mathds{P}_{\betheta} \| \mathds{P}_{\betheta_0})^{\frac{1}{\alpha}}.
\end{equation*}
For fixed $n$ and $\delta \in (0,1)$, the first term on the right-hand side is decreasing with $r$, so setting $r=2$ gives the result.
$\hfill\Box$
\endproof

\section{Proof of Proposition \ref{theorem-IS-concentration-lower}}\label{appx:P3-proof}
\proof{Proof:}
The proof is inspired by the proof of Theorem 3.1 in \cite{Metelli2021}.
We construct a function $h$ and two probability measures $\mathds{P}_{\betheta}$ and $\mathds{P}_{\betheta_0}$ that satisfy the inequality. Consider a constant $a>0$, and let the random variables $X$ and $\tilde{X}$ take values in the support $\mathcal{X} = \{-a,0, a\}$, with $h(x)=x$. It is straightforward to observe that $ \| h \|_{\infty, \mathds{P}_{\betheta_0}}  = a$, which ensures the existence of the infinity norm.

We now define the two probability measures as follows, for $\theta, \theta_0 \in [0,1]$,
\begin{eqnarray*}
   \Pr( X=a) & = \Pr(X=-a ) = {\theta}/{2},\quad \Pr( X=0 )=1-\theta, \\
   \Pr( \tilde{X}=a ) & = \Pr( \tilde{X}=-a ) = {\theta_0}/{2}, \quad \Pr( \tilde{X}=0)=1-\theta_0.
\end{eqnarray*}
It immediately follows that $\mathds{E}[h(X)]=\mathds{E}[h(\tilde{X})]=0$, so $\bar{h}(\betheta)=0$. We can compute the divergence as
\begin{eqnarray*}
  I_{\alpha}(\mathds{P}_{\betheta} \|  \mathds{P}_{\betheta_0}) & = & 2\left( \frac{\theta/2}{\theta_0/2}\right)^{\alpha} \frac{\theta_0}{2}+\left( \frac{1-\theta}{1-\theta_0} \right)^{\alpha} (1-\theta_0) \\
   &=& \left( \frac{\theta}{\theta_0}\right)^{\alpha} \theta_0 + \left( \frac{1-\theta}{1-\theta_0} \right)^{\alpha} (1-\theta_0) \\
   & \geq &  \left( \frac{\theta}{\theta_0} \theta_0 + \frac{1-\theta}{1-\theta_0}  (1-\theta_0)  \right)^{\alpha} = 1.
\end{eqnarray*}
The last inequality holds due to the convexity of $x^{\alpha} $ for any $\alpha \in (1,2]$ when $x \geq 0$ and Jensen's inequality.

We choose $\theta$ and $\theta_0$ as follows. For any value $\alpha \in (1,2]$, we define
\begin{eqnarray*}
   \theta & = & \left( \frac{a}{n \epsilon}\right)^{\alpha-1} \xi, \\
   \theta_0 & = & \left( \frac{a}{n \epsilon}\right)^{\alpha} \xi.
\end{eqnarray*}
where $\xi = I_{\alpha}(\mathds{P}_{\betheta} \|  \mathds{P}_{\betheta_0})-1 \geq 0$, and $\epsilon>0$ will be specified later. By definition, we have $a {\theta}/{\theta_0} = n \epsilon$.

Now, consider the classical LR estimator $ \bar {h}_n^{\LR}$. Recall that, based on our specific construction, its expectation is zero.
Given the i.i.d. sample $\{\tilde{X}_1,\cdots, \tilde{X}_n\}$ drawn from measure $\mathds{P}_{\betheta_0}$ and by the construction, $ \bar {h}_n^{\LR}$ is expressed as
\begin{eqnarray*}
   \bar {h}_n^{\LR} &=& \frac{1}{n}\sum_{i=1}^n h(\tilde{X}_i) \frac{d\mathds{P}_{\betheta}}{d\mathds{P}_{\betheta_0}} \\
   &=& \frac{1}{n} \sum_{i=1}^n  {\bf 1} \{\tilde{X}_i=a\} a \frac{\theta}{\theta_0}  - {\bf 1} \{\tilde{X}_i=-a\}a \frac{\theta}{\theta_0}  \\
    &=& \epsilon \sum_{i=1}^n  {\bf 1} \{\tilde{X}_i=a\}  -  {\bf 1} \{\tilde{X}_i=-a\}.
\end{eqnarray*}
The last equation holds due to $a {\theta}/{\theta_0} = n \epsilon$.
For any deviation level $\epsilon>0$,
\begin{equation*}
  \Pr(  \bar {h}_n^{\LR}\geq \epsilon  ) = \Pr(  \bar {h}_n^{\LR}\leq - \epsilon  ) \geq  \Pr(  \bar {h}_n^{\LR} = \epsilon  ) .
\end{equation*}
Obviously, the special sample composed of zeros except for a single $a$ results in $\bar {h}_n^{\LR} = \epsilon$. Thus,
\begin{eqnarray*}
  \Pr(  \bar {h}_n^{\LR}\geq \epsilon  ) &=& \Pr( \bar {h}_n^{\LR}\leq - \epsilon  ) \geq  \Pr( \bar {h}_n^{\LR} = \epsilon  ) \\
   & \geq & n  \frac{\theta_0 }{2} (1-\theta_0)^{n-1}\\
   &=&  \frac{n}{2} \left( \frac{a}{n \epsilon}\right)^{\alpha} \xi \left(1- \left( \frac{a}{n \epsilon}\right)^{\alpha} \xi \right)^{n-1}.
\end{eqnarray*}

We take the following value of $\epsilon$,
\begin{equation*}
  \epsilon = \epsilon^{\ast} =  a \left( \frac{ \xi}{\delta n^{\alpha-1}}\right)^{\frac{1}{\alpha}} \left(1- \frac{ e \delta}{n}\right)^{\frac{n-1}{\alpha}}.
\end{equation*}
Then we check that
\begin{equation*}
  \theta_0 =  \left(\frac{a}{ n \epsilon^{\ast}}\right)^{\alpha}\xi  =  \frac{a^{\alpha} \xi}{n^{\alpha}  a^{\alpha} \frac{ \xi}{\delta n^{\alpha-1}} \left(1- \frac{ e \delta}{n}\right)^{n-1} } = \frac{\delta}{n} \left(1- \frac{ e \delta}{n}\right)^{-(n-1)}.
\end{equation*}
It is straightforward to verify that for $x\in (0,1)$,  $\left(1- { x}/{n}\right)^{-(n-1)} $ is an increasing function. Consequently, $\left(1- { x}/{n}\right)^{-(n-1)}  \leq \left(1- { 1}/{n}\right)^{-(n-1)} \equiv c_n $. Notice that
\begin{equation*}
  c_n = \left(\frac{n}{n-1}\right)^{n-1}\cdot 1 \leq \left(\frac{\frac{n}{n-1}(n-1)+1}{n-1+1}\right)^{n-1+1}=\left(\frac{n+1}{n}\right)^n = c_{n+1}.
\end{equation*}
Then, $c_n$ is an increasing sequence with a limit value $\lim_{n \rightarrow \infty} c_n = e$.
Thus, as long as $e \delta \in(0,1)$, we have $\theta_0 \leq  {e \delta}/{n} \leq {1}/{n} \leq 1$, which ensures that the probability measure $ \mathds{P}_{\betheta_0}$ is valid. To ensure
that the probability measure $ \mathds{P}_{\betheta}$ is valid, we enforce $\theta  \leq 1$. Recall that $\alpha \in (1,2]$, or alternatively, we enforce $\theta^{{\alpha}/{(\alpha-1)} } \leq 1$.

Notice that
\begin{equation*}
  \theta^{\frac{\alpha}{\alpha-1} } = \left(\frac{a}{ n \epsilon^{\ast}}\right)^{\alpha}\xi  \xi ^{\frac{1}{\alpha-1}}  = \theta_0 \xi ^{\frac{1}{\alpha-1}} \leq  \frac{e \delta}{n}\xi ^{\frac{1}{\alpha-1}},
\end{equation*}
thus $n \geq e\delta \xi ^{{1}/{(\alpha-1)}} $ makes $\theta \leq 1$. Then the conditions for parameters are $e \delta \in(0,1)$ and $n \geq \max(1,e\delta \xi ^{{1}/{(\alpha-1)}} )$.

For the selected parameter $\epsilon^{\ast}$, recall that $\theta_0 ={\delta}/{n}   \left(1-  {e\delta}/{n} \right)^{n-1} \leq  {e \delta}/{n}$, and we check that
\begin{eqnarray*}
  \Pr(  \bar {h}_n^{\LR}\geq \epsilon^{\ast}  )   & \geq  &   \frac{n}{2} \theta_0  (1-\theta_0)^{n-1}  \\
  & \geq & \frac{n}{2} \theta_0  \left(1-  \frac{e\delta}{n} \right)^{n-1} \\
  & = & \frac{n}{2} \frac{\delta}{n} \left(1- \frac{ e \delta}{n}\right)^{-(n-1)}  \left(1-  \frac{e\delta}{n} \right)^{n-1} =\frac{\delta}{2}.
\end{eqnarray*}
Then, we can conclude that
\begin{equation*}
  \Pr( | \bar {h}_n^{\LR} | \geq \epsilon^{\ast}  ) = 2  \Pr(  \bar {h}_n^{\LR}\geq \epsilon^{\ast}  ) \geq \delta.
\end{equation*}
$\hfill\Box$
\endproof

\section{Proof of Proposition \ref{theorem-IS-bound-concentration-lower-1}}\label{appx:P4-proof}
\proof{Proof:}
We construct a function $h$ and two probability measures $\mathds{P}_{\betheta}$ and $\mathds{P}_{\betheta_0}$ that satisfy the inequality. Consider a constant $a>0$, and let the random variables $X$ and $\tilde{X}$ take values in the support $\mathcal{X} = \Re$.

We now define the two probability measures as follows. Denote the density functions of $\mathds{P}_{\betheta}$ and $\mathds{P}_{\betheta_0}$ as $f_{\betheta}(x)$ and $f_{\betheta_0}(x)$, respectively. For $\theta, \theta_0 \in [0,1]$, define
\begin{equation*}
  {f_{\betheta}(x)} =
\left\{
    \begin{array}{lc}
        \frac{1-\theta}{2a}, & \quad |x|<a,  \\
        \frac{\theta}{2}e^{a } e^{-  |x| },  &  \quad |x| \geq   a.   \\
    \end{array}
\right.
\end{equation*}
\begin{equation*}
  {f_{\betheta_0}(x)} =
\left\{
    \begin{array}{lc}
        \frac{1-\theta_0}{2a}, & \quad |x|<a,  \\
        \frac{\theta_0}{2}e^{a } e^{-  |x| },  &  \quad |x| \geq   a.   \\
    \end{array}
\right.
\end{equation*}
By definition, it is easy to check that
\begin{eqnarray*}
   \Pr( X \geq a ) & = \Pr( X \leq -a ) = \frac{\theta}{2},\quad \Pr( |X|<a)=1-\theta, \\
   \Pr(  \tilde{X} \geq a  ) & = \Pr(  \tilde{X} \leq -a  ) = \frac{\theta_0}{2}, \quad \Pr( |\tilde{X}|<a )=1-\theta_0.
\end{eqnarray*}
Thus, $\int_\Re f_{\betheta}(x) dx = \int_\Re f_{\betheta_0}(x) dx = 1$.

The likelihood ratio is given by
\begin{equation*}
  {\l(x)} = \frac{f_{\betheta}(x)}{f_{\betheta_0}(x)} =
\left\{
    \begin{array}{lc}
        \frac{1-\theta}{1-\theta_0}, & \quad |x|<a,  \\
        \frac{\theta}{\theta_0},  &  \quad |x| \geq   a.   \\
    \end{array}
\right.
\end{equation*}
Notice that the divergence $I_{\alpha}(\mathds{P}_{\betheta} \| \mathds{P}_{\betheta_0})$ is valid because
\begin{eqnarray*}
  I_{\alpha}(\mathds{P}_{\betheta} \| \mathds{P}_{\betheta_0}) & = & \mathds{E}[\l^{\alpha}(\tilde{X})] \\
  & = &  2\left( \frac{\theta}{\theta_0}\right)^{\alpha} \frac{\theta_0}{2}+\left( \frac{1-\theta}{1-\theta_0} \right)^{\alpha} (1-\theta_0) \\
   &=& \left( \frac{\theta}{\theta_0}\right)^{\alpha} \theta_0 + \left( \frac{1-\theta}{1-\theta_0} \right)^{\alpha} (1-\theta_0) \\
   & \geq &  \left( \frac{\theta}{\theta_0} \theta_0 + \frac{1-\theta}{1-\theta_0}  (1-\theta_0)  \right)^{\alpha} = 1.
\end{eqnarray*}
The last inequality holds due to the convexity of $x^{\alpha} $ for any $\alpha >1$ when $x \geq 0$ and Jensen's inequality.

We now define the performance function $h(\cdot)$ as
\begin{equation*}
  h(x) =
\left\{
    \begin{array}{lc}
        0, & \quad |x|<a,  \\
        x,  &  \quad |x| \geq   a.   \\
    \end{array}
\right.
\end{equation*}
It immediately follows that $\mathds{E}[h(X)]=\mathds{E}[h(\tilde{X})]=0$, so that $\bar{h}(\betheta)=0$. For any $|\lambda| <1$, the moment generating function of $h(x)$ exists. In fact,
\begin{equation*}
  M_F(\lambda) =   \mathds{E}[  e^{\lambda h(\tilde{X})  } ]  = (1-\theta_0) + \frac{\theta_0}{2} \frac{e^{ -\lambda a }}{ 1+\lambda} + \frac{\theta_0}{2} \frac{e^{ \lambda a }}{ 1-\lambda} < \infty.
\end{equation*}
Thus, the $p$th norm $\|h\|_{p,\mathds{P}_{\betheta_0}}$ exists for any $p \geq 1$. We select $a=\|h\|_{p,\mathds{P}_{\betheta_0}} >0$.

The parameters $\theta$ and $\theta_0$ are chosen as follows. For any value $r \in (1,2]$, define
\begin{eqnarray*}
   \theta & = & \left( \frac{a}{n \epsilon}\right)^{r-1} \xi, \\
   \theta_0 & = & \left( \frac{a}{n \epsilon}\right)^{r} \xi,
\end{eqnarray*}
where $\xi =  I_{\alpha}(\mathds{P}_{\betheta} \| \mathds{P}_{\betheta_0})^{{r}/{\alpha}} > 0$, and $\epsilon>0$ will be specified later. By definition, we have $a {\theta}/{\theta_0} = n \epsilon$.

Consider the classical LR estimator $\bar {h}_n^{\LR}$. Recall that its expectation is zero in this specific setting. Given the i.i.d. sample $\{\tilde{X}_1,\cdots, \tilde{X}_n\}$ drawn from measure $\mathds{P}_{\betheta_0}$ and by the construction, $\bar {h}_n^{\LR}$ is expressed as
\begin{eqnarray*}
  \bar{h}_n^{\LR} &=& \frac{1}{n}\sum_{i=1}^n h(\tilde{X}_i) \l(\tilde{X}_i) \\
   &=& \frac{1}{n} \sum_{i=1}^n {\bf 1} \{|\tilde{X}_i| \geq a\} \tilde{X}_i \frac{\theta}{\theta_0}   \\
    &=& \epsilon \sum_{i=1}^n {\bf 1}\{|\tilde{X}_i| \geq a\} \frac{ \tilde{X}_i}{a}.
\end{eqnarray*}
The last equation holds due to $a {\theta}/{\theta_0} = n \epsilon$. For the deviation level $\epsilon>0$, it is straightforward to see that $\Pr( \bar {h}_n^{\LR}\geq \epsilon  ) =\Pr( \bar {h}_n^{\LR}\leq - \epsilon  )$. Consider a specific sample where only one point takes value in $[a,\infty)$ and the remaining points fall within $(-a,a)$. Clearly, this sample results in $\bar {h}_n^{\LR} \geq \epsilon$. Therefore,
\begin{equation*}
   \Pr( \bar {h}_n^{\LR}\geq \epsilon  )  \geq  n  \frac{\theta_0 }{2} (1-\theta_0)^{n-1}\\
   =  \frac{n}{2} \left( \frac{a}{n \epsilon}\right)^{r} \xi \left(1- \left( \frac{a}{n \epsilon}\right)^{r} \xi \right)^{n-1}.
\end{equation*}

We take the following value of $\epsilon$
\begin{equation*}
  \epsilon = \epsilon^{\ast} =  a \left( \frac{ \xi}{\delta n^{r-1}}\right)^{\frac{1}{r}} \left(1- \frac{ e \delta}{n}\right)^{\frac{n-1}{r}}.
\end{equation*}
If $e \delta \in(0,1)$ and  $n \geq \max(1,e\delta \xi ^{{1}/{(r-1)}} )$, following the similar steps in the proof of Proposition \ref{theorem-IS-concentration-lower}, we can check that
\begin{equation*}
  \Pr( |\bar {h}_n^{\LR} | \geq \epsilon^{\ast}  ) = 2  \Pr( \bar {h}_n^{\LR}\geq \epsilon^{\ast}  ) \geq \delta.
\end{equation*}
Notice that
\begin{equation*}
  \epsilon = \epsilon^{\ast} =  aI_{\alpha}(\mathds{P}_{\betheta} \| \mathds{P}_{\betheta_0})^{\frac{1}{\alpha}}   \left( \frac{ 1}{\delta n^{r-1}}\right)^{\frac{1}{r}} \left(1- \frac{ e \delta}{n}\right)^{\frac{n-1}{r}}
\end{equation*}
is decreasing with $r$. Thus its the minimal value is obtained by taking $r=2$, which is
\begin{equation*}
  \epsilon = \epsilon^{\ast} = aI_{\alpha}(\mathds{P}_{\betheta} \| \mathds{P}_{\betheta_0})^{\frac{1}{\alpha}}  \frac{1}{\sqrt{n \delta}}  \left(1- \frac{ e \delta}{n}\right)^{\frac{n-1}{2}}.
\end{equation*}
Finally, we have $n \geq \max(1,e\delta \xi ^{{1}/{(r-1)}} )=\max(1, e\delta  I_{\alpha}(\mathds{P}_{\betheta} \| \mathds{P}_{\betheta_0})^{{2}/{\alpha}})$.
$\hfill\Box$
\endproof

\section{Proof of Proposition \ref{theorem-TR-consistency}}\label{appx:P5-proof}
\proof{Proof:} The proof is similar to \cite{Ionides2008}. First, we consider the bias $B_n$ of $\bar {h}_n^{\TR}$, which is given by
\begin{eqnarray}
  B_n &=& \mathds{E}\left[\bar {h}_n^{\TR}\right] - \bar{h}(\btheta)  \nonumber \label{eq-TR-consistency-bias-1} \\
   & = & \mathds{E}\left[ \frac{1}{n}\sum_{i=1}^n h(\tilde{\bX}_i)\l'_n(\tilde{\bX}_i) \right] - \bar{h}(\btheta) \nonumber \label{eq-TR-consistency-bias-2} \\
   & = & \mathds{E}[ h(\tilde{\bX}_i)\l'_n(\tilde{\bX}_i) ]  - \bar{h}(\btheta) \nonumber \label{eq-TR-consistency-bias-3} \\
   & = &  \int_{\mathcal{X}} h(\bx)\l'_n(\bx) d\mathds{P}_{\betheta_0}(\bx)  -  \int_{\mathcal{X}} h(\bx)d\mathds{P}_{\betheta}(\bx)  \nonumber \label{eq-TR-consistency-bias-4}\\
   & = & \int_{\mathcal{X}} h(\bx)\l'_n(\bx) d\mathds{P}_{\betheta_0}(\bx)  -  \int_{\mathcal{X}} h(\bx)\l(\bx)d\mathds{P}_{\betheta_0}(\bx)  \nonumber \label{eq-TR-consistency-bias-5} \\
   & = & \int_{\mathcal{X}} h(\bx)\left[ \min(\l(\bx),\tau_n) - \l(\bx) \right] d\mathds{P}_{\betheta_0}(\bx)     \label{eq-TR-consistency-bias-6} \\
   & = & \int_{\bx: \l(\bx)>\tau_n}h(\bx)\left(\tau_n - \l(\bx) \right)d\mathds{P}_{\betheta_0}(\bx) , \nonumber \label{eq-TR-consistency-bias-7}
\end{eqnarray}
where \eqref{eq-TR-consistency-bias-6} follows from the definition of $\l'_n(\bx)$. Observing that $\l(\bx)>\tau_n$ over the region of integration and $\l(\bx) \geq 0$, we have
\begin{equation} \nonumber \label{eq-TR-consistency-bias-8}
  |B_n|  \leq  \int_{\bx: \l(\bx)>\tau_n} |h(\bx)|  \l(\bx) d\mathds{P}_{\betheta_0}(\bx)  =  \int_{\bx: \l(\bx)>\tau_n} |h(\bx)|  d\mathds{P}_{\betheta}(\bx) .
\end{equation}
$\mathds{E}[h^2(\bX)]<\infty$ implies that $\mathds{E}[|h(\bX)|] <\infty$. As $\tau_n \to \infty$, the dominated convergence gives $B_n \to 0$, i.e., $\bar {h}_n^{\TR}$ is asymptotically unbiased.

Next, we consider the variance of $\bar {h}_n^{\TR}$, which is given by
\begin{eqnarray}
  \Var(\bar {h}_n^{\TR}) 
   & \leq &  \mathds{E}[\left(\bar {h}_n^{\TR}\right)^2]   = \frac{1}{n} \mathds{E} \left[ h^2(\tilde{\bX})\l'_n(\tilde{\bX})^2 \right] \nonumber \label{eq-TR-consistency-variance-2}\\
   & = & \frac{1}{n} \int_{\mathcal{X}} h^2(\bx) \left(\min\{\l(\bx),\tau_n\} \right)^2 d\mathds{P}_{\betheta_0}(\bx) \nonumber \label{eq-TR-consistency-variance-3}\\
   & \leq & \frac{\tau_n}{n} \int_{\mathcal{X}} h^2(\bx) \l(\bx)  d\mathds{P}_{\betheta_0}(\bx)    \nonumber \label{eq-TR-consistency-variance-4}\\
   & = & \frac{\tau_n}{n}  \int_{\mathcal{X}} h^2(\bx)  d\mathds{P}_{\betheta}(\bx)   \nonumber \label{eq-TR-consistency-variance-5}\\
   & = &\frac{\tau_n}{n}\mathds{E}[h^2(\bX)] \nonumber \label{eq-TR-consistency-variance-6}.
\end{eqnarray}
Thus, $\Var(\bar {h}_n^{\TR}) \to 0$ as $n \to \infty$ since $\tau_n/n\to 0$.

Now, for any $\varepsilon >0$, due to Markov's inequality, we have
\begin{equation} \nonumber \label{eq-TR-consistency-bias-9}
  \Pr(|\bar {h}_n^{\TR} - \bar{h}(\btheta)|\geq \varepsilon) \leq \frac{\Var(\bar {h}_n^{\TR} - \bar{h}(\btheta))}{\varepsilon^2} =\frac{\Var(\bar {h}_n^{\TR} )}{\varepsilon^2} =  \frac{\tau_n}{n \varepsilon^2 }\mathds{E}[h^2(\bX)]  .
\end{equation}
Then, as $n \to \infty$, $\Pr(|\bar {h}_n^{\TR} - \bar{h}(\btheta)|\geq \varepsilon) \to 0$, so $\bar {h}_n^{\TR}$ is a weakly consistent estimator.
$\hfill\Box$
\endproof

\section{Proof of Lemma \ref{theorem-TR-bound}}\label{appx:L1-proof}
\proof{Proof:} The proof is similar to Lemma 2 of \cite{papini2019optimistic}. Starting with the bias, note that the LR estimator $\bar {h}_n^{\LR}$  is unbiased for $ \bar{h}(\btheta)$, so
\begin{eqnarray}
  B_n & = &  \mathds{E}[\bar {h}_n^{\TR}] - \bar{h}(\betheta)  \nonumber   \\
   & = &  \mathds{E}[\bar {h}_n^{\TR}] - \mathds{E}[\bar {h}_n^{\LR}]   \nonumber  \\
   & = &  \mathds{E}\left[ \frac{1}{n}\sum_{i=1}^n h(\tilde{\bX}_i)\l'_n(\tilde{\bX}_i) \right]  - \mathds{E}\left[ \frac{1}{n}\sum_{i=1}^n h(\tilde{\bX}_i)\l(\tilde{\bX}_i) \right]  \nonumber  \\
   & = &  \mathds{E}[ h(\tilde{\bX})\l'_n(\tilde{\bX}) ]  - \mathds{E}[ h(\tilde{\bX})\l(\tilde{\bX}) ]   \nonumber\\
   & = & \mathds{E}\left[  h(\tilde{\bX})\left( \l'_n(\tilde{\bX}) - \l(\tilde{\bX})\right)\right]  \nonumber\\
   & = &  \mathds{E}\left[  h(\tilde{\bX})  (\tau_n - \l(\tilde{\bX})) {\bf 1} \{ \l(\tilde{\bX}) >\tau_n \}   \right] \label{eq-TR-bound-bias-0}.
\end{eqnarray}
Note that $|\tau_n - \l(\tilde{\bX})| <  \l(\tilde{\bX})$ over the region $ \l(\tilde{\bX}) >\tau_n  $, thus
\begin{eqnarray*}
  |B_n| & \leq & \mathds{E}\left[ | h(\tilde{\bX}) | |\tau_n - \l(\tilde{\bX})|{\bf 1}\{ \l(\tilde{\bX}) >\tau_n \}   \right]   \\
   & \leq &\| h \|_{\infty, \mathds{P}_{\betheta_0}}  \mathds{E}\left[   \l(\tilde{\bX}) {\bf 1} \{ \l(\tilde{\bX}) >\tau_n \}    \right]  \\
   & = &\| h \|_{\infty, \mathds{P}_{\betheta_0}} \mathds{E}\left[ \l^{\alpha}(\tilde{\bX}) \l^{1-\alpha} (\tilde{\bX})  {\bf 1}\{ \l(\tilde{\bX}) >\tau_n \}   \right]   \\
   & \leq & \| h \|_{\infty, \mathds{P}_{\betheta_0}}{\tau_n}^{1-\alpha}\mathds{E}\left[  \l^{\alpha} (\tilde{\bX})  {\bf 1} \{ \l(\tilde{\bX}) >\tau_n  \}    \right]   \\
   & \leq & \| h \|_{\infty, \mathds{P}_{\betheta_0}} {\tau_n}^{1-\alpha}  \mathds{E}[  \l^{\alpha} (\tilde{\bX})] \nonumber \\
   & = & \| h \|_{\infty, \mathds{P}_{\betheta_0}} {\tau_n}^{1-\alpha}  I_{\alpha}(\mathds{P}_{\betheta} \| \mathds{P}_{\betheta_0}),
   \end{eqnarray*}
where the third inequality holds by observing that $\l(\tilde{\bX}) >\tau_n $ over the area and $\alpha > 1$.

For the variance, the argument is similar, and we have
\begin{eqnarray*}
  V_n &=& \Var(\bar {h}_n^{\TR}) \leq  \mathds{E}[ (\bar {h}_n^{\TR} )^2]\\
   & = & \frac{1}{n} \mathds{E}\left[ h^2(\tilde{\bX})\l'_n(\tilde{\bX})^2 \right]  \\
   & \leq & \frac{1}{n}\| h \|_{\infty, \mathds{P}_{\betheta_0}}^2 \mathds{E}\left[ \l'_n(\tilde{\bX})^2 \right]  \nonumber \\
   & = & \frac{1}{n}\| h \|_{\infty, \mathds{P}_{\betheta_0}}^2 \mathds{E} \left[ \l'_n(\tilde{\bX})^{\alpha} \l'_n(\tilde{\bX})^{2-\alpha} \right]  \\
   & \leq & \frac{1}{n} \| h \|_{\infty, \mathds{P}_{\betheta_0}}^2  {\tau_n}^{2-\alpha} \mathds{E}[ \l^{\alpha}(\tilde{\bX}) ] \\
   & = & \frac{1}{n} \| h \|_{\infty, \mathds{P}_{\betheta_0}}^2  {\tau_n}^{2-\alpha} I_{\alpha}(\mathds{P}_{\betheta} \| \mathds{P}_{\betheta_0}),\end{eqnarray*}
where the last inequality holds by observing that $\alpha \leq 2$, and by definition $\l'_n(\tilde{\bX}) \leq \tau_n$ and $\l'_n(\tilde{\bX}) \leq \l(\tilde{\bX})$.
$\hfill\Box$
\endproof



\section{Proof of Lemma \ref{theorem-TR-bound-moment}}\label{appx:L2-proof}
\proof{Proof.} Again starting with the bias, recalling equation \eqref{eq-TR-bound-bias-0}, we have
\begin{eqnarray*}
  |B_n| & \leq & \mathds{E}\left[ | h(\tilde{\bX}) | |\tau_n - \l(\tilde{\bX})|{\bf 1} \{ \l(\tilde{\bX}) >\tau_n \}    \right] \\
   & \leq & \mathds{E}\left[ | h(\tilde{\bX})|   \l(\tilde{\bX}) {\bf 1}\{ \l(\tilde{\bX}) >\tau_n \}    \right]  \\
   & \leq & \left( \mathds{E}[ |h^p(\tilde{\bX})|  ] \right)^{\frac{1}{p}}
   \left(\mathds{E}[  \l^q(\tilde{\bX}) {\bf 1}\{ \l(\tilde{\bX}) >\tau_n \}  ] \right)^{\frac{1}{q}} \\
   & = & \| h \|_{p,\mathds{P}_{\betheta_0}} \left(\mathds{E}[  \l^{q-\alpha}(\tilde{\bX})  \l^{\alpha}(\tilde{\bX}) {\bf 1} \{ \l(\tilde{\bX})) >\tau_n \} ] \right)^{\frac{1}{q}}   \\
   & \leq & \| h \|_{p,\mathds{P}_{\betheta_0}} \left( {\tau_n}^{q-\alpha} \mathds{E}[    \l^{\alpha}(\tilde{\bX}) {\bf 1} \{ \l(\tilde{\bX}) >\tau_n \}  ] \right)^{\frac{1}{q}}  \\
   & \leq & \| h \|_{p,\mathds{P}_{\betheta_0}} \left( {\tau_n}^{q-\alpha} \mathds{E} [    \l^{\alpha}(\tilde{\bX}) ] \right)^{\frac{1}{q}} \\
   & = & \|h \|_{p,\mathds{P}_{\betheta_0}} \left( {\tau_n}^{q-\alpha} I_{\alpha}(\mathds{P}_{\betheta} \| \mathds{P}_{\betheta_0})\right)^{\frac{1}{q}}
\\
   & = & \| h\|_{p,\mathds{P}_{\betheta_0}}  {\tau_n}^{1-\alpha+\frac{\alpha}{p}} I_{\alpha}(\mathds{P}_{\betheta} \| \mathds{P}_{\betheta_0})^{1-\frac{1}{p}},
\end{eqnarray*}
where the second inequality follows from $|\tau_n - \l(\tilde{\bX})| <  \l(\tilde{\bX})$ over the region $\l(\tilde{\bX}) >\tau_n$, the third inequality holds by applying H\"{o}lder inequality with $p>1,q>1$, and $\frac{1}{p} + \frac{1}{q} = 1$, the fourth inequality is obtained by picking suitable $p$ such that $q \leq \alpha$ and the fact $\l(\tilde{\bX}) > \tau_n$ over the region $\l(\tilde{\bX}) >\tau_n$, the last equality derives from substitution $\frac{1}{q} = 1- \frac{1}{p}$, where recall that $q = \frac{p}{p-1}\leq \alpha$.

For the variance, we have
\begin{eqnarray*}
  V_n &=& \Var(\bar {h}_n^{\TR}) = \frac{1}{n} \Var ( h(\tilde{\bX})\l'_n(\tilde{\bX}) )\\
  & \leq &  \frac{1}{n}  \mathds{E}\left[ h^{2} (\tilde{\bX})  \l'_n(\tilde{\bX})^{2} \right] \\
   & \leq &  \frac{1}{n} \left( \mathds{E} [ { |h^{2\tilde{p}} (\tilde{\bX})  |}  ] \right)^{\frac{1}{\tilde{p}}}
   \left(\mathds{E} [  \l'_n(\tilde{\bX})^{2\tilde{q}}  ] \right)^{\frac{1}{\tilde{q}}} \\
   & = & \frac{1}{n} \| h \|_{p,\mathds{P}_{\betheta_0}}^2 \left(\mathds{E}  [  \l'_n(\tilde{\bX})^{2\tilde{q}-\alpha}  \l'_n(\tilde{\bX})^{\alpha}  ] \right)^{\frac{1}{\tilde{q}}}  \\
   & \leq & \frac{1}{n} \| h \|_{p,\mathds{P}_{\betheta_0}}^2 \left( {\tau_n}^{2\tilde{q}-\alpha} \mathds{E} [  \l^{\alpha}(\tilde{\bX})  ] \right)^{\frac{1}{\tilde{q}}}   \\
   & = & \frac{1}{n} \| h \|_{p,\mathds{P}_{\betheta_0}}^2 \left( {\tau_n}^{2\tilde{q}-\alpha} I_{\alpha}(\mathds{P}_{\betheta} \| \mathds{P}_{\betheta_0})\right)^{\frac{1}{\tilde{q}}} \\
   & = & \frac{1}{n} \| h \|_{p,\mathds{P}_{\betheta_0}}^2 {\tau_n}^{2-\alpha+\frac{2\alpha}{p}} I_{\alpha}(\mathds{P}_{\betheta} \| \mathds{P}_{\betheta_0})^{1-\frac{2}{p}},
\end{eqnarray*}
where the second inequality holds by applying H\"{o}lder inequality with $\tilde p>1$, $\tilde q>1$, and $\frac{1}{\tilde{p}} + \frac{1}{\tilde{q}} = 1$, the second equality is obtained by choosing $\tilde{p} = \frac{p}{2}$, the third inequality follows from carefully choosing $p$ such that $\alpha \leq 2\tilde{q}$  and by definition $\l'_n(\tilde{\bX}) \leq \tau_n$ and $\l'_n(\tilde{\bX}) \leq \l(\tilde{\bX})$, the last equality derives from substitution $\frac{1}{\tilde{q}} = 1- \frac{1}{\tilde{p}} = 1 - \frac{2}{p} $, where recall that $\alpha \leq 2\tilde{q} = \frac{2\tilde{p}}{\tilde{p}-1} = \frac{2p}{p-2}$, and $p=2\tilde{p} >2$.

Combing both conditions on bias and variance, for $p > 2$ necessitates $\frac{p}{p-1} \leq \alpha \leq \frac{2p}{p-2}$.
$\hfill\Box$
\endproof

\section{Discussion of Parameters in Theorem \ref{theorem-TR-bound-moment-concentration-MGF}}\label{appx:remark}
For simplicity, we denote $Y=h(\tilde{\bX})l'_n(\tilde{\bX})$ here. The condition \eqref{eq-TR-bound-moment-concentration-MGF-1-0} in Theorem \ref{theorem-TR-bound-moment-concentration-MGF} changes into
\begin{equation*}
  \mathds{E}\left[  e^{ \lambda (Y -\mathds{E}[Y])}\right]
    \leq  e^{\frac{1}{2} \lambda^2  \sigma^2}, \quad \forall|\lambda| \leq \frac{1}{\Lambda }, \label{eq:remark-1}
\end{equation*}
for some $\sigma^2 \geq \Var(Y)$ and $\Lambda \geq 0$. It is obvious that larger $\sigma^2$ can relax the constraints on $\Lambda$, leading to potentially smaller $\lambda$, and vice versa. Considering the upper bound in \eqref{eq-TR-bound-moment-concentration-MGF-1}, there should be a balance between $\sigma^2$ and $\Lambda$.

In the following, we will give some typical choice of the parameters $\sigma^2$ and $\Lambda$ in some special settings.

If $Y$ follows a normal distribution, it is easy to obtain that 
\begin{equation*}
  \mathds{E} \left[   e^{ \lambda (Y -\mathds{E}[Y])} \right] = e^{\frac{1}{2} \lambda^2  \Var(Y)}, \quad \forall \lambda \in \mathds{R},
\end{equation*}
so we can take $\sigma^2=\Var(Y)$ and $\Lambda=0$. 

If $Y$ is a bounded random variable taking values in $[a,b]$, Hoeffding's inequality gives 
\begin{equation*}
  \mathds{E} \left[   e^{ \lambda (Y -\mathds{E}[Y])} \right] \leq e^{\frac{1}{8} \lambda^2  (b-a)^2}, \quad \forall \lambda \in \mathds{R},
\end{equation*}
so we can set $\sigma^2=\frac{1}{4}(b-a)^2$ and $\Lambda=0$. Furthermore, if $|Y| \leq B$, then $b-a \leq 2B$, so we can take $\sigma^2=B^2$ and $\Lambda=0$.  

If $Y$ follows an exponential distribution with expectation $\mu=\mathds{E}[Y] >0$, then 
\begin{equation*}
  \mathds{E} \left[   e^{ \lambda (Y -\mathds{E}[Y])} \right] = \frac{e^{-\lambda \mu}}{1-\lambda \mu}, \quad \forall \lambda <\frac{1}{\mu}.
\end{equation*}
Notice that 
\begin{equation*}\label{eq:remark}
  \frac{e^{-t}}{\sqrt{1-2t}}\leq e^{2 t^2}, \quad \forall |t| \leq \frac{1}{4},
\end{equation*}
then if $|\lambda| <1/(2\mu)$, we have 
\begin{eqnarray*}
  \mathds{E} \left[   e^{ \lambda (Y -\mathds{E}[Y])} \right] &=& \frac{e^{-\lambda \mu}}{1-\lambda \mu} =  \left( \frac{e^{-\frac{\lambda \mu}{2}}}{\sqrt{1-2\frac{\lambda \mu}{2}}} \right)^2\\
   & \leq & \left( e^{2\left(\frac{\lambda \mu}{2}\right)^2} \right)^2=e^{\lambda^2 \mu^2}<e^{\frac{1}{2}\lambda^2 (2\mu)^2}.
\end{eqnarray*}
So we can take $\sigma^2=(2\mu)^2$ and $\Lambda=2\mu$.


\section{Proof and Discussion of Proposition \ref{theorem-TR-bound-moment-concentration-min1}}\label{appx:P6-proof}
Bernstein's moment condition (Definition \ref{def:bernstein}) characterizes a random variable by how quickly its moments grow, and it is often used to verify the sub-exponential property. A typical sufficient condition is that $h$ is bounded, a scenario which corresponds to the settings discussed in Section \ref{sc:concentration-bound-inf}. Definition \ref{def:bernstein} is also satisfied by various unbounded variables (e.g., sub-Gaussian variables), enhancing its applicability across a wider range of contexts.

Next, we prove Proposition \ref{theorem-TR-bound-moment-concentration-min1}, and need the following lemma.
 \begin{lemma}\label{theorem-TR-bound-moment-concentration}
If $h(\tilde{\bX}), \tilde{\bX}\sim \mathds{P}_{\betheta_0}$, satisfies Bernstein's moment condition for parameter $b_h>0$, $\| h \|_{p,\mathds{P}_{\betheta_0}} < \infty$ for some $p > 2 $, and $I_{\alpha}(\mathds{P}_{\betheta} \| \mathds{P}_{\betheta_0})$ is finite for some $\alpha >1$ ($ \frac{p}{p-1} \leq \alpha \leq \frac{2p}{p-2}$), then
\begin{equation}
   \Pr\left(\begin{aligned}
    & |\bar {h}_n^{\TR} - \bar{h}(\btheta)|\leq   \| h \|_{p,\mathds{P}_{\betheta_0}} \sqrt{\frac{2  {\tau_n}^{2-\alpha+\frac{2\alpha}{p}}   \ln\frac{2}{\delta} }{n} I_{\alpha}(\mathds{P}_{\betheta} \| \mathds{P}_{\betheta_0})^{1-\frac{2}{p}} } ~\\
    &   ~~+ \| h \|_{p,\mathds{P}_{\betheta_0}}  \frac{b \tau_n^{1+\frac{\alpha}{p}}I_{\alpha}(\mathds{P}_{\betheta} \| \mathds{P}_{\betheta_0})^{-\frac{1}{p}}  \ln\frac{2}{\delta}}{n}  + \| h \|_{p,\mathds{P}_{\betheta_0}}  {\tau_n}^{1-\alpha+\frac{\alpha}{p}} I_{\alpha}(\mathds{P}_{\betheta} \| \mathds{P}_{\betheta_0})^{1-\frac{1}{p}}
    \end{aligned}
     \right) \geq 1-\delta,~~\delta \in (0,1). \label{eq-TR-bound-moment-concentration}
\end{equation}
where $b=b_h I_{\alpha}(\mathds{P}_{\betheta} \| \mathds{P}_{\betheta_0})^{\frac{1}{p}} / \|h\|_{p,\mathds{P}_{\betheta_0}}$.
\end{lemma}
\proof{Proof:} The proof is a straightforward application of Bernstein's inequality together with Lemma \ref{theorem-TR-bound-moment}.
Since $h(\tilde{\bX})$ satisfies Bernstein's moment condition with parameter $b_h$ and $\l'_n(\tilde{\bX}) \leq \tau_n$, then for $h(\tilde{\bX})\l'_n(\tilde{\bX})$, $\forall k \geq 2$
\begin{eqnarray*}
  \mathds{E}[ |h(\tilde{\bX})\l'_n(\tilde{\bX})|^k] & \leq & \mathds{E}[ |h(\tilde{\bX})|^k |\l'_n(\tilde{\bX})|^k] \\
   & \leq &  \frac{1}{2} \mathds{E}[ h^2(\tilde{\bX}) ] \tau_n^2 k!  ( b_h \tau_n) ^{k-2}.
\end{eqnarray*}
Choose a constant $c \geq 1$ such that $c\mathds{E}[ |h(\tilde{\bX})\l'_n(\tilde{\bX})|^2] \geq \mathds{E}[ h^2(\tilde{\bX}) ] \tau_n^2$, and notice that $\frac{\alpha}{p} > 0$, then
\begin{eqnarray*}
  \mathds{E}[ |h(\tilde{\bX})\l'_n(\tilde{\bX})|^k] & \leq & \frac{1}{2} c\mathds{E}[ |h(\tilde{\bX})\l'_n(\tilde{\bX})|^2] k! ( b_h \tau_n) ^{k-2} \\
   & \leq & \frac{1}{2} \mathds{E}[ |h(\tilde{\bX})\l'_n(\tilde{\bX})|^2] k! (c b_h \tau_n^{1+\frac{\alpha}{p}}) ^{k-2}.
\end{eqnarray*}
Then $h(\tilde{\bX})\l'_n(\tilde{\bX})$ satisfies Bernstein's moment condition with parameter $cb_h \tau_n^{1+\frac{\alpha}{p}}$. We observe that
$cb_h$ is also a parameter ensuring that $h(\tilde{\bX})$ meets Bernstein's moment condition. So we can pick a relatively large parameter $b_h$ in Bernstein's moment condition for $h(\tilde{\bX})$, and omit $c$ in the following for simplicity. For convenience, set $b_h \tau_n^{1+\frac{\alpha}{p}} = b\| h \|_{p,\mathds{P}_{\betheta_0}}  \tau_n^{1+\frac{\alpha}{p}}I_{\alpha}(\mathds{P}_{\betheta} \| \mathds{P}_{\betheta_0})^{-\frac{1}{p}}$, so $b=b_h I_{\alpha}(\mathds{P}_{\betheta} \| \mathds{P}_{\betheta_0})^{\frac{1}{p}} / \| h \|_{p,\mathds{P}_{\betheta_0}} $.
Applying Bernstein's inequality ({\citealt{boucheron2013concentration}}), we can state that with probability at least $1-\delta$ it holds that
\begin{eqnarray}
  |\bar {h}_n^{\TR} - \mathds{E}[\bar {h}_n^{\TR}]| & \leq & \sqrt{2 \mathds{E}[(\bar {h}_n^{\TR})^2]    \ln\frac{2}{\delta}} + \frac{b_h \tau_n^{1+\frac{\alpha}{p}}\ln\frac{2}{\delta}}{n} \nonumber \\
   & \leq &  \|h \|_{p,\mathds{P}_{\betheta_0}} \sqrt{\frac{2  {\tau_n}^{2-\alpha+\frac{2\alpha}{p}}   \ln\frac{2}{\delta} }{n} I_{\alpha}(\mathds{P}_{\betheta} \| \mathds{P}_{\betheta_0})^{1-\frac{2}{p}} } + \| h \|_{p,\mathds{P}_{\betheta_0}}  \frac{b \tau_n^{1+\frac{\alpha}{p}}I_{\alpha}(\mathds{P}_{\betheta} \| \mathds{P}_{\betheta_0})^{-\frac{1}{p}} \ln\frac{2}{\delta}}{n}, \nonumber \\ \label{eq-TR-bound-moment-concentration-2}
\end{eqnarray}
where the last line is obtained by equation \eqref{eq-TR-bound-moment-variance} in Lemma \ref{theorem-TR-bound-moment}. Combined with the bias term, i.e., equation \eqref{eq-TR-bound-moment-bias} in Lemma \ref{theorem-TR-bound-moment}, we have
\begin{eqnarray}
  |\bar {h}_n^{\TR} - \bar{h}(\betheta)| &=& |\bar {h}_n^{\TR} - \mathds{E}[\bar {h}_n^{\TR}]  + \mathds{E}[\bar {h}_n^{\TR}] - \bar{h}(\betheta)| \nonumber \\
   & \leq &  \| h \|_{p,\mathds{P}_{\betheta_0}} \sqrt{\frac{2  {\tau_n}^{2-\alpha+\frac{2\alpha}{p}}   \ln\frac{2}{\delta} }{n} I_{\alpha}(\mathds{P}_{\betheta} \| \mathds{P}_{\betheta_0})^{1-\frac{2}{p}} } + \|h \|_{p,\mathds{P}_{\betheta_0}}  \frac{b \tau_n^{1+\frac{\alpha}{p}}I_{\alpha}(\mathds{P}_{\betheta} \| \mathds{P}_{\betheta_0})^{-\frac{1}{p}} \ln\frac{2}{\delta}}{n}  \nonumber \\
   & &  +\| h \|_{p,\mathds{P}_{\betheta_0}}  {\tau_n}^{1-\alpha+\frac{\alpha}{p}}I_{\alpha}(\mathds{P}_{\betheta} \| \mathds{P}_{\betheta_0})^{1-\frac{1}{p}}.  \label{eq-TR-bound-moment-concentration-3}
\end{eqnarray}
$\hfill\Box$
\endproof

\proof{Proof of Proposition \ref{theorem-TR-bound-moment-concentration-min1}:}
We can minimize the right-hand side term in \eqref{eq-TR-bound-moment-concentration-3} by vanishing its first derivative, which is shown as following
\begin{equation}\label{eq-TR-bound-moment-concentration-5}
   \sqrt{\frac{2 \ln\frac{2}{\delta} }{n}I_{\alpha}(\mathds{P}_{\betheta} \| \mathds{P}_{\betheta_0})} (1-\frac{\alpha}{2} +\frac{\alpha}{p}) {\tau_n}^{\frac{\alpha}{2}} +  \frac{b \ln\frac{2}{\delta}}{n} (1+\frac{\alpha}{p}) \tau_n^{\alpha} + (1-\alpha+\frac{\alpha}{p}) I_{\alpha}(\mathds{P}_{\betheta} \| \mathds{P}_{\betheta_0}) =0.
\end{equation}
This quadratic function in ${\tau_n}^{\frac{\alpha}{2}}$ has a positive solution given by:
\begin{equation}\label{eq-TR-bound-moment-concentration-6}
    \tau_n^{\dag} = \left( \frac{-\sqrt{2}(1-\frac{\alpha}{2}+\frac{\alpha}{p}) + \sqrt{2(1-\frac{\alpha}{2}+\frac{\alpha}{p})^2 - 4b(1+\frac{\alpha}{p})(1-\alpha+\frac{\alpha}{p}))}}{2b(1+\frac{\alpha}{p})}\right)^{\frac{2}{\alpha}}  \left( \frac{n I_{\alpha}(\mathds{P}_{\betheta} \| \mathds{P}_{\betheta_0}) }{\ln\frac{2}{\delta}}\right)^{\frac{1}{\alpha}},
\end{equation}

Now, we show that the solution $\tau_n^{\dag}$ given in Equation \eqref{eq-TR-bound-moment-concentration-6} gives the unique minimum value of the upper bound in Equation \eqref{eq-TR-bound-moment-concentration-3}. Suppose a general truncation threshold has a form as
$\tau_n^{\frac{\alpha}{2}} = x \left( \frac{n I_{\alpha}(\mathds{P}_{\betheta} \| \mathds{P}_{\betheta_0}) }{\ln\frac{2}{\delta}}\right)^{\frac{1}{2}}$, $x>0$. Substituting it into Equation \eqref{eq-TR-bound-moment-concentration-3}, its right-hand side becomes $ g(x) \| h \|_{p,\mathds{P}_{\betheta_0}}\left( \frac{\ln\frac{2}{\delta}}{n} \right)^{1-\frac{1}{\alpha}-\frac{1}{p}} I_{\alpha}(\mathds{P}_{\betheta} \| \mathds{P}_{\betheta_0})^{\frac{1}{\alpha}}$, where
\begin{equation}\label{eq-TR-bound-moment-concentration-8}
  g(x) = g(x;\alpha,p) = \sqrt{2} x^{\frac{2}{\alpha}+\frac{2}{p}-1} +b x^{\frac{2}{\alpha}+\frac{2}{p}} +x^{\frac{2}{\alpha}+\frac{2}{p}-2}.
\end{equation}
For simplicity, denote $k=\frac{2}{\alpha}+\frac{2}{p}$. Because $p > 2$ and $\frac{p}{p-1} \leq \alpha \leq \frac{2p}{p-2}$, then $k \in [1,2] $ and
\begin{equation*}\label{eq-TR-bound-moment-concentration-9}
  g(x) = g(x;k) = \sqrt{2} x^{k-1} +b x^{k} +x^{k-2}.
\end{equation*}
It is easy to obtain its second derivative as
\begin{equation*}\label{eq-TR-bound-moment-concentration-10}
  g''(x) =x^{k-4}\left( \sqrt{2}(k-1)(k-2) x + b k(k-1)x^2 +(k-2)(k-3)\right) \equiv x^{k-4} g_1(x).
\end{equation*}
Note the $g_1(x)$ is a quadratic function of $x$. Recall that $k \in [1,2]$ and $x>0$, thus the minimum value of $g_1(x)$ over region $(0,\infty)$ is obtained at its axis of symmetry with value
\begin{eqnarray*}
  \min_{x>0} g_1(x) &=& (k-2)(k-3) -\frac{2(k-1)^2(k-2)^2 }{4 b k(k-1)} \nonumber \\
   &=& \frac{2-k}{2bk}\left( (1-2b)k^2-3(1-2b)k+2\right).
\end{eqnarray*}
Note that $k \in [1,2]$, and the second term is a quadratic function of $k$ with a axis $k=\frac{3}{2}$. Then it is easy to verify that the above term is non-negative provided $b \geq \frac{1}{18}$. Thus $g(x)$ is convex and has a unique minimum value at point $x^{\dag} = \frac{-\sqrt{2}(1-\frac{\alpha}{2}+\frac{\alpha}{p}) + \sqrt{2(1-\frac{\alpha}{2}+\frac{\alpha}{p})^2 - 4b(1+\frac{\alpha}{p})(1-\alpha+\frac{\alpha}{p}))}}{2b(1+\frac{\alpha}{p})} $.
$\hfill\Box$
\endproof

\section{Finding a Parameter Value to Satisfy Bernstein's Condition}\label{appx:para}
We discuss how to find the parameter in Bernstein's condition, i.e., $b >0$ such that $\mathds{E}[ |Y|^k] \leq \frac{1}{2}\mathds{E}[ Y^2]   k! b^{k-2}$ for all $k\geq 2$. For simplicity, we denote $v=\mathds{E}[ Y^2]$ in the following.

If $Y$ satisfies Bernstein's moment condition, Bernstein's inequality gives a concentration inequality for sample mean $\bar{Y}_n^{\MC}=1/n \sum_{i=1}^{n}Y_i$ as $\Pr( |\bar{Y}_n^{\MC} - \mathds{E}[\bar{Y}_n^{\MC}]| \geq \sqrt{2vt/n} + bt/n ) \leq 2e^{-t}$, for any $t>0$ and sample size $n$. Consider the bound of deviation term $\sqrt{2vt/n} + bt/n $ in the concentration inequality, $v$ determines the slow convergent term, and $b$ is a coefficient of fast convergent term. So roughly speaking, the useful part of this bound is contributed by $v$, and is not very sensitive to the parameter $b_h$. Here, we seek some information for $b$.

(1) For some typical settings of $Y$, we can directly compute its absolute moment and get the corresponding parameters. If $Y$ is bounded by $B$, then $b=B/3$ (\citealt{boucheron2013concentration}). If $Y\sim N(0,\sigma^2)$, $b=\sigma$ (\citealt{zhang2021concentration}). If $Y$ follows an exponential distribution with rate parameter $\lambda$, $b=1/\lambda$. We can use the exact value of $b$ or $cb(c \geq 1)$ as a possible parameter in our tests.

(2) For a general case, $Y$ may lack a MGF or fail to satisfy Bernstein's moment condition. Our idea comes from the sub-exponential norm of a sup-exponential random variable, $\|Y\|_{\psi_1} = \sup_{k \geq 1} k^{-1}\mathds{E}[ |Y|^k]^{\frac{1}{k}}$ (\citealt{vershynin2010introduction}, \citealt{zhang2021concentration}). By Stirling's approximation $k! \geq (k/e)^k$, $\mathds{E}[ |Y|^k] \leq e^k \|Y\|_{\psi_1}^k k! \leq \frac{1}{2} k! (2e^2\|Y\|_{\psi_1}^2)(e\|Y\|_{\psi_1})^{k-2}$. So a possible choice of parameter is $b=e\| Y\|_{\psi_1}$. If the random variable $Y$ follows a sub-exponential distribution, we believe the few lower order moments will give a good estimate to the true norm $\| Y\|_{\psi_1}$. In our test, we run a Monte Carlo simulation to estimate this norm, and take $b=2 \| Y\|_{\psi_1}$ as a estimated parameter to derive ``TruLR-E'' estimator.

\section{More Simulation Results for Normal Distribution.}\label{appx:normal}
\begin{figure}[htpb]
\begin{minipage}[t]{0.48\linewidth}
\centering
\includegraphics[width=8.2cm]{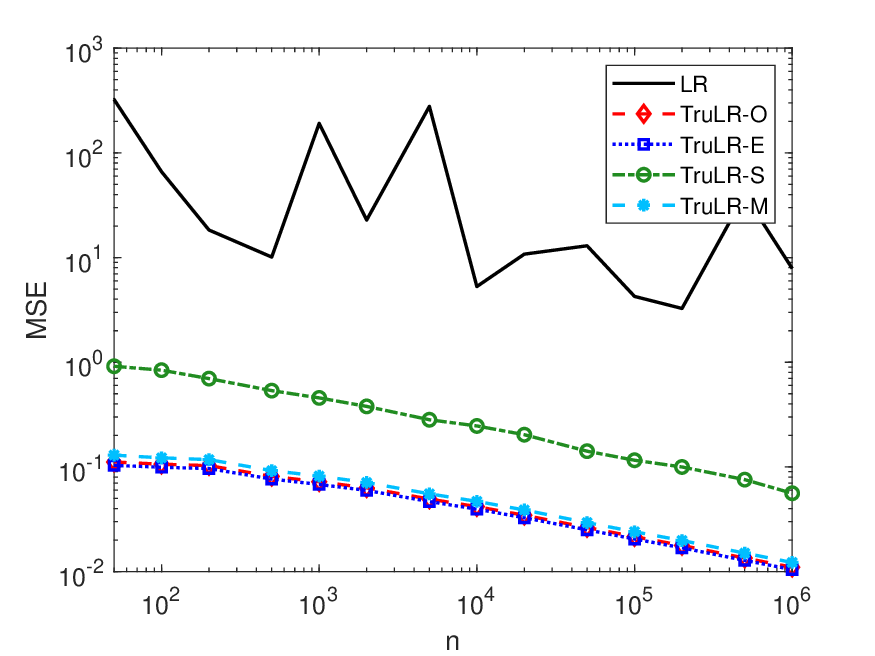}
\end{minipage}
\begin{minipage}[t]{0.48\linewidth}
\centering
\includegraphics[width=8.2cm]{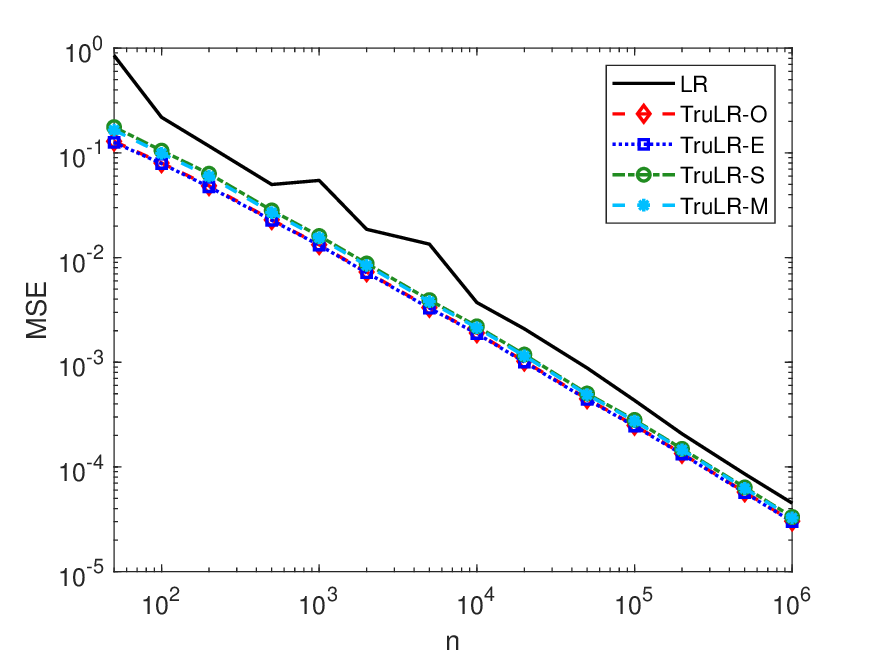}
\end{minipage}
\caption{\baselineskip10pt MSEs of LR and TruLR estimators for normal distribution. Left graph is for case (i) $(\mu_0,\sigma_0)=(0,1.7)$, $(\mu,\sigma)=(0.2,4)$, $\alpha=1.2$, $p=40$; right graph is for case (ii) $(\mu_0,\sigma_0)=(1,1.5)$, $(\mu,\sigma)=(0.6,2)$, $\alpha=2.1$, $p=4$. Each MSE is computed based on $5000$ independent replications ($\delta=0.01$).}
\label{fig:normal_all}
\end{figure}

We can use the Bernstein's constant $b_h$ founded in \ref{appx:para} to construct another TruLR estimator (denoted as ``TruLR-E''). Under the same settings as in \ref{test:normal}, the MSEs of all the estimators are shown in Figure \ref{fig:normal_all}.

We find that the TruLR-O and TruLR-E estimators perform nearly identically in both settings. For example, when $n=50000$, the MSEs of the TruLR-O and TruLR-E are 0.026 and 0.025, respectively in setting (i), are 0.00045 and 0.00044, respectively in setting (ii). Recall that we take Bernstein's constant $b_h=\sigma_0$ for TruLR-O estimators, that is 1.7 in setting (i) and 1.5 in setting (ii), while we use the estimated Bernstein's constant $b_h$ for TruLR-E estimators, i.e., 2.6850 in setting (i) and 2.8888 in setting (ii). This demonstrates that our truncation boundary method is not sensitive to the choice of Bernstein's constant $b_h$ .

\begin{figure}[htpb]
\begin{minipage}[t]{0.48\linewidth}
\centering
\includegraphics[width=8.2cm]{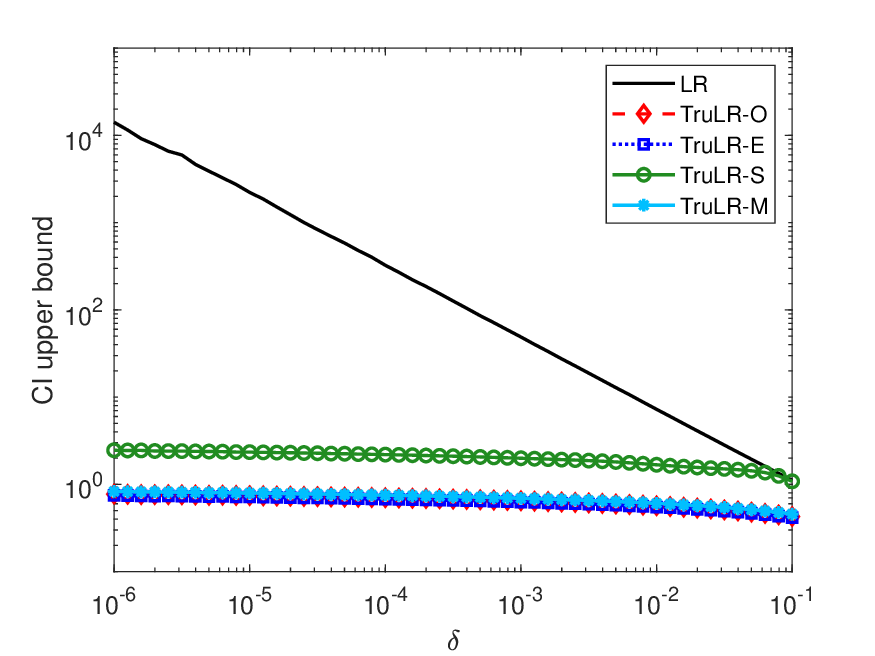}
\end{minipage}
\begin{minipage}[t]{0.48\linewidth}
\centering
\includegraphics[width=8.2cm]{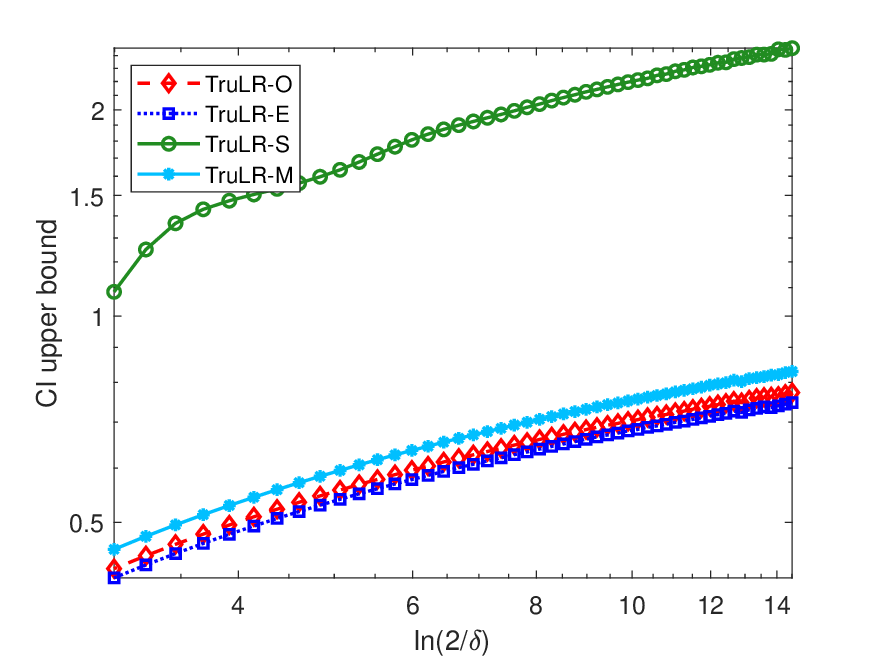}
\end{minipage}
\caption{\baselineskip10pt Error upper bounds for LR and TruLR estimators in Case (i) of the normal distribution: $(\mu_0,\sigma_0)=(0,1.7)$, $(\mu,\sigma)=(0.2,4)$, $\alpha=1.2$, $p=40$, $n=5000$. Each upper bound is computed based on $10^8$ independent replications. The left graph is the log-log plot of the error upper bound versus $\delta$, whereas the right graph is the log-log plot of the error upper bound versus $\ln(2/\delta)$.}
\label{fig:norm_CI_all}
\end{figure}

Figure \ref{fig:norm_CI_all} shows the concentration upper bounds, from which we also find that the TruLR-O and TruLR-E estimators perform similarly.

\section{Synthetic Example for Chi-Squared Distribution}\label{appx:chi2}

\begin{figure}[htpb]
\begin{minipage}[t]{0.48\linewidth}
\centering
\includegraphics[width=8.2cm]{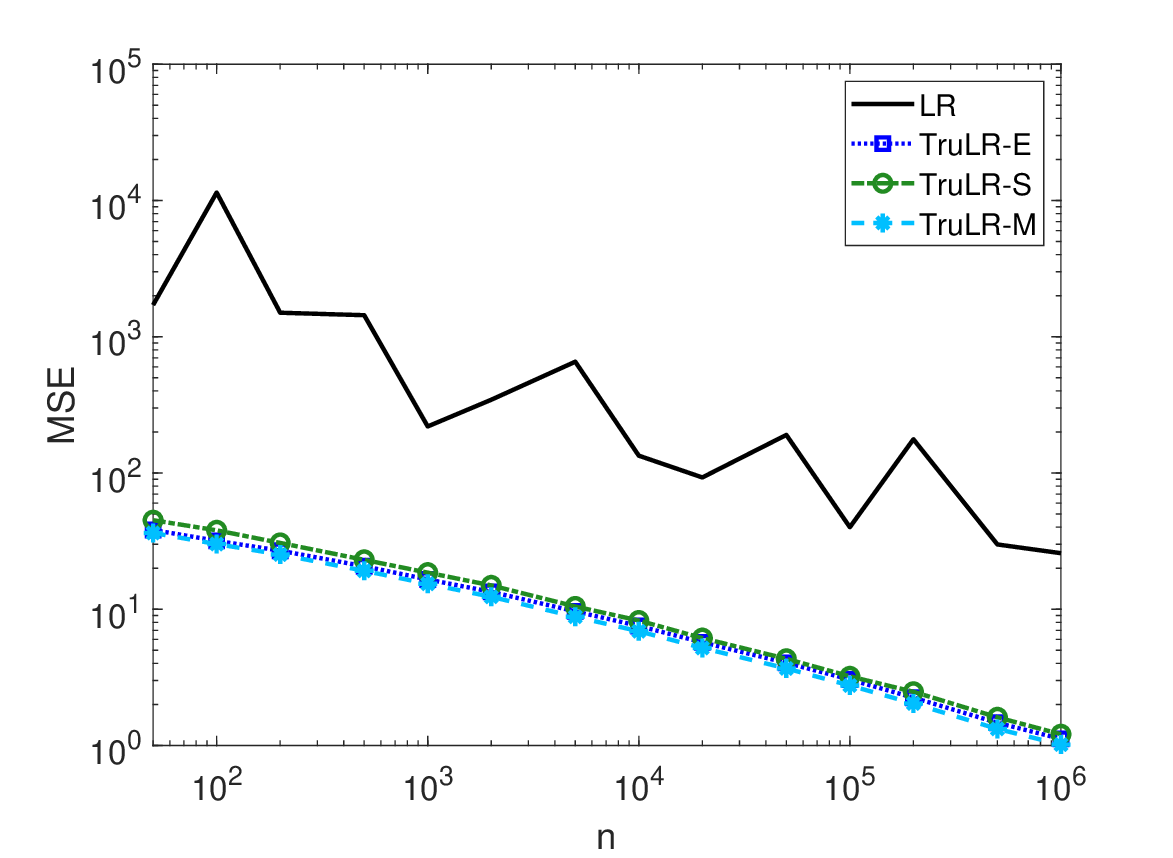}
\end{minipage}
\begin{minipage}[t]{0.48\linewidth}
\centering
\includegraphics[width=8.2cm]{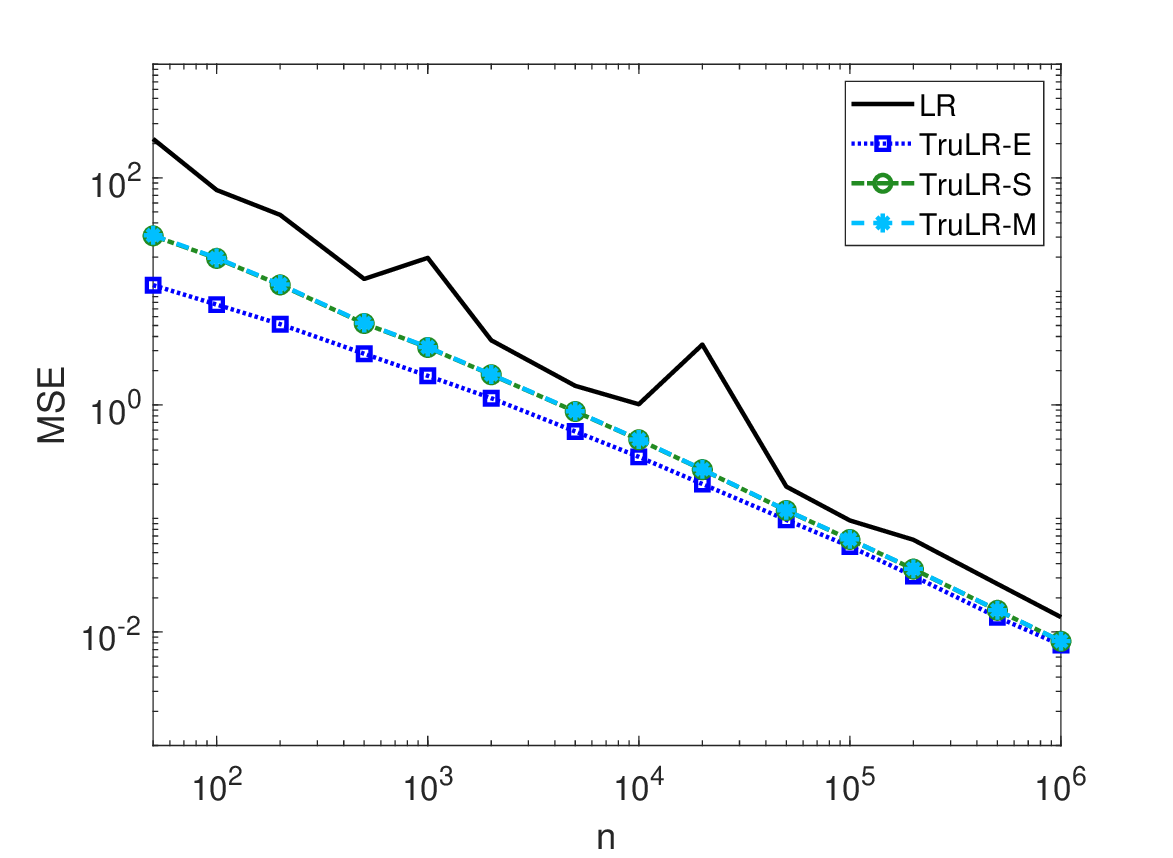}
\end{minipage}
\caption{\baselineskip10pt MSEs of LR and TruLR estimators for chi-squared distribution. Left graph is for case (i) $k_0=12$, $k=3$, $\alpha=1.3$, $p=40$; right graph is for case (ii) $k_0=18$, $k=10$, $\alpha=2.2$, $p=4$. Each MSE is estimated based on $5000$ independent replications ($\delta=0.01$).}
\label{fig:chisq}
\end{figure}

\begin{figure}[htpb]
\begin{minipage}[t]{0.48\linewidth}
\centering
\includegraphics[width=8.2cm]{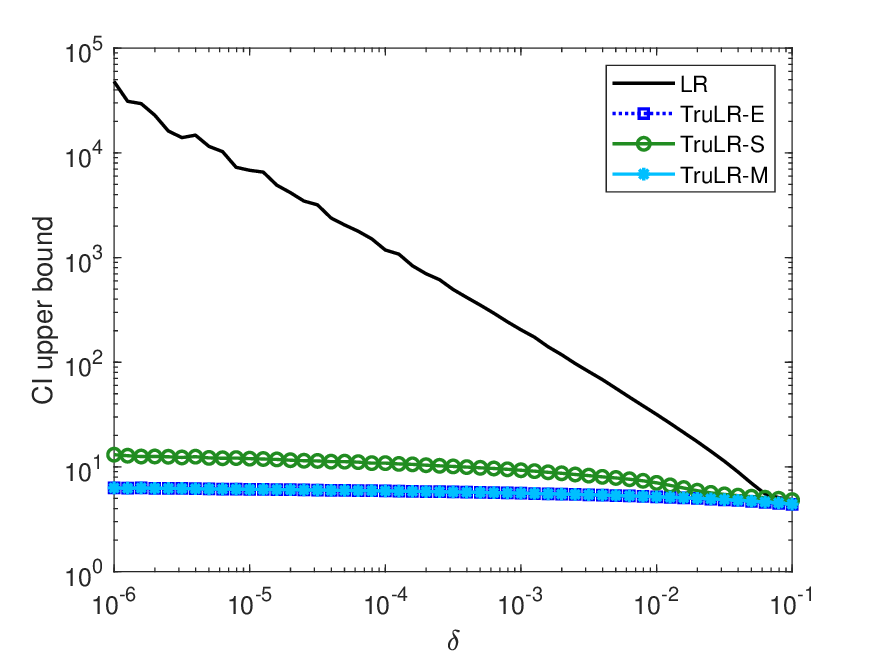}
\end{minipage}
\begin{minipage}[t]{0.48\linewidth}
\centering
\includegraphics[width=8.2cm]{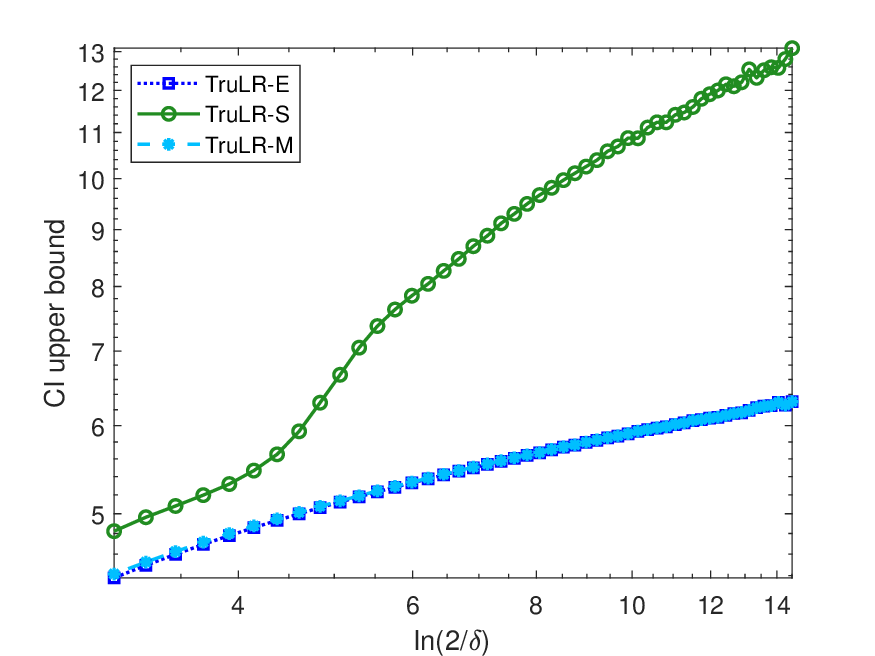}
\end{minipage}
\caption{\baselineskip10pt Error upper bounds for LR and TruLR estimators in Case (i) of the chi-squared distribution: $k_0=12$, $k=3$, $\alpha=1.3$, $p=40$, $n=5000$. Each upper bound is estimated based on $10^7$ independent replications. The left graph is the log-log plot of the error upper bound versus $\delta$, whereas the right graph is the log-log plot of the error upper bound versus $\ln(2/\delta)$.}
\label{fig:chisq_CI}
\end{figure}

Here we consider a synthetic example for chi-squared random variable $X \sim \chi^2(k)$ with performance function $h(X)=X$.
Similar to the normal distributed random variable, its infinite norm does not exist, but Assumptions \ref{assump-MGF} and Definition \ref{def:bernstein} are satisfied.
Let $\btheta=k$, and we can calculate that
$I_{\alpha}(\mathds{P}_{\betheta} \| \mathds{P}_{\betheta_0}) =  \left( \Gamma(k_0/2)/\Gamma(k/2) \right)^{\alpha-1}\left( \Gamma(k_{\alpha}/2)/\Gamma(k/2) \right) $, where $k_\alpha = (1-\alpha)k_0 + \alpha k>0$, and the role of $\alpha$ is the analogous to the previous two examples, and we consider two settings: (i) $k_0=12$, $k=3$, $\alpha=1.3$, $p=40$; (ii) $k_0=18$, $k=10$, $\alpha=2.2$, $p=4$. Similar to the normal distribution example, the choice of $p$ needs to satisfy the constraint relationship between $\alpha$ and $p$ in Theorem \ref{theorem-TR-bound-moment-concentration-min1} and Corollary \ref{theorem-TR-bound-moment-concentration-MGF-min2}. {Although the chi-squared random variable satisfies the Bernstein's condition, here we still use the value estimated by the approximation method proposed in \ref{appx:para} and substitute it for the Bernstein's constant $b_h$ to illustrate  the effectiveness of our approach}, and such a truncation boundary method is also denoted as ``TruLR-E''. We vary the number of samples $n$ from $50$ to $10^6$, and the MSEs of these estimators are shown in Figure \ref{fig:chisq}. In addition, we vary $\delta$ from $10^{-1}$ to $10^{-6}$ in Case (i), and the CI upper bounds of the LR estimator and TruLR estimators are shown in Figure \ref{fig:chisq_CI}.

From these figures, we observe results similar to those in the first and second examples: When the nominal $\btheta_0$ and the target $\btheta$ are far apart, the accuracy of the TruLR method is much better than that of the LR method. However, when $\btheta_0$ and $\btheta$ are close, the improvement in TruLR is not significant. In this example, the TruLR-E estimator remains effective and performs comparably to the TruLR-S and TruLR-M estimators, or even slightly better. In Figure \ref{fig:chisq_CI}, the error upper bounds of the LR estimator appear as a straight line in the log-log plot against $\delta$, while those of the TruLR estimators show linear relationship in the log-log plot against $\ln(2/\delta)$, which shows that the LR estimator and TruLR estimators satisfy the polynomial concentration and exponential concentration, respectively.

\section{Problem Setting for Portfolio Pricing}\label{appx:port_pricing}
We first provide the details for TruBLR estimators.
Recall that $X^{(i)}_{m}=\ln(S_i(t_m))$, $\bX^{(i)} = (X^{(i)}_{1},X^{(i)}_{2},\cdots,X^{(i)}_{M})^{\top}$, and $\btheta^{(i)}=(S_i(0), \sigma_i)^{\top}$. Then, $\bX^{(i)}$ follows a $M$-dimensional normal distribution $ \mathcal{N}(\bmu^{(i)},\bSigma^{(i)})$, where $\bmu^{(i)}$ and $\bSigma^{(i)}$ are determined by parameters $\btheta^{(i)}$ with
\begin{equation}\label{eq-portfolio-X}
  \bmu^{(i)} = \ln S_i(0)+(r-\frac{1}{2}\sigma_i^2)\Delta t \ba ~~\text{and}~~\bSigma^{(i)} = \sigma_i^2 \Delta t \bA,
\end{equation}
where
\begin{equation}\label{eq-portfolio-X-c}
  \ba =  \left(\begin{array}{c}
                                                                              1 \\
                                                                              2 \\
                                                                              \vdots \\
                                                                              M
                                                                            \end{array}\right)~~\text{and}~~
  \bA =\left(
                              \begin{array}{cccc}
                                1 & 1 & \cdots & 1 \\
                                1 & 2 & \cdots & 2 \\
                                \vdots & \vdots & \ddots & \vdots \\
                                1 & 2 & \cdots & M \\
                              \end{array}
                            \right)  .
\end{equation}

Notice that we have recalibrated parameters of input random variables once a week, and this process continues for $k$ weeks. Denote these parameters for the $i$th Asian option at the $j$th week as $\btheta_j^{(i)}$, and the simulation data collected at the $j$th week for the $i$th Asian option are denoted by $\{\tilde{\bX}^{(i)}_{j,s}\}_{s=1}^{n}$, where $n$ is the sample size of the historical dataset and $j=0,1,\ldots,k-1$.
The parameter $\btheta_j^{(i)}$ determines the corresponding multi-dimensional normal distribution $\mathcal{N}(\bmu_j^{(i)}, \bSigma_j^{(i)})$ by equation \eqref{eq-portfolio-X}.
Notice that the subscript ``$j$'' mean the $j$th week's data and the superscript ``$(i)$'' means the $i$th Asian option.
Our goal is to estimate the price for the $i$th Asian option at the $k$th week, i.e., we want to estimate $v^{(i)}_{k}(\betheta^{(i)}_{k})$. For simplicity, we denote $\betheta^{(i)}_{k}$ as $\betheta^{(i)}$.

Based on the historical data, the LR estimator is given by 
\begin{equation}\label{eq-portfolio-ILR}
 \bar v^{\LR,(i)}_n = \frac{1}{k}\sum_{j=0}^{k-1} \frac{1}{n}\sum_{s=1}^{n}h^{(i)}\left(\tilde{\bX}^{(i)}_{j,s}\right)\l\left(\tilde{\bX}^{(i)}_{j,s}\right),
\end{equation}
where $\l(\tilde{\bX}^{(i)}_{j,s}) = {f_{\betheta^{(i)}}(\tilde{\bX}^{(i)}_{j,s})}/{f_{\betheta_j^{(i)}}(\tilde{\bX}^{(i)}_{j,s})}$ is the likelihood ratio, and
\begin{equation}\label{eq-portfolio-pdf}
  f_{\betheta_j^{(i)}}(\bx) = \frac{1}{(2\pi)^{M/2} \det(\bSigma_j^{(i)})} \exp\left(-\frac{(\bx-\bmu_j^{(i)})^{\top}(\bSigma_j^{(i)})^{-1}(\bx-\bmu_j^{(i)})}{2}\right), \quad j=0,1,\cdots, k-1.
\end{equation}

Notice that the performance function $h^{(i)}(\bX^{(i)})$ given by \eqref{eq-portfolio-performance} has any moments and satisfies Definition \ref{def:p-norm}, however it does not satisfy the moment generating function condition (Assumption \ref{assump-MGF}), and we cannot directly use our truncation method for likelihood ratio.
However, we can use an approximation approach. Notice that $h^{(i)}(\bX^{(i)})$ can also be expressed as
\begin{equation}\label{eq-portfolio-performance-1}
  h^{(i)}(\bX^{(i)}) = e^{-rT} \max\left(\frac{K_i}{M}  \sum_{m=1}^M \left(e^{X^{(i)}_{m}-\ln K_i} - 1\right), 0\right), \quad i=1,2,\cdots, N.
\end{equation}
If the price of underlying asset $S_i(t_m)$ is near the strike price $K_i$, then $X^{(i)}_{m}-\ln K_i$ is near zero, thus we expect that a Taylor series expansion could approximate the original performance function reasonably well. We take the following expansion,
\begin{equation}\label{eq-portfolio-performance-2}
   h^{(i)}(\bX^{(i)}) \approx e^{-rT} \max\left(\frac{K_i}{M}  \sum_{m=1}^M \left((X_{m}^{(i)}-\ln K_i )+ \frac{1}{2}(X_{m}^{(i)}-\ln K_i )^2 \right), 0\right), \quad i=1,2,\cdots, N.
\end{equation}
This approximation is heuristic, but from the simulation examples as we will show, this method gives effective estimators. Notice that the right-hand side in \eqref{eq-portfolio-performance-2} is bounded by a polynomial of $\bX^{(i)}$, and satisfies the moment generating function condition (Assumption \ref{assump-MGF}). So we can use this expansion to determine the optimal truncation boundary for the likelihood ratio in Theorem \ref{theorem-TR-bound-moment-concentration-MGF-min1}.

In addition, we can calculate that
\begin{equation}\label{eq-portfolio-I}
  I_{\alpha}\left(\mathds{P}_{\betheta^{(i)}} \| \mathds{P}_{\betheta^{(i)}_j}\right) =  e^{\frac{1}{2} \alpha(\alpha-1)\left(\bemu^{(i)}-\bemu^{(i)}_j\right)^{\top}\left(\bSigma^{(i)}_{\alpha}\right)^{-1}\left(\bemu^{(i)}-\bemu^{(i)}_j\right)}
   \frac{\det\left(\bSigma^{(i)}\right)^{(1-\alpha)/2}\det\left(\bSigma^{(i)}_j\right)^{\alpha/2}}{\det\left(\bSigma^{(i)}_{\alpha}\right)^{1/2}},
\end{equation}
where $\bSigma^{(i)}_{\alpha} = (1-\alpha)\bSigma^{(i)} + \alpha\bSigma^{(i)}_j$ is positive.
In our framework, $\bSigma^{(i)}_{\alpha}$ is positive is equivalent to $(1-\alpha)\sigma_{i}^2 + \alpha\sigma_{i,j}^2$ is positive, where $\sigma_{i}$ and $\sigma_{i,j}$ are the volatilities of underlying asset $S_i(t)$ under measure $\mathds{P}_{\betheta^{(i)}}$ and measure $\mathds{P}_{\betheta_j^{(i)}}$, respectively.


The initial value and volatilities of underlying assets are shown in Table \ref{Tab:para-AsianPort}. We collect the historical simulation data under the parameters of input random variables, $\btheta_0^{(i)}$, $\btheta_1^{(i)}$, $\btheta_2^{(i)}$, and $\btheta_3^{(i)}$, which are shown in Table \ref{Tab:para-AsianPort}. The $\alpha_i$ in Table \ref{Tab:para-AsianPort} from the first to the fourth columns are picked up by condition $(1-\alpha)\sigma_{i}^2 + \alpha\sigma_{i,j}^2>0, j=0,1,2,3$ to make the divergence $I_{\alpha}(\mathds{P}_{\betheta^{(i)}} \| \mathds{P}_{\betheta_j^{(i)}}) $ convergent, which is applied for all TruLR estimators.


\begin{table}[htpb!]
\scriptsize
\centering
		\caption{\baselineskip10pt Mean and standard error, ratio of MSEs of the LR and TruLR estimators for pricing portfolio of Asian options (based on $100000$ independent replications). Other parameters are reported in Table \ref{Tab:para-AsianPort}. ``ratio'' means the ratio of the MSE of LR estimator to each TruLR estimator.}\label{Tab:Asian}
\begin{tabular}{lcrrrrrrrrrr}
\toprule
\multicolumn{2}{c}{$n$}         & 500      & 1000      & 2000      & 5000      & 10000      & 20000      & 50000      & 100000      & 200000      & 500000     \\
\midrule
\multirow{2}{*}{LR}    & mean & 19929.21 & 124485.43 & 22532.22 & 172197.09 & 186084.51 & 4196.15 & 1656536.11 & 6998.01 & 9790.20 & 1750.74  \\
                       & std err & 7818.36 	&109486.89	&14744.51	&168366.55	&128695.65	&1809.35 	&1648822.13	&4697.53 	&4781.47	&568.10 \\
                       \hline
\multirow{3}{*}{TruLR-M} & mean & 249.92 	&202.80 	&166.56 	&125.38	&101.46 	&80.55 	&61.06 	&49.90 	&41.92 	&34.03  \\
                       & std err  & 2.63 	&1.52 	&0.97 	&0.53 	&0.40 	&0.27	&0.20 	&0.14  &0.12	&0.09 \\
                       &ratio & 79.74 	&613.84 	&135.28 	&1373.36 	&1834.12	&52.09	&21731.60 	&140.25 	&233.53 	&51.44\\
                       \hline
\multirow{3}{*}{TruLR-S} & mean & 339.54 	&283.99	&244.85 	&211.13 	&194.86 	&147.95 	&121.75 	&113.89	&89.19	&86.78  \\
                       & std err & 14.61 	&12.27 	&8.37 	&8.60	&11.92 	&5.39 	&5.20 	&6.39 	&3.66	&9.28 \\
                       &ratio &58.70	&438.34 	&92.02 	&815.60 	&954.97	&28.36 	&13606.23 	&61.44 	&109.77 	&20.17  \\
                       \bottomrule
\end{tabular}
\end{table}

\end{document}